\documentclass[prd,twocolumn,showpacs,preprintnumbers,amsmath,amssymb,nofootinbib]{revtex4-1}

\usepackage{graphicx}
\usepackage{dcolumn}
\usepackage{xspace}
\usepackage[utf8]{inputenc}
\usepackage{float}
\usepackage{subfigure}
\usepackage{url}

\usepackage{color}
\usepackage{comment}

\newcommand{\eq}[1]{Eq.~(\ref{#1})}

\newcommand{\bbc}{\text{BR}(B_c\to \tau\nu)}

\newcommand{\real}{\mathrm{Re}\,}

\newcommand{\gev}{\mbox{GeV}}

\newcommand{\dline}[1]{{\parbox{3.5em}{\rule{0cm}{2.3ex}{#1}\strut}}}

\definecolor{BlueViolet}{rgb}{0.2, 0.00, 0.7}
\definecolor{Blue}{rgb}{0.15, 0.00, 0.9}
\definecolor{halayaube}{rgb}{0.4, 0.22, 0.33}
\definecolor{sanddune}{rgb}{0.59, 0.44, 0.09}
\usepackage[colorlinks=true,linkcolor=Blue,citecolor=Blue,urlcolor=BlueViolet,hyperfootnotes=false]{hyperref}

\begin{document}

\preprint{PSI-PR--18--16}
\preprint{TTP--18--42}
\title{\boldmath Impact of polarization observables and $ B_c\to \tau \nu$
  on new physics explanations of the $b\to c \tau \nu$ anomaly}

\author{Monika Blanke}
\email{monika.blanke@kit.edu}
\affiliation{Institut f\"ur Kernphysik (IKP), Karlsruher
  Institut f\"ur Technologie (KIT),
  76021 Karlsruhe, Germany\\
 Institut f\"ur Theoretische Teilchenphysik (TTP), Karlsruher
  Institut f\"ur Technologie (KIT),
  76131 Karlsruhe, Germany}

\author{Andreas Crivellin}
\email{andreas.crivellin@cern.ch}
\affiliation{Paul Scherrer Institut, CH--5232 Villigen PSI, Switzerland}

\author{Stefan de Boer}
\email{stefan.boer@kit.edu}
\affiliation{Institut f\"ur Theoretische Teilchenphysik (TTP), Karlsruher
  Institut f\"ur Technologie (KIT), 76131 Karlsruhe, Germany}

\author{Teppei Kitahara}
\email{teppeik@kmi.nagoya-u.ac.jp}
\affiliation{Institute for Advanced Research, Nagoya University, Furo-cho Chikusa-ku, Nagoya, 464-8602 Japan}
\affiliation{Kobayashi-Maskawa Institute for the Origin of Particles and the Universe, Nagoya University, Nagoya 464-8602, Japan}
\affiliation{Physics Department, Technion--Israel Institute of Technology, Haifa 3200003, Israel}

\author{Marta Moscati}
\email{marta.moscati@kit.edu}
\affiliation{Institut f\"ur Theoretische Teilchenphysik (TTP), Karlsruher Institut f\"ur Technologie (KIT), 76131 Karlsruhe, Germany}

\author{Ulrich Nierste}
\email{ulrich.nierste@kit.edu}
\affiliation{Institut f\"ur Theoretische Teilchenphysik (TTP), Karlsruher Institut f\"ur Technologie (KIT), 76131 Karlsruhe, Germany}
\author{Ivan Ni\v{s}and\v{z}i\'c}
\email{ivan.nisandzic@kit.edu}
\affiliation{Institut f\"ur Theoretische Teilchenphysik (TTP), Karlsruher Institut f\"ur Technologie (KIT), 76131 Karlsruhe, Germany}

\date{November 23, 2018}

\begin{abstract}
The combined analysis of the BaBar, Belle, and LHCb data on
$B\to D\tau\nu$, $B\to D^*\tau\nu$ and $B_c\to J/\Psi\tau\nu$ decay
observables shows evidence of physics beyond the Standard Model
(SM). In this article, we study all the one- and two-dimensional
scenarios which can be generated by adding a single new particle to the SM. We put special emphasis on the model-discriminating power of $F_L(D^*)$ and of the $\tau$ polarizations, and especially  on the constraint from the branching fraction ${\rm BR}(B_c\to\tau\nu)$. We  critically review this constraint and do not support the aggressive limit of ${\rm BR}(B_c\to\tau\nu)<10\%$ used in some analyses. While the impact of $F_L(D^*)$ is currently still limited, the ${\rm BR}(B_c\to\tau\nu)$ constraint has a  significant impact: depending on whether one uses a  limit of $60\%$, $30\%$ or $10\%$, the pull for new physics (NP) in scalar operators  changes drastically. More specifically, for a conservative $60\%$ limit a scenario  with scalar operators gives the best fit to data, while for an aggressive $10\%$ limit this scenario is strongly disfavored and the best fit is obtained in a scenario in which only a left-handed vector operator is generated. We find a sum rule for the  branching ratios of $B\to D\tau\nu$, $B\to D^*\tau\nu$ and $\Lambda_b\to \Lambda_c\tau\nu$ which holds for any NP contribution to the Wilson coefficients. This sum rule entails an enhancement of ${\rm BR}(\Lambda_b\to \Lambda_c\tau\nu)$ over its SM prediction by $(24\pm 6)\%$ for the current $\mathcal{R}(D^{(*)})$ data.
\end{abstract}
\maketitle

\vspace{-1mm}
\maketitle
\renewcommand{\thefootnote}{\#\arabic{footnote}}
\setcounter{footnote}{0}
\section{\label{sec:level1}Introduction}

Low-energy precision flavor observables probe new physics (NP) in a complementary way to direct searches for new particles at high energies.  In this respect, tauonic $B$ meson decays are an excellent window into NP: in combination with the well-studied $B$ decays to light leptons ($\ell=\mu,e$) they test lepton flavor universality (LFU). Within the Standard Model (SM), LFU is only broken by the small Higgs Yukawa interactions and it manifests itself (to a very good approximation) only via the masses entering  the phase space of the different decay modes.

The theory predictions for the individual semileptonic decay rates suffer from hadronic uncertainties related to the form factors and from parametric uncertainties stemming
from the errors in the CKM elements (e.g., see
Refs.~\cite{Ricciardi:2016pmh,Grinstein:2016xpg,DeFazio:2017spv} for recent reviews). However, in normalizing the branching ratios
${\rm BR}(B\to D^{(*)} \tau\nu)$ to ${\rm BR}(B\to D^{(*)} \ell \nu)$,  $\ell=\mu,\,e$, and analogously also their counterparts for other $b$-flavored hadrons,
\begin{align}
\begin{aligned} {\cal R}(D^{(*)})\,&\equiv\,{
\rm BR}(B\to D^{(*)} \tau
  \nu)/{
\rm BR}(B\to D^{(*)} \ell \nu)\,,\\
{\cal R}(J/\Psi)\,&\equiv\,{
\rm BR}(B_c\to J/\Psi \tau
  \nu)/{
\rm BR}(B_c\to J/\Psi \ell \nu)\,,\\
  {\cal R}(\Lambda_c)\,&\equiv\,{\rm BR}(\Lambda_b \to\Lambda_c \tau\nu)/{\rm BR}(\Lambda_b \to\Lambda_c \ell\nu)\,,
  \end{aligned}
\end{align}
the dependence on the CKM elements drops out
and the uncertainties originating from the form factors are significantly reduced~\cite{Nierste:2008qe,Kamenik:2008tj,Fajfer:2012vx,Alonso:2017ktd}.

Experimentally, the BaBar collaboration performed an analysis of ${\cal R}(D)$ and ${\cal R}(D^*)$ using the full available data set
\cite{Lees:2012xj,Lees:2013uzd}. The same ratios were also measured by the Belle collaboration~\cite{Huschle:2015rga,Sato:2016svk,Hirose:2016wfn,Hirose:2017dxl},
while the LHCb collaboration has measured
${\cal R}(D^*)$~\cite{Aaij:2015yra,Aaij:2017uff,Aaij:2017deq}. Combining these data, the HFLAV collaboration~\cite{Amhis:2016xyh} determines the ratios
\begin{eqnarray}
\begin{aligned}
{\cal R}(D)\,=\,0.407\pm0.039\pm0.024  \,, \label{rdhflav}\\
{\cal R}(D^*)\,=\,0.306\pm0.013\pm0.007\,. 
\end{aligned}
\end{eqnarray}
Here, the first error is statistical and the second one is systematic. Comparing these measurements to the corresponding
SM predictions  \cite{Bigi:2016mdz,Bernlochner:2017jka,Bigi:2017jbd,Jaiswal:2017rve}
\begin{eqnarray}
\begin{aligned}
{\cal R}_{\rm SM}(D)\,=\,0.299\pm0.003 \,, \\
{\cal R}_{\rm SM}(D^*) \,=\,0.258\pm0.005 \,,
\end{aligned} 
\end{eqnarray}
reveals a tension at the level of
$3.8\,\sigma$~\cite{Amhis:2016xyh}.\footnote{Recent discussions of long-distance electromagnetic effects in ${\cal R}(D)$ can be found in Refs.~\cite{Becirevic:2009fy, deBoer:2018ipi}.}  This is also consistent with the previous evaluations of ${\cal R}(D)$ in Refs.~\cite{Nierste:2008qe, Kamenik:2008tj, Becirevic:2009fy, Lattice:2015rga,Na:2015kha} and of ${\cal R}(D^\ast)$ in Ref.~\cite{Fajfer:2012vx}.

The observed anomaly receives further support from the LHCb analysis of
${\cal R}(J/\Psi)$~\cite{Aaij:2017tyk} which also finds an experimental value
significantly above the SM prediction. Unfortunately, the relevant
form factors are poorly known in this case~\cite{Murphy:2018sqg,Cohen:2018dgz,Tran:2018kuv}. Hence we do not include this measurement in our analysis. For a discussion of NP effects in ${\cal R}(J/\psi)$, see Refs.~\cite{Watanabe:2017mip,Chauhan:2017uil}.

For later use we further quote the SM prediction for the ratio
${\cal R}(\Lambda_c)$~\cite{Detmold:2015aaa}:
\begin{equation}
	{\cal R}_{\rm SM}(\Lambda_c) = 0.33 \pm 0.01\,.
\end{equation} 

The Belle collaboration has measured the $\tau$ polarization asymmetry along the longitudinal directions of the $\tau$ lepton in $B\to D^\ast\tau\nu$, defined as
\begin{equation}
	P_\tau(D^\ast) =\frac{\Gamma(B \to D^{*} \tau^{\lambda=+1/2} \nu)-\Gamma(B \to D^{*} \tau^{\lambda=-1/2} \nu)}{\Gamma(B \to
  D^{*} \tau \nu)},
\end{equation}
where $\lambda$ denotes the $\tau$ helicity, obtaining \cite{Hirose:2016wfn,Hirose:2017dxl}
\begin{align}
P_\tau(D^\ast) = & -0.38\pm 0.51^{+0.21}_{-0.16} \label{eq:ptau}\,.
\end{align}
This observable turns out to be interesting for discriminating NP models, especially if the accuracy is improved in the future by the Belle II experiment. 
	
Recently, the Belle collaboration has also measured the longitudinal $D^{\ast}$ polarization in $B\to D^*\tau\nu$, defined as 
\begin{equation}
	F_{L}(D^{\ast})=\frac{\Gamma(B \to D^{*}_L \tau \nu)}{\Gamma(B \to
  D^{*} \tau \nu)}\,.
\end{equation}
Like the $\tau$ polarization, also the $D^{\ast}$ polarization can distinguish between different Lorentz structures; i.e., NP in scalar, tensor or vector operators affects the $D^{\ast}$ polarization in a complementary way to the overall rate. The preliminary Belle result is~\cite{Adamczyk}
\begin{align}
F_L(D^{\ast})= 0.60\pm 0.08 \pm 0.035\,, \label{flds}
\end{align}
which agrees with the SM prediction of
\begin{align}
F_{L,\,\textrm{SM}}(D^{\ast})=0.46 \pm 0.04\,,
\end{align}
at the $1.5\,\sigma$ level \cite{Alok:2016qyh}.  Nonetheless, this
result can still favor or disfavor specific NP scenarios.

Similarly, the $\tau$ polarization in $B\to D\tau\nu$ can provide information about the Lorentz structure of NP~\cite{Nierste:2008qe,Alonso:2017ktd}. However, $P_\tau(D)$ has not been measured yet. The reason for this is that the $\tau$ is reconstructed in decay modes with at least one neutrino, and the missing energy blurs the information on the $\tau$ momentum. One can deal with this problem by considering differential decay distributions involving only kinematic variables of the \emph{visible}\ final state particles, for instance the $D$ and $\pi$ energies, and the angle between the $D$ and $\pi$ tracks in the decay chain $B\to D \nu \tau[\to \pi \nu]$. These decay distributions have a high sensitivity to NP~\cite{Nierste:2008qe,Alonso:2017ktd}.

Furthermore, the $B_c$ lifetime has a significant impact on
possible NP solutions~\cite{Celis:2016azn,Alonso:2016oyd},
because it constrains the yet unmeasured branching ratio ${\rm BR}(B_c\to \tau \nu)$.
The lifetime measurement is very precise~\cite{Tanabashi:2018oca},
\begin{align}
\tau (B_c) = & (0.507\pm 0.009)\, \mbox{ps}\,, \label{eq:bcexp}
\end{align}
while a theory prediction is quite challenging (we will return to
this issue in detail later).

Even though many model independent analyses in this context have been
performed~\cite{Fajfer:2012jt,Sakaki:2012ft,Tanaka:2012nw,
  Becirevic:2012jf,Datta:2012qk,Duraisamy:2013kcw,Dutta:2013qaa,
  Duraisamy:2014sna,Sakaki:2014sea,Freytsis:2015qca, Alok:2016qyh,
  Alonso:2016gym,Bardhan:2016uhr,
  Bhattacharya:2016zcw,Celis:2016azn,Alok:2017qsi,Dutta:2017wpq,Azatov:2018knx,Bifani:2018zmi,Huang:2018nnq,Asadi:2018sym,Hu:2018veh,Feruglio:2018fxo,Angelescu:2018tyl,Iguro:2018fni,Bhattacharya:2018kig,Aebischer:2018iyb},
it is important to reconsider the situation in light of the recent
$F_L(D^*)$ measurement and to critically revise and examine the
treatment of the $B_c\to \tau \nu$ decay. Furthermore, we will highlight the future
potential of the polarization observables $F_L(D^*)$, $P_\tau(D^\ast)$,
and (the yet unmeasured) $P_\tau(D)$ to discriminate between different
scenarios of NP. We will also highlight the interplay among
$\mathcal{R}({D{^{(*)}}})$ and ${\cal R}(\Lambda_c)$, where
$\mathcal{R}(\Lambda_c)$ provides a consistency check of the
measurements.
  
The paper is organized as follows: In Sec.~\ref{sec:eft}, we fix our notation for the relevant effective Hamiltonian.  In Sec.~\ref{sec:obs}, we discuss theoretical and phenomenological aspects of ${\rm BR} (B_c \to \tau \nu ) $ and list compact analytic formulas for the considered observables. In Sec.~\ref{sec:sc}, we present our phenomenological studies in scenarios with one and two nonzero NP Wilson coefficients. The chosen scenarios correspond to the cases in which the NP coefficients are generated by the exchange of a single heavy spin-0 or spin-1 particle. {Section~\ref{sec:corr} is devoted to the study of correlations between the ratios ${\cal R}(D^{(*)})$ and ${\cal R}(\Lambda_c)$ and the polarization observables $F_L(D^*)$ and $P_\tau(D^{(*)})$. Finally, we conclude in Sec.~\ref{sec:conc}.}

\section{Effective Field Theory\label{sec:eft}}

We are interested in NP which is realized above the $B$ meson mass
scale. Especially in the case at hand, this is a reasonable assumption, since modifying a charged current obviously requires a new charged particle for which light masses are experimentally excluded. Therefore, we can integrate out the heavy degrees of freedom, and the SM as well as the NP physics contributions are parametrized by the effective
Hamiltonian
\begin{equation}
\renewcommand{\arraystretch}{1.8}
\begin{array}{r}
 {\cal H}_{\rm eff}=  2\sqrt{2} G_{F} V^{}_{cb} \big[(1+C_{V}^{L}) O_{V}^L +   C_{S}^{R} O_{S}^{R} 
 \\   +C_{S}^{L} O_{S}^L+   C_{T} O_{T}\big] \,,\quad\quad
\end{array}
\label{Heff}
\end{equation}
with
\begin{equation}
\renewcommand{\arraystretch}{1.8}
\begin{array}{l}
   O_{V}^L  = \left(\bar c\gamma ^{\mu } P_L b\right)  \left(\bar\tau \gamma_{\mu } P_L \nu_{\tau}\right)\,, \\ 
   O_{S}^R  = \left( \bar c P_R b \right) \left( \bar\tau P_L \nu_{\tau}\right)\,, \\
   O_{S}^L  = \left( \bar c P_L b \right) \left( \bar\tau P_L \nu_{\tau}\right)\,,   \\
   O_{T}  = \left( \bar c \sigma^{\mu\nu}P_L  b \right) \left( \bar\tau \sigma_{\mu\nu} P_L \nu_{\tau}\right)\,,   \\
\end{array}
\label{Oeff}
\end{equation}
where we assumed the absence of both (light) right-handed
neutrinos,\footnote{For studies of right-handed neutrino effects in
  ${\cal R}(D^{(*)})$, see~\cite{Iguro:2018qzf,Greljo:2018ogz,Robinson:2018gza,Azatov:2018kzb}.} and of NP couplings to the light lepton generations (as studied in Ref.~\cite{Jung:2018lfu}). Note that we have factored out the SM contribution such that all Wilson coefficients
$C_{S,V,T}^{L,R}$ originate from NP only. Here we do not include a
vector operator with a right-handed coupling to quarks, because such an operator (with the desired LFU violation) does not arise at the dimension-six level in the SU(2$)_L$-invariant effective theory~\cite{Buchmuller:1985jz,Grzadkowski:2010es,
  Aebischer:2015fzz}.

The Wilson coefficients in \eq{Heff} depend on the
renormalization scale. We will quote our results for the coefficients defined at the scale of the heavy NP particle, which we take as 1$\,$TeV. The coefficients at the scale $\mu=m_b$ are related to those defined at 1$\,$TeV as \cite{Gonzalez-Alonso:2017iyc}
\begin{align}
C_V^L (m_b)&=  C_V^L (1\,\mbox{TeV})\,, \label{wcrun}\\[1mm]
C_S^R(m_b) &= 1.737\, C_S^R (1\,\mbox{TeV}) \,,\nonumber \\[1mm]
  \left( \begin{array}{c}
      C_S^L(m_b) \\ C_T(m_b)           
   \end{array}\right) &=  \left( \begin{array}{rr}
           1.752 & -0.287 \\
           -0.004 & 0.842 
   \end{array}\right)
  \left( \begin{array}{c}
         C_S^L(1\,\mbox{TeV}) \\ C_T(1\,\mbox{TeV})    
   \end{array}\right) . \nonumber
\end{align}

\section{Observables\label{sec:obs}}
While the theory predictions for ${\cal R}(D^{(*)})$ in \eq{rdhflav} as well as the polarization observables like $F_L({D^{*}})$ in \eq{flds} are quite straightforward, the $B_c$ lifetime constraint in \eq{eq:bcexp} warrants some discussion. In principle, the decay width of $B_c\to \tau\nu$ places a powerful constraint on the scalar operators in \eq{Heff}. However, the branching ratio  ${
\rm BR}(B_c\to \tau\nu)$ has not been measured yet. Therefore,  one only has the option of  comparing the measured $B_c$ lifetime with the theoretical calculations of Refs.~\cite{Gershtein:1994jw,Bigi:1995fs,Beneke:1996xe,Chang:2000ac,Kiselev:2000pp}. In this way the authors of Ref.~\cite{Alonso:2016oyd} have set an upper limit of 30\% on the contribution from $B_c\to \tau \nu$ to the total $B_c$ decay width. Furthermore, the authors of Ref.~\cite{Akeroyd:2017mhr} even advocate that the NP contribution to $\bbc$ can be at most 10\%.

\subsection{\boldmath Constraints from $\bbc$}

For the estimate of $\bbc<10\%$ from Ref.~\cite{Akeroyd:2017mhr}, LEP data on a mixture of $B_c\to \tau \nu$ and $B^-\to \tau \nu$ decays (with $b$ quarks from $Z$ boson decays) are used as an input. In order to extract information on ${
\rm BR}(B_c\to \tau \nu)$ from these data one must know the probability $f_c$ that a $b$ quark hadronizes into a $B_c$ meson.  $f_c$ is a small number, of the order of $10^{-2}$ or less. In Ref.~\cite{Akeroyd:2017mhr} the ratio of the $b\to B_c$ and $b\to B_u$ fragmentation functions, $f_c/f_u$, is extracted from data accumulated at hadron colliders. As a first critical remark, we recall that fragmentation functions depend on the kinematics. In the case of the $b\to B_s$ and $b\to B_d$ fragmentation functions the LHCb collaboration indeed finds evidence for a decrease of $f_s/f_d$ with the transverse momentum $p_T$ of the $B_{d,s}$ meson~\cite{Aaij:2013qqa}.  The authors of Ref.~\cite{Akeroyd:2017mhr}  infer $f_c/f_u$ from an average of CMS and LHCb measurements of
  \begin{align}
  R\equiv & \frac{f_c}{f_u} \,
     \frac{{
\rm BR}(B_c^-\to J/\psi \pi^-)}{{
\rm BR}(B^- \to J/\psi  K^-)}  \label{eq:r}\,.
\end{align}    
The individual measurements are~\cite{Khachatryan:2014nfa,Aaij:2014ija,Amhis:2016xyh}
\begin{equation}
\begin{aligned}
R&= (4.8 \pm 0.5 \pm 0.6)\times 10^{-3}\;[{\rm CMS}]\,,\\
R&= (6.83 \pm 0.18 \pm 0.09)\times 10^{-3}\;[{\rm LHCb}]\,.
\end{aligned}
\end{equation}
Since CMS data are taken for $p_T> 15\, \gev$ while LHCb employs
$0< p_T< 20\, \gev$, the data seemingly support a decrease of $R$ and
thereby of $f_c/f_u$ with $p_T$, in qualitative agreement with the LHCb
finding for $f_s/f_d$. Furthermore, the $p$-$p$ collisions at CMS and
LHCb or $p$-$\bar p$ collisions at the Tevatron produce $B_c$ mesons
through mechanisms which have no counterpart in $Z$ decays: A prominent
production process at hadron machines involves a $\bar b$ quark from one
(anti)proton and a $c$ quark from the other one, i.e.\ mechanisms
involving heavy-quark parton distribution functions or gluon splittings
into heavy-quark pairs.  We therefore doubt that values for $f_c/f_u$
extracted from Tevatron and LHC data can directly be used for $Z$ peak
analyses.

Moreover, even the 30\% limit from Ref.~\cite{Alonso:2016oyd} has to
be taken with a grain of salt. Recall that the dominant contribution to
the $B_c$ decay rate comes from the decay of the charm quark within the
$B_c$ meson.  The applicability of the calculational method (expansion
in inverse powers of the heavy-quark masses combined with
nonrelativistic QCD) to this charm decay is not clear and the result
found in Ref.~\cite{Beneke:1996xe} exhibits a large dependence on the
value of the charm mass, which moreover is not well defined in a
leading-order QCD calculation. To constrain NP effects in the $B_c$
lifetime the upper bound of the SM prediction
$0.4\,\rm{ps} \leq \tau(B_c)\leq 0.7\,\rm{ps}$ \cite{Beneke:1996xe} is
relevant, because it corresponds to the smallest possible SM
contribution to the total $B_c$ decay width. Lowering the charm mass by
only 0.05 GeV below the value of 1.4 GeV used as the lower limit in
Ref.~\cite{Beneke:1996xe}, the allowed NP contribution to the total
$B_c$ width increases to 40\%. Taking into account all uncertainties the
assumption of up to 60\% room for NP in the $B_c$ decay width is not too
conservative. Therefore, we will show our results for three different
limits on the $B_c\to\tau\nu$ branching ratio: 10\%, 30\%, and 60\%.

\begin{widetext}
\subsection{Numerical formulas}

The observables of interest are given by
  \begin{eqnarray}
    {\cal R}(D) & \simeq &  {\cal R}_{\rm SM}(D)\Big\{\vert 1+C^L_V\vert^2+1.54 \,\real[(1+C^L_V)(C^{L\ast}_S+C^{R\ast}_S)]+1.09 \vert C^L_S+C^R_S\vert^2+1.04 \,\real[(1+C^L_V)C_T^\ast]\nonumber\\
                &&+0.75 \vert C_T\vert^2\Big\}\,, \label{eq:rd}  \\
    {\cal R}(D^*)&\simeq&{\cal R}_{\rm SM}(D^*)\Big\{\vert 1+C^L_V\vert^2 + 0.13\,\real[(1+C^L_V)(C^{R\ast}_S-C^{L\ast}_S)]+0.05\vert C^R_S-C^L_S\vert^2-5.0 \,\real[(1+C^L_V)C_T^\ast]\nonumber\\
                &&+16.27 \vert C_T\vert^2\Big\} \,,\label{eq:rds} \\
    P_\tau(D) & \simeq & \left(\frac{{\cal R}(D)}{{\cal R}_{\rm SM}(D)}\right)^{-1}\Big\{0.32 \vert 1+C^L_V\vert^2 + 1.54 \,\real[(1+C^L_V)(C^{L\ast}_S+C^{R\ast}_S)] + 1.09\vert C^{L}_S+C^{R}_S\vert^2 \nonumber\\
                &&-0.35 \,\real[(1+C^L_V)C^\ast_T]+0.05 \vert C_T\vert^2\Big\}\,,\\
    P_\tau(D^*) & \simeq &\left(\frac{{\cal R}(D^*)}{{\cal R}_{\rm SM}(D^*)}\right)^{-1}\Big\{-0.49 \vert1 + C_V^L\vert^2 + 0.13 \,\real[(1 + C_V^L) (C_S^{R\ast} - C_S^{L\ast})] + 0.05 \vert C_S^R - C_S^L\vert^2\nonumber\\
                &&+1.67 \,\real[(1 + C_V^L) C_T^\ast] + 0.93 \vert C_T\vert^2\Big\}\,,\\
F_L({D^\ast}) & \simeq & \left(\frac{{\cal R}(D^*)}{{\cal R}_{\rm SM}(D^*)}\right)^{-1}\Big\{0.46 \vert1 + C_V^L\vert^2 + 0.13 \,\real[(1 + C_V^L) (C_S^{R\ast} - C_S^{L\ast})] + 0.05 \vert C_S^R - C_S^L\vert^2\nonumber\\
&&- 1.98 \,\real[(1 + C_V^L) C_T^\ast] + 3.2 \vert C_T\vert^2\Big\}\,,\\
{\cal R}(\Lambda_c) & \simeq & {\cal R}_{\rm SM}(\Lambda_c)\Big\{\vert 1 + C_V^L\vert^2 + 0.34 \,\real[(1 + C_V^L) C_S^{L\ast}] + 0.50 \,\real[(1 + C_V^L) C_S^{R\ast}] + 0.53 \,\real[C_S^L C_S^{R\ast}]\nonumber\\
&&+0.33 (\vert C_S^L\vert^2 + \vert C_S^R\vert^2) - 3.10 \,\real[(1 +
    C_V^L) C_T^\ast] + 10.44 \vert C_T\vert^2\Big\}, \\
\bbc & \simeq & 0.02\bigg(\frac{f_{B_c}}{0.43\,\text{GeV}}\bigg)^2 \Big\vert 1+C_V^L + 4.3\,(C_S^R-C_S^L)\Big\vert ^2,
 \label{eq:rlc}
\end{eqnarray}
\end{widetext}
in terms of the Wilson coefficients defined at the low scale $\mu=m_b$.

\begin{figure*}[t]
	\begin{center}
		\includegraphics[width=0.9\textwidth, bb= 0 0 260 130]{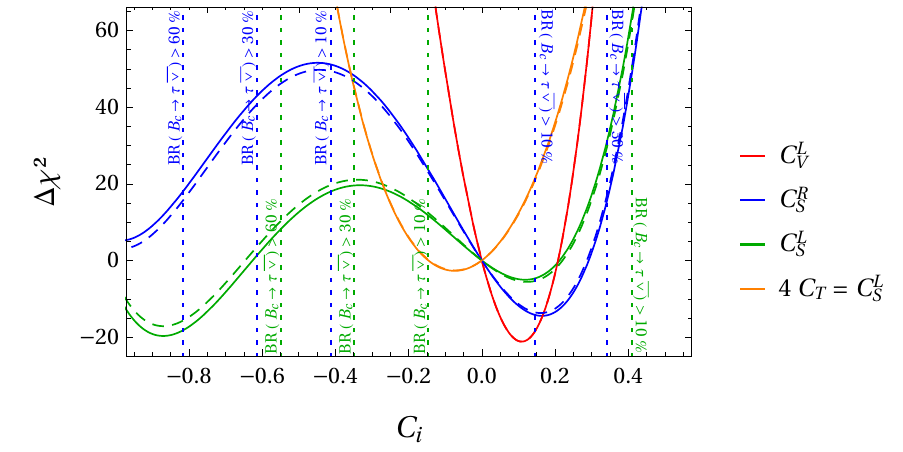}
	\end{center}
	\vspace{-0.5cm}
\caption{$\Delta \chi^2$ for the four one-dimensional scenarios in which only a single real Wilson coefficient (at the TeV scale) receives a NP contribution. The dashed lines show the situation before the $F_L({D^*})$ measurement, while the solid lines include the latter. The dotted vertical lines correspond to the limit on $C_S^{L,R}$ from ${\rm BR}(B_c\to \tau \nu)$ assuming a maximal value of $10\%$, $30\%$ and $60\%$ (i.e., the outer side of these lines is excluded by the corresponding constraint). Thus, only a 10\% limit on $\bbc$ can exclude the best-fit point for $C_S^R$ while for $C_S^L$ this point is always excluded and only positive values can provide a good fit to data.}
	\label{WCsingle}
\end{figure*}

The numerical coefficients correspond to the central values of the
form factors. Concerning our choice of the form factors, we use the average of Ref.~\cite{Aoki:2016frl} (obtained from two lattice QCD evaluations from Refs.~\cite{Lattice:2015rga, Na:2015kha}) for the vector and scalar form factors entering $B\to D$ transitions.  In the case of $B \to D^\ast$ we adopt the fit results from Ref.~\cite{Amhis:2016xyh} for $V, A_1, A_2$, while for $A_0$ we employ the result from Ref.~\cite{Bernlochner:2017jka} using $A_1$ from Ref.~\cite{Amhis:2016xyh} for the normalization. The tensor form factors for both decay processes are taken from Ref.~\cite{Bernlochner:2017jka}. We take the value for the $B_c$ meson decay constant, {$f_{B_c} = 0.427\,\text{GeV}$}, from Ref.~\cite{McNeile:2012qf}, {neglecting the small uncertainty}. Finally, the complete set of the baryonic form factors for $\Lambda_b \to \Lambda_c\tau\nu$ has recently been provided in Refs.~\cite{Detmold:2015aaa, Datta:2017aue}, see also Ref.~\cite{Bernlochner:2018kxh}.

\section{Analysis of different NP scenarios
\label{sec:sc}}

In our statistical analysis we follow the same approach as outlined in Ref.~\cite{Descotes-Genon:2015uva}, with a further caveat regarding the $\bbc$ constraint (to be discussed below). We build the $\chi^2$ function as
\begin{equation}
	\chi^2(C_k) = \sum_{ij}^{N_\text{obs}} [{\mathcal O}^\text{exp}_i - {\mathcal O}^\text{th}_i(C_k)] \mathcal C^{-1}_{ij} [{\mathcal O}^\text{exp}_j - {\mathcal O}^\text{th}_j(C_k)]\,,
	\label{eq:chi2}
\end{equation}
where $\mathcal O^{\textrm{exp(th)}}_i$ are the measured (predicted) observables and $C_k$ are the Wilson coefficients of the effective Hamiltonian in \eq{Heff}. In the covariance matrix $ \mathcal C$, the correlation of ${\cal R}(D)$ and ${\cal R}(D^*)$~\cite{Amhis:2016xyh} is taken into account. For $F_L({D^{*}})$ and $P_\tau (D^*)$ we add the statistical and systematic errors in quadrature.

The best-fit point is obtained by minimizing the $\chi^2$ function in the region of parameter space that is compatible with the $\bbc$ constraint. In other words, this constraint is imposed as a hard cut on the parameter space. For this reason, in the scenarios in which a best-fit point is compatible with the ${\rm BR}(B_c\to\tau\nu)<60\%$ constraint, but predicts $10\%<{\rm BR}(B_c\to\tau\nu)<60\%$, imposing the $10\%$ constraint moves the best-fit point to the boundary of the new allowed region in parameter space.

\begin{table*}[t]
	\begin{tabular}{|c||c|c|c|c|c||c|c|c|c|c|c|c|}\hline
1D hyp.   & best-fit & $1\,\sigma$ range & $2\,\sigma$ range & $p$-value (\%)
& pull$_{\rm SM}$  & ${\cal R}(D)$ & ${\cal R}(D^*)$& $F_L({D^{*}})$ & $P_\tau (D^*)$ & $P_\tau (D)$ & ${\cal R}(\Lambda_c) $\\
\hline\hline    		
$C_V^L$    &  0.11 &   [0.09, 0.13]  & [0.06, 0.15] &35&  4.6   &
\dline{0.371  $-0.8\,\sigma$} &  \dline{0.312  +0.4$\,\sigma$}    &  \dline{0.46  $-1.6\,\sigma$}   & \dline{$-0.49$  $-0.2\,\sigma$}& \dline{0.32 \\}&\dline{0.40 \\}\\
\hline 		
$C_S^R|_{10\%}$    & 0.15  &   [0.13, 0.15]   &  [0.08, 0.15]  & 1.7 &  3.8 & \dline{0.440 $+0.7\,\sigma$} & \dline{0.263 $-2.8\,\sigma$}& \dline{0.48 $-1.4\,\sigma$}& \dline{$-0.44$  $-0.1\,\sigma$}& \dline{0.53 \\}&\dline{0.38 \\}\\
\hline  
$C_S^R|_{30\%,60\%}$    & 0.16  &   [0.13, 0.20]   &  [0.08, 0.23]  & 1.8 &  3.8 & 
\dline{0.460  +1.2$\,\sigma$}& \dline{0.265 $-2.8\,\sigma$}& \dline{0.48 $-1.3\,\sigma$}& \dline{$-0.43$ $-0.1\,\sigma$}& \dline{0.55 \\}&\dline{0.39 \\}\\ \hline	
$C_S^L$    &  0.12 &  [0.07, 0.16]    &  [0.01, 0.20]  &0.02 &  2.2   &  \dline{0.412 $+0.1\,\sigma$}& \dline{0.247 $-4.0\,\sigma$}&   \dline{0.45 $-1.8\,\sigma$}  &\dline{$-0.53$ $-0.3\,\sigma$}& \dline{0.50 \\ }&\dline{0.36 \\ }\\
\hline
$C_S^L=4C_T$ & $-0.07$  &   [$-0.12$, $-0.03$]   &  [$-0.15$, 0.02]  & 0.01& 1.6    & 
\dline{0.242 $-3.6\,\sigma$}& \dline{0.280 $-1.7\,\sigma$}  & \dline{0.46 $-1.6\,\sigma$}     &\dline{$-0.45$ $-0.1\,\sigma$} & \dline{0.18 \\}&\dline{0.34 \\}\\
\hline
\end{tabular}
\caption{Fit results for the 1D hypotheses (hyp.) defined in Sec.~\ref{sec:1D} including all available data. The best-fit points and ranges for the Wilson coefficients are quoted for $\mu=1\,$TeV. Note that these results are independent of the choice of the three different limits on $\bbc$. The single exception is the $C_S^R$ scenario, for which the 10\% limit leads to a slightly worse fit than the other two. The last six columns show the predictions for the corresponding observable at the best-fit point.  For the quantities already measured we list the discrepancy (see Eq.~\eqref{discrepancy}) between the predicted and the experimental value (e.g. for $C_S^L$ the predicted value of ${\cal R}(D^*)=0.247$ at the best-fit point is $4.0\,\sigma$ below the measured value). Note that the predicted observables are at the same time included in the fit.}
	\label{tab:results1D}
\end{table*}
 
We quantify the goodness-of-fit as a $p$-value expressing the probability that the remaining differences between theory and experiment are due to statistical fluctuations. This probability corresponds to the one for a $\chi^2$-distributed random variable (having central values in the values predicted at the best-fit point) to reach a higher value than the one obtained from the data, assuming as number of degrees of freedom the difference between the number of observables included in the fit and the number of free parameters fitted. Namely,
\begin{equation}
	p\text{-value} = 1-\text{CDF}_{N_\text{obs}-N_\text{par}} (\chi^2_\text{min})\,,
\end{equation} 
where $\text{CDF}_{n}$ stands for the cumulative distribution function of a $\chi^2$-distributed random variable with $n$ degrees of freedom, $N_\text{obs}=4$ is the number of observables included in the fit, $N_\text{par}$ is the number of fitted parameters (i.e., $N^\text{1D}_\text{par}=1$, $N^\text{2D}_\text{par}=2$) and $\chi^2_\text{min}$ is the value of the $\chi^2$ at the best-fit point.

For the SM ($N_\text{par}=0$) the $p$-value is
\begin{equation}
	p\text{-value}_\text{SM} \sim 7\cdot10^{-5}\,,
\end{equation} 
which corresponds to a deviation of data at the $4\,\sigma$ level.

For each scenario, we perform a likelihood ratio test between the best-fit point and a generic point $x$ in parameter space under the assumption that the variables are normally distributed. This test quantifies how much the best-fit point is favored over the other points in the parameter space. In other words, the $s$-sigma intervals in the 1D and 2D scenarios to be studied correspond to the points $x_s$ in the parameter space such that  
\begin{equation}
   x_s: s(x_s) = \sqrt{\text{CDF}^{-1}_1(\text{CDF}_{N_\text{par}}(\chi^2(x_s) - \chi^2_\text{min}))}\,,
\end{equation}
where $N_\text{par}=1,2$ again stands for the number of fitted parameters. The likelihood ratio test between the best-fit point
and the SM, i.e., the SM-pull, is defined as the $p$-value
corresponding to $\chi^2_{\text{SM}} - \chi^2_{\text{min}}$, with
$\chi^2_{\text{SM}}=\chi^2(0)$, and is then expressed in terms of
standard deviations ($\sigma$).
 
The discrepancies of the measured observables in Tables~\ref{tab:results1D} and~\ref{tab:results2D} are defined as the difference between the predicted value at the best-fit point and data, expressed as multiples of the experimental error ($\sigma^{\mathcal O^\text{exp}_i}$), i.e.,
\begin{equation}
	\text{d}_{{\mathcal O}_i} =  \frac{\mathcal O^\text{NP}_i - \mathcal O^\text{exp}_i}{\sigma^{\mathcal O^\text{exp}_i}}\,.
	\label{discrepancy}
\end{equation}

\subsection{One-dimensional scenarios}\label{sec:1D}

In a first step, we consider one-dimensional scenarios (with real
Wilson coefficients) which can be generated by a single new particle added to the SM:

\begin{itemize}
\item {\boldmath $C_V^L$}: This setup arises in models with vector
  leptoquarks (LQs) like the SU(2$)_L$-singlet vector LQ of the
  Pati-Salam model
  ($U_1$)~\cite{Alonso:2015sja,Calibbi:2015kma,Fajfer:2015ycq,Barbieri:2015yvd,Barbieri:2016las,Hiller:2016kry,Bhattacharya:2016mcc,Buttazzo:2017ixm,Kumar:2018kmr,Assad:2017iib,DiLuzio:2017vat,Calibbi:2017qbu,Bordone:2017bld,Barbieri:2017tuq,Blanke:2018sro,Greljo:2018tuh,Bordone:2018nbg,Matsuzaki:2018jui,Crivellin:2018yvo,DiLuzio:2018zxy,Biswas:2018snp},
  the scalar SU(2$)_L$-triplet and/or scalar SU(2$)_L$-singlet
  LQ~\cite{Deshpande:2012rr,Tanaka:2012nw,Sakaki:2013bfa,Freytsis:2015qca,Bauer:2015knc,Cai:2017wry,Crivellin:2017zlb,Altmannshofer:2017poe,Marzocca:2018wcf}
  (with left-handed couplings only) or in models with left-handed
  $W^\prime$
  bosons~\cite{He:2012zp,Greljo:2015mma,Boucenna:2016wpr,He:2017bft}.
\item {\boldmath $C_S^R$}: This operator is generated in models with extra charged scalars. In particular it is the dominant operator in 2HDMs of type II in the large $\tan\beta$ region (see, e.g., Refs.~\cite{Kalinowski:1990ba,Hou:1992sy} for an early account) and can be generated with the SU(2)$_L$-doublet vector LQ  ($V_2$)~\cite{Kosnik:2012dj,Biswas:2018iak}.
\item \begin{boldmath}$C_S^L$:\end{boldmath} This setup is again motivated by models with extra charged scalars. However, here a generic flavor structure is needed to make $O_S^L$ the dominant
operator~\cite{Crivellin:2012ye,Crivellin:2013wna,Celis:2012dk,Ko:2012sv,Crivellin:2015hha,Dhargyal:2016eri,Chen:2017eby,Iguro:2017ysu,Martinez:2018ynq,Biswas:2018jun}.
\item \begin{boldmath}$C_S^L=4C_T$\end{boldmath}: $C_S^L=4C_T$ at the NP
  scale is generated by the scalar SU(2$)_L$-doublet $S_2$ (also called
  $R_2$) LQ~\cite{Becirevic:2016yqi,Becirevic:2018afm}. However, QCD
  renormalization-group (RG) effects from the NP scale down to the $m_b$
  scale change this relation. Furthermore, electroweak RG effects mix
  the left-handed scalar and tensor operators above the electroweak
  symmetry breaking scale
  \cite{Alonso:2013hga,Gonzalez-Alonso:2017iyc}. Taking into account
  these effects for NP of $\mathcal O(\text{TeV})$ we use
  $C_S^L \simeq 8.1 C_T$ at the scale
  $\mu = m_b$~\cite{Gonzalez-Alonso:2017iyc}.
\end{itemize}
\medskip 

In Fig.~\ref{WCsingle}, we show the $\Delta \chi^2 (C_i)\equiv \chi^2 (C_i) - \chi^2_{\text{SM}}$ (i.e., the
difference compared to the $\chi^2$ in the SM as a function of the Wilson coefficients) for these four cases. The
dashed lines correspond to the situation before the $F_L({D^*})$ measurement and the solid lines depict the situation once
$F_L({D^*})$ is included. One can see from the plot that while the vector operator still gives the best fit to the data, $F_L({D^*})$ slightly improves the agreement of the $C_S^R$ scenario with data in the vicinity of the best-fit point. The dotted vertical lines delimit the area allowed by different bounds on ${\rm BR}(B_c\to \tau \nu)$, which is only relevant for the $C_S^L$ and $C_S^R$ scenarios. One observes that even for the conservative limit  ${\rm BR}(B_c\to \tau \nu)\leq 60\%$ 
negative solutions for $C_S^L$ and $C_S^R$ are disfavored with respect to the SM point.

Table~\ref{tab:results1D} summarizes the results for the four 1D
scenarios. Here we give the best-fit point, the corresponding
$1\,\sigma$ and $2\,\sigma$ ranges around this point, as well as the  $p$-value (characterizing the goodness of the fit) and the pull with respect to  the SM. The last six columns show the predictions for the  observables under consideration at the best-fit point. In addition,  the discrepancy (defined in Eq.~\eqref{discrepancy}) between the predicted value for the observables and the  current measurement is given for those observables for which a measurement is available.
	
Let us illustrate this with the $C_S^L$ scenario as an example. Here, if the best-fit point $C_S^L=0.12$ is realized in nature, the probability that statistical fluctuations would account for the remaining discrepancy between theory and data is 0.02\%; i.e., the scenario describes the data poorly. This can be attributed to the fact that the predicted values of ${\cal R}(D^*)$ and $F_L({D^*})$ are below their measured values by 4.0 and 1.8 standard deviations, respectively. $\bbc$ is important for this scenario because it excludes the otherwise favored value $C_S^L\sim -0.9$, as can be seen in Fig.~\ref{WCsingle}, independent of which of the three limits we choose. The value of the SM pull, $\text{pull}_\text{SM}=2.2\sigma$, shows that $C_S^L=0.12$ describes the data only moderately better than $C_S^L=0$.

\begin{table*}
\begin{tabular}{|c||c|c|c||c|c|c|c|c|c|c|c|c|}\hline
  	2D hyp.  & best-fit& \scalebox{0.85}{$p$-value {(\%)}} 
& pull$_{\rm SM}$  & ${\cal R}(D)$ & ${\cal R}(D^*)$ & $F_L({D^{*}})$ &  $P_\tau (D^*)$ & $P_\tau (D)$ & ${\cal R}(\Lambda_c) $\\\hline\hline
  	$(C_V^L,\,C_S^L=-4C_T)$&($0.08,0.05)$&22.0&4.2&
  	\dline{0.394 $-0.3$ $\,\sigma$}& \dline{0.308 $+0.2\,\sigma$}&\dline{0.45 $-1.7$ $\,\sigma$} &\dline{$-0.50$ $-0.2$ $\,\sigma$}& \dline{0.40 \\}& \dline{0.41 \\}\\ \hline
  	$\left(C_S^R,\,C_S^L)\right|_{60 \%}$& \parbox{7em}{\rule{0pt}{2.3ex}($-0.19,-0.74)$ $(0.34,-0.22)$\strut}&68.5&4.5&
  	\dline{0.412 $+0.1\,\sigma$} &\dline{0.299 $-0.5$ $\,\sigma$}& \dline{0.54 $-0.7$ $\,\sigma$}&\dline{$-0.27$ $+0.2\,\sigma$}& \dline{0.50 \\}&\dline{0.40 \\}\\ \hline
  	$\left(C_S^R,\,C_S^L)\right|_{30 \%}$&\parbox{7em}{\rule{0pt}{2.3ex}($-0.30,-0.64)$ $(0.24,-0.11)$\strut}&11.8&4.1&
  	\dline{0.423 $+0.4\,\sigma$}&\dline{0.280 $-1.8$ $\,\sigma$}&\dline{0.51  $-1.0 $ $\,\sigma$}& \dline{$-0.35$ 0.0 $\,\sigma$}&\dline{0.51 \\}&\dline{0.39\\}\\ \hline
  	$\left(C_S^R,\,C_S^L)\right|_{10 \%}$
  	&\parbox{7em}{\rule{0pt}{2.3ex}($0.14,0.00)$ $(-0.40,-0.55)$\strut}&0.6&3.4&\dline{0.433 $+0.6\,\sigma$} &\dline{0.263 $-2.9$ $\,\sigma$}&\dline{0.48 $-1.4$ $\,\sigma$} &\dline{$-0.44$ $-0.1\,\sigma$} &\dline{0.53 \\}&\dline{0.38\\}\\ \hline 
  	$(C_V^L,\,C_S^R)$&($0.09,0.06)$&30.8&4.3&
  	\dline{0.413 $+0.1\,\sigma$}&\dline{0.305 $-0.1$ $\,\sigma$}&\dline{0.47 $-1.5$ $\,\sigma$} &\dline{$-0.47$ $-0.2$ $\,\sigma$}&\dline{0.41 \\}&\dline{0.42 \\}\\ \hline 
\scalebox{0.85}{ 		
  	$\left.({\rm Re}[C_S^L=4C_T],\,{\rm
  Im}[C_S^L=4C_T])\right|_{60,30\%}$}
&($-0.06,\pm 0.40)$&22.0&4.2&
  	\dline{0.404 $-0.1$ $\,\sigma$}&\dline{0.306 0.0 $\,\sigma$}
  	&\dline{0.45 $-1.7\,\sigma$} 
  	&\dline{$-0.39$ 0.0$\,\sigma$}&\dline{0.50 \\}&\dline{0.41 \\}
  	\\ \hline
\scalebox{0.85}{$\left. ({\rm Re}[C_S^L=4C_T],\,{\rm
  Im}[C_S^L=4C_T])\right|_{10\%}$} 
&($-0.02,\pm 0.24)$&0.3&3.2&
  	\dline{0.339 $-1.5$ $\,\sigma$}& \dline{0.274 $-2.2$ $\,\sigma$}&\dline{0.46  $-1.7$ $\,\sigma$}& \dline{$-0.45$ $-0.1$ $\,\sigma$} &\dline{0.40 \\}&\dline{0.36 \\}\\ \hline
\end{tabular}
 \caption{Results of the fit for the Wilson coefficients (given at the matching scale of 1 TeV) for the 2D hypotheses (hyp.) defined in Sec.~\ref{sec:2D} including all available data with ${\rm BR}(B_c\to\tau\nu)<60$\%, $ {\rm BR}(B_c\to\tau\nu)<30$\% and $ {\rm BR}(B_c\to\tau\nu)<10$\%, respectively. In case there is no label for the constraint on $\bbc$ used, the fit is valid for all three benchmark scenarios.}	
  	\label{tab:results2D}
  \end{table*}
  
The hypothesis of NP entering through $C_V^L$ has a favorable $p$-value
of 35\% and the $C_S^L=4 C_T$ scenario gives the worst fit. As a caveat,
we recall that we have restricted ourselves to real values of the
coefficients. Therefore, if complex values for $C_S^L=4 C_T$ are
permitted the situation will change. However, we do not consider complex
values for the Wilson coefficients in the other three scenarios. For
$C_V^L$ this would not change the predictions and for $C_S^L$ and
$C_S^R$ complex values are constrained by ${\rm BR}(B_c\to\tau \nu)$. 

Note that the results are quite independent of the bound used for the contribution to $\bbc$. The significance of the four one-dimensional scenarios does not change depending on whether one uses the conservative bound of 60\% or the most commonly used one of 30\% for $\bbc$. Furthermore, only the $C_S^R$ scenario is slightly affected once the hypothetical future limit of 10\% is used; the $p$-value changes slightly from $1.8\%$ to $1.7\%$. Also note that in the $C_V^L$-scenario polarization observables $F_L (D^\ast), P_\tau(D^\ast)$ and $P_\tau(D)$ are unchanged with respect to the SM. Therefore, a significant deviation in these observables would automatically disfavor (or potentially exclude) this scenario.
  
\subsection{Two-dimensional scenarios} \label{sec:2D}

Let us now consider several two-dimensional hypotheses. Again, we consider only scenarios which can be generated by adding a single new field to the SM particle content.

\begin{itemize}
\item {\boldmath $(C_V^L,\,C_S^L=-4C_T)$}:
This setup is obtained in models with an  SU(2$)_L$-singlet scalar LQ ($S_1$). Here the relation $C_S^L=-4C_T$ is again assumed at the NP scale. Through the RG running mentioned above, starting from an $\mathcal O(\text{TeV})$ matching scale, the relation becomes $C_S^L \simeq- 8.5 C_T$ at the low scale~\cite{Gonzalez-Alonso:2017iyc}.
\item {\boldmath $(C_S^R,\,C_S^L)$}: As for the 1D cases, this scenario is motivated by charged Higgs exchange.
\item {\boldmath $(C_V^L,\,C_S^R)$}:
This setup is generated by models with vector LQs like the SU(2$)_L$-singlet LQ $U_1$.
\item  {\boldmath $({\rm Re}[C_S^L=4C_T],\,{\rm Im}[C_S^L=4C_T])$}: At the high scale, this relation is generated by the scalar SU(2$)_L$-doublet LQ $S_2$. As in the 1D case, RG effects modify this relation to $C_S^L \simeq 8.1 C_T$ at the scale $\mu  = m_b$. Here we consider complex couplings because, as seen in the previous subsection, real parameters do not give a good fit to the data. On the other hand, as shown in Ref.~\cite{Becirevic:2018afm}, complex Wilson coefficients are able to reproduce the $\mathcal{R}({D^{(*)}})$ data.
\end{itemize}

\begin{figure*}[th]
\begin{center}
{
\includegraphics[width=0.4\textwidth, bb= 0 0 260 250]{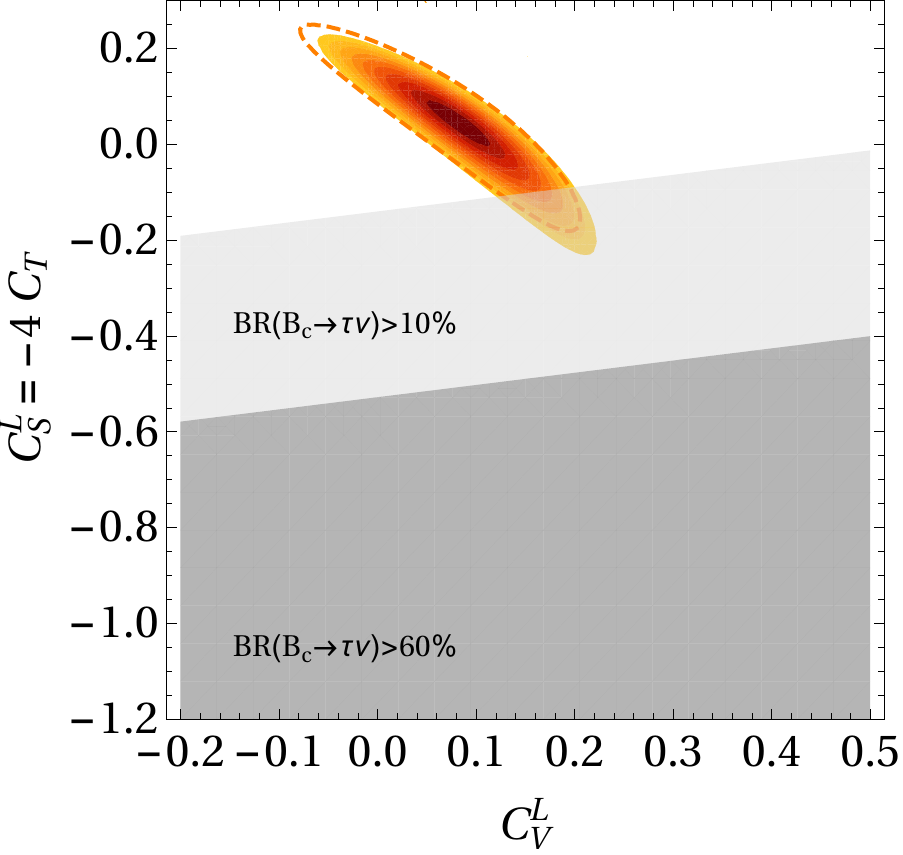}
}
{
\includegraphics[width=0.47\textwidth, bb=0 0 304 253]{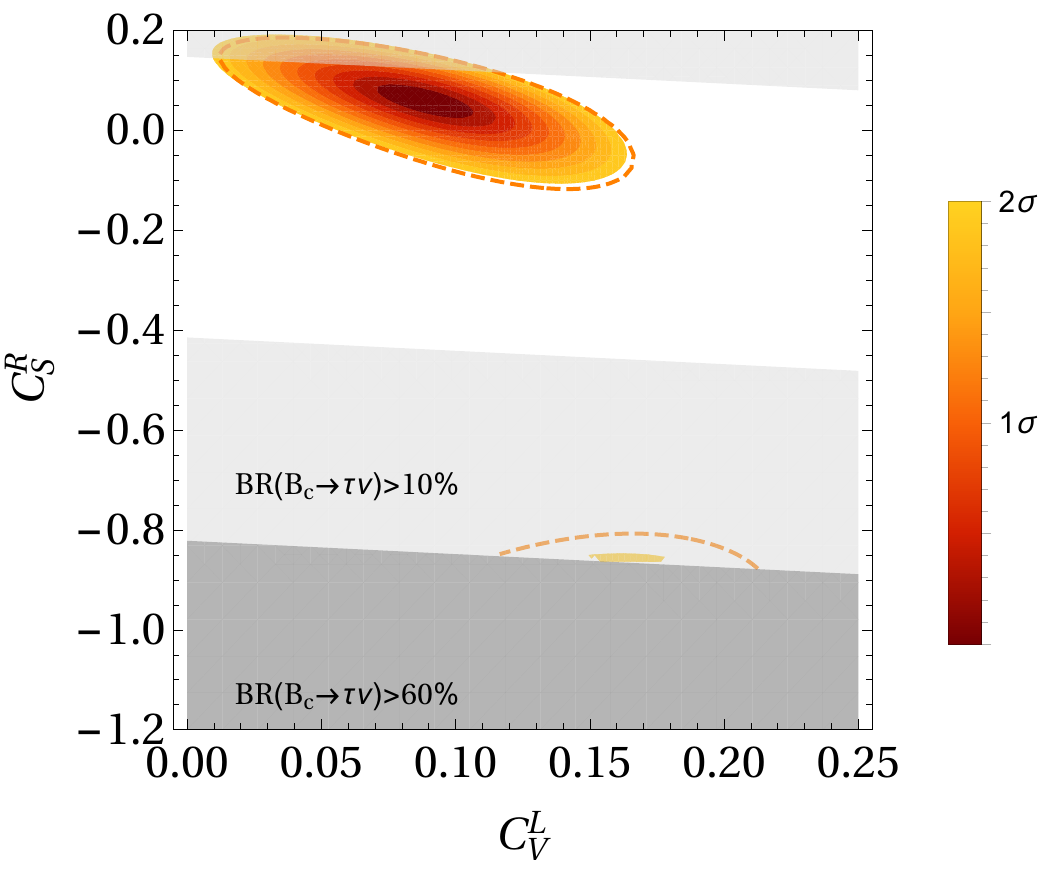}
}
\\
\vspace{0.1cm}
\hspace{0.5cm}
{
\includegraphics[width=0.4\textwidth, bb= 0 0 260 249]{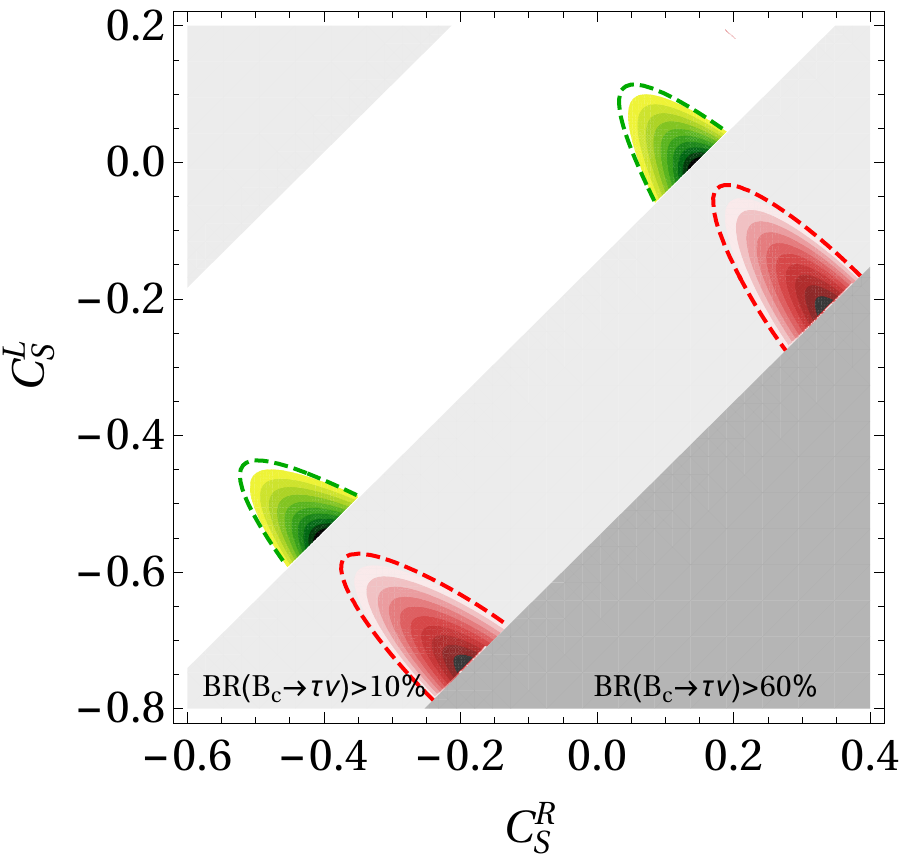} 
}
{
\includegraphics[width=0.51\textwidth, bb=0 0 336 250]{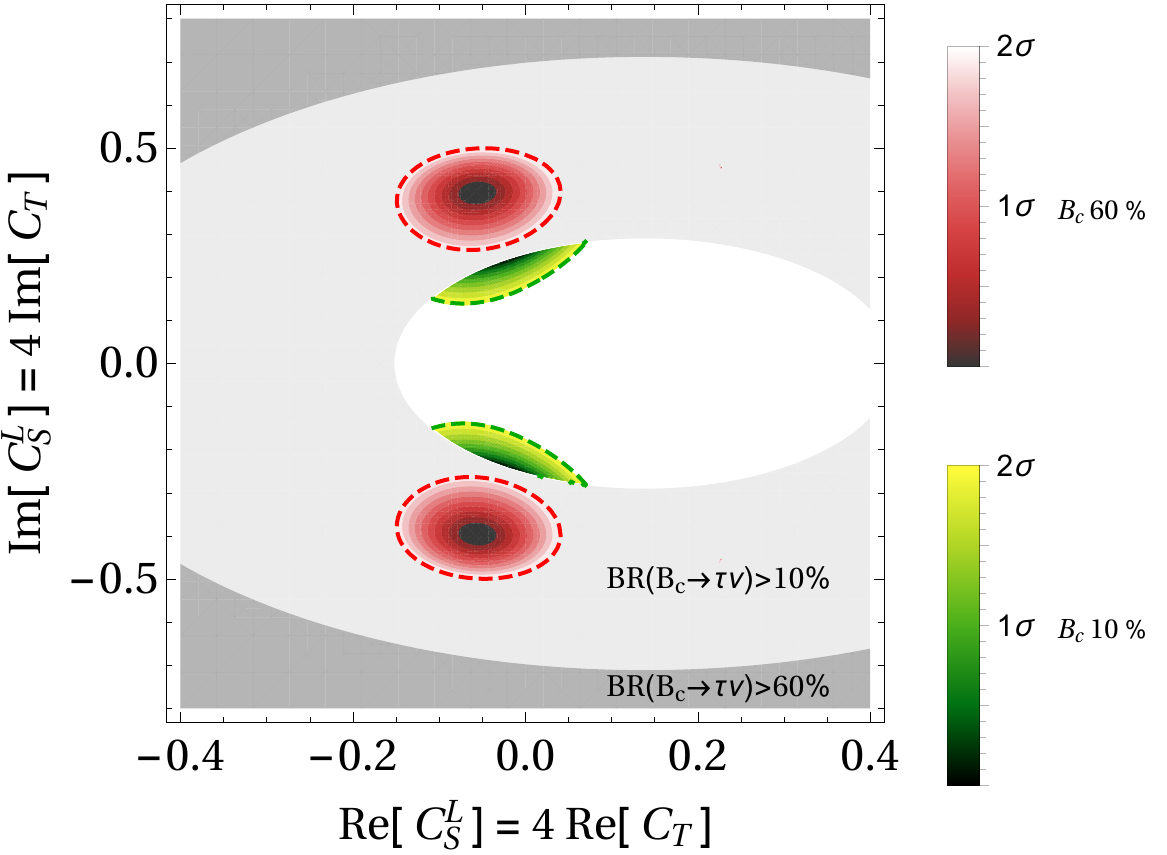}
}
\caption{Results of the fits in the two-dimensional NP scenarios, with Wilson coefficients given at the matching scale of $1\,$TeV. The $p$-values of the best fit are given in Table~\ref{tab:results2D}. The dashed ellipses show the situation before the $F_L({D^*})$ measurement at $2\,\sigma$ level, while the colored regions include $F_L({D^*})$. We impose either a $60\%$ or a $10\%$ limit on $\bbc$. The scenarios shown in the upper plots (orange color coding) are not affected by either of these constraints. In the scenarios shown in the lower plots the best-fit points and the corresponding $\sigma$-regions move when we consider a $10\%$ (green color coding) constraint instead of the $60\%$ one.}
\label{WCdouble}
\end{center}
\end{figure*}

The results of these fits are given in Table~\ref{tab:results2D}, for a
limit on $\bbc$ of 60\%, 30\% and 10\%, respectively. We treat again the
$\bbc$ constraint as a hard limit. Note that the $\bbc$ constraint has
no impact on the best-fit points of the $(C_V^L,\,C_S^L=-4C_T)$ and
$(C_V^L,\,C_S^R)$ scenarios. For the
$({\rm Re}[C_S^L=4C_T],\, {\rm Im}[C_S^L=4C_T])$ scenario, only the
hypothetical future bound of 10\% significantly reduces the goodness of
the fit (from 22\% to 0.3\%). The impact on the $(C_S^R,\,C_S^L)$
scenario is very significant: While for the most conservative limit of
60\% this scenario gives the best fit among all the scenarios
considered, the agreement with data is only moderate for the 30\% limit
and even very bad for the 10\% one. Note that the tension arises
  only in $\mathcal{R}(D^*)$ which is governed by the same coupling
  $C_S^R-C_S^L$ as $\bbc$.
	
The content of Table~\ref{tab:results2D} translates to the plots shown in Fig.~\ref{WCdouble}. Here, one can see that if the overall best-fit point is excluded by the $\bbc$ limit, the point with the minimum $\chi^2$ compatible with this bound is taken instead. Thus, the new best-fit point lies on the boundary of the region excluded by $\bbc$, and it is surrounded by the corresponding confidence region. Therefore, different limits for $\bbc$ lead to different preferred regions, and the best-fit points are also distinct concerning the overall goodness (the $p$-value) of the fit.  In the last six columns of Table~\ref{tab:results2D} we give again the predictions of the observables and their discrepancy (defined in Eq.~\eqref{discrepancy}) with the experimental value.

\section{Correlations between observables} \label{sec:corr}

Let us now assess the future discriminatory power of the various $b\to c\tau\nu$ observables and evaluate the correlations among the observables within our two-dimensional scenarios of Sec.~\ref{sec:2D}. 
	
\begin{figure*}[tp]
	\subfigure{
		\includegraphics[width=0.39\textwidth, bb= 0 0 350 306]{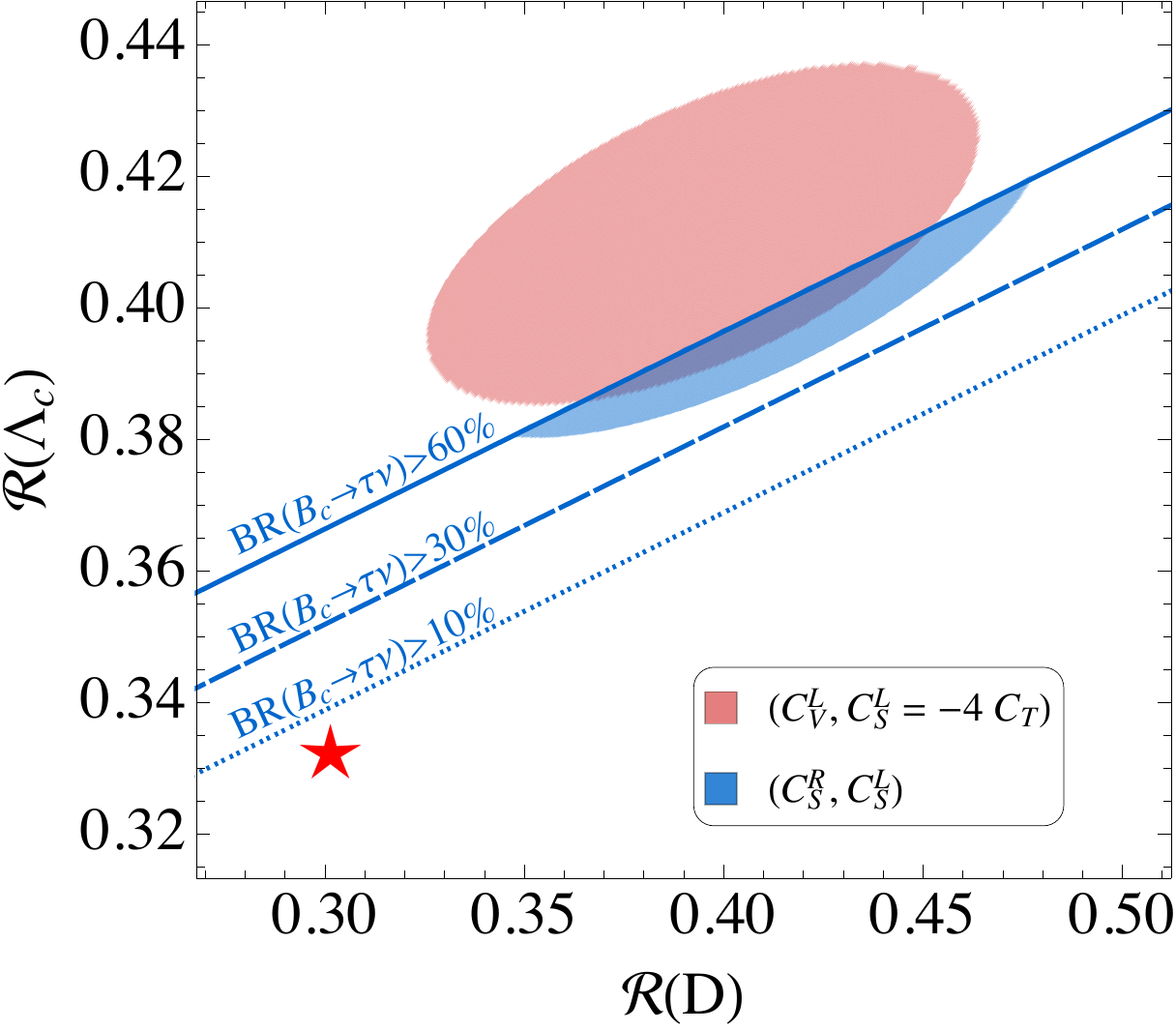}			}\quad
	\subfigure{	
		\includegraphics[width=0.405\textwidth, bb= 0 0 350 293]{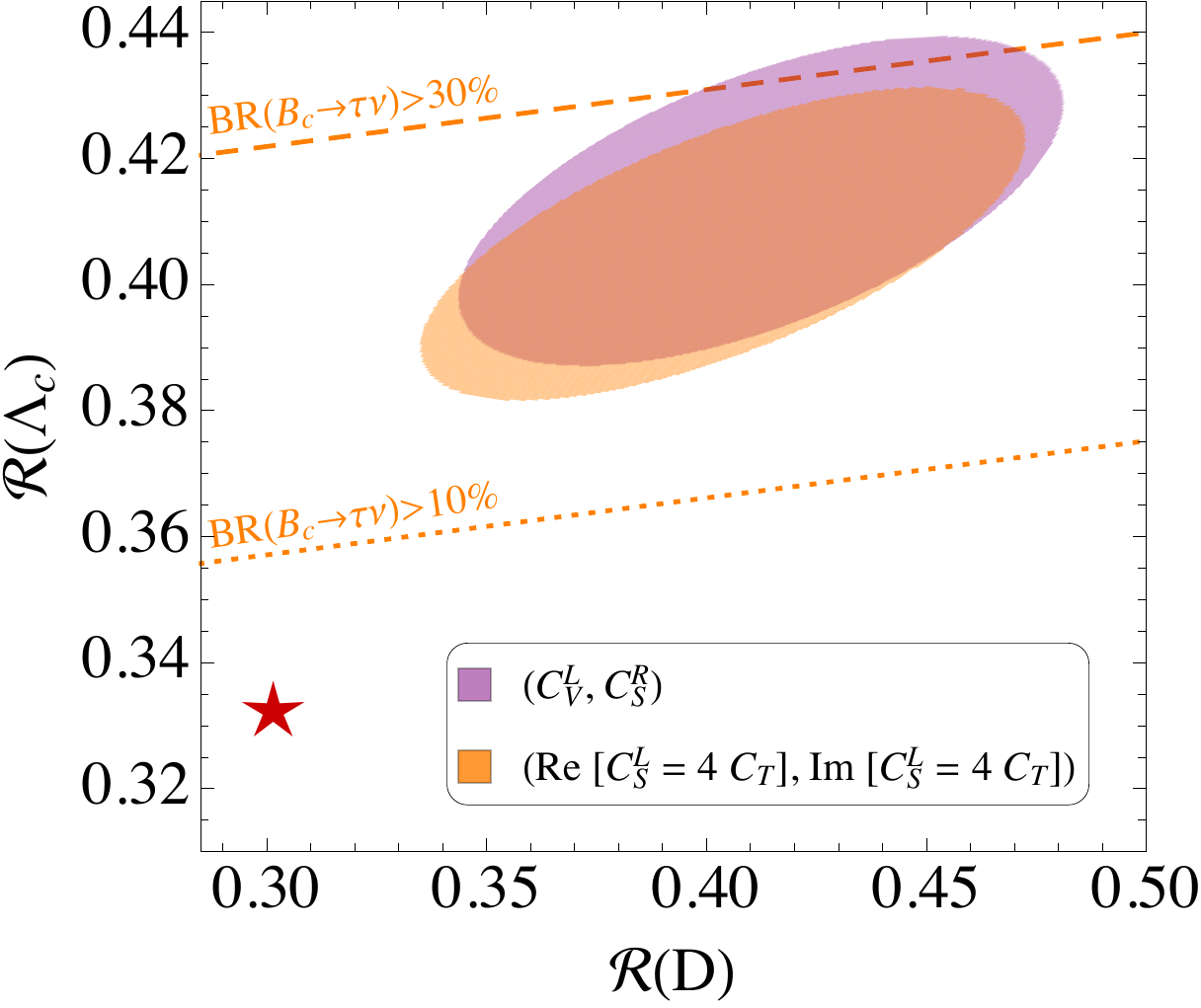}          }\\
	\subfigure{
		\includegraphics[width=0.40\textwidth, bb= 0 0 350 300]{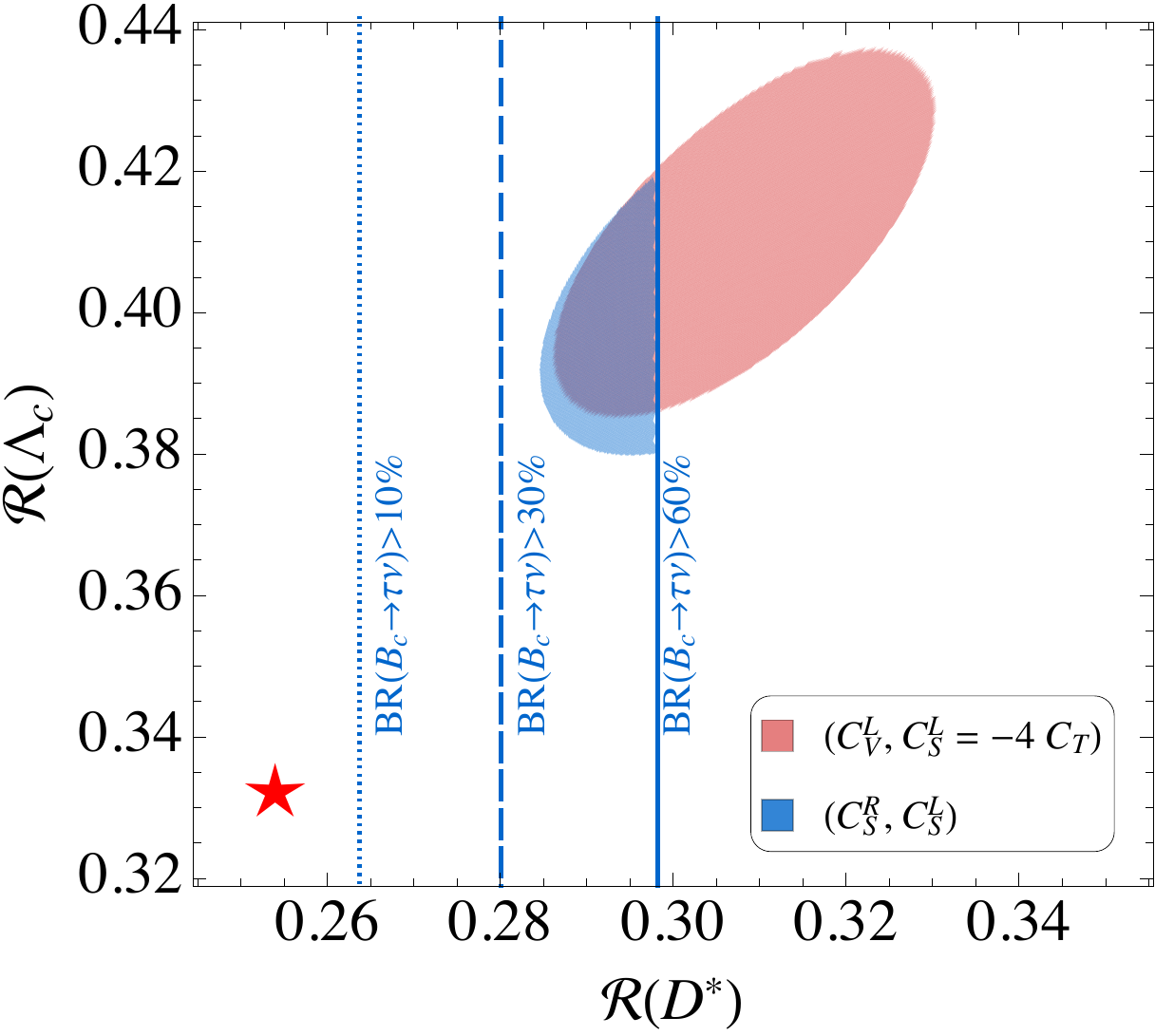}			}\quad
	\subfigure{	
		\includegraphics[width=0.38\textwidth, bb= 0 0 350 306]{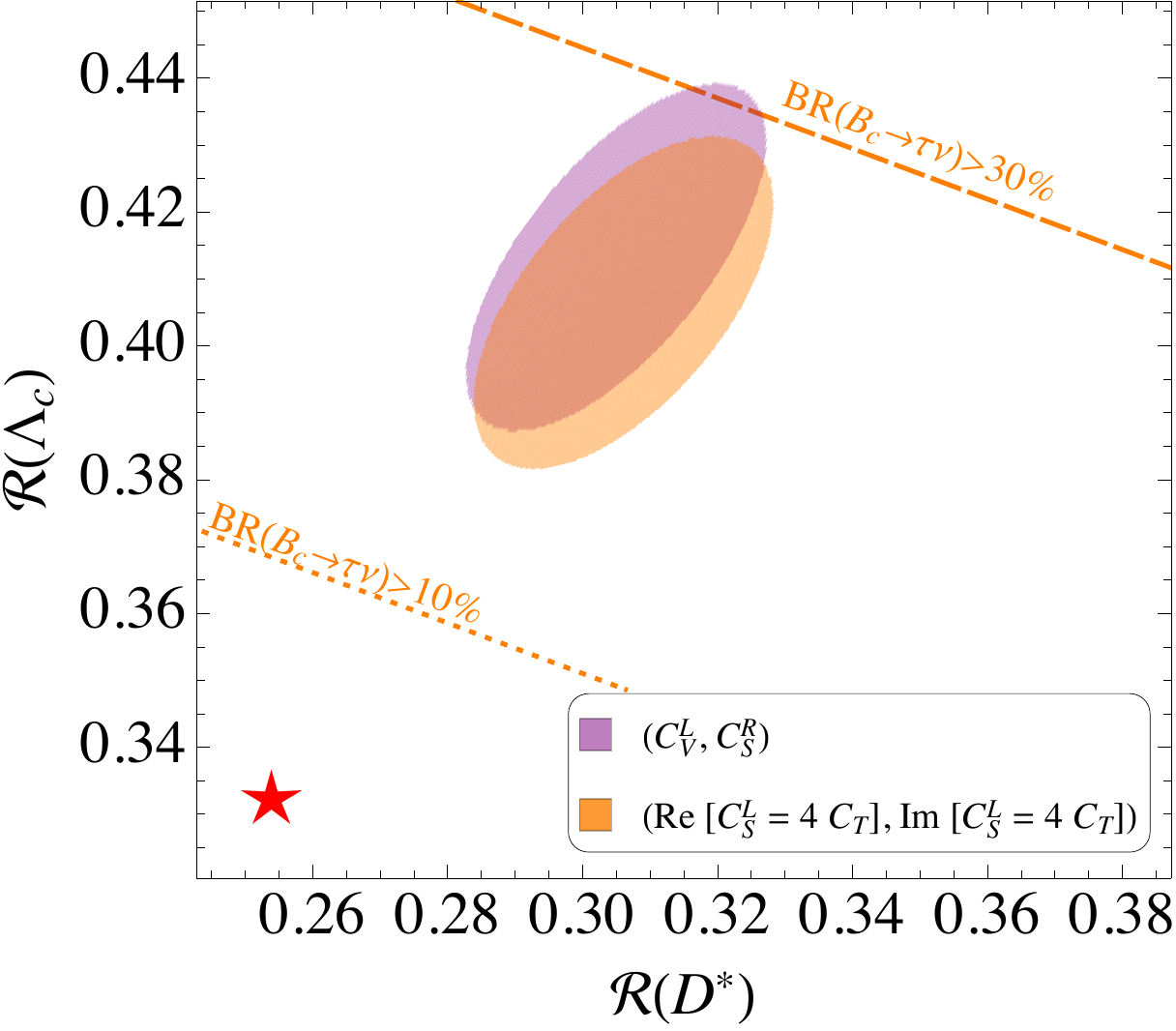}          }\\
\caption{Preferred $1\,\sigma$ regions in the four two-dimensional scenarios in the $\mathcal{R}(D^{(\ast)})$--$\mathcal{R}(\Lambda_c)$ plane for $\bbc<60\%$. The regions of the plot in the left panel correspond to the scenarios $(C^L_V, C^L_S=-4C_T)$ (red) and $(C^R_S, C^L_S)$ (blue), while the plots on the right side correspond to $(C^L_V, C^R_S)$ (purple) and $(C^L_S = 4\,C_T)$ (orange). The solid, dashed and dotted lines refer to a limit on $\bbc$ of $60\%, 30\%$ and $10\%$, respectively. The stars represent the SM predictions.}
\label{Fig:Correl-First}
\end{figure*}

Let us start with the correlations among $\mathcal{R}(D^{(\ast)})$ and $\mathcal{R}(\Lambda_c)$ as shown in Fig.~\ref{Fig:Correl-First}. The colored regions  in the $\mathcal{R}(D^{(\ast)})$--$\mathcal{R}(\Lambda_c)$ plane are   allowed at the $1\,\sigma$ level as obtained by  the fit (see Fig.~\ref{WCdouble}). In addition, the different bounds from ${\rm BR}(B_c\to \tau\nu)$ are shown. As seen in the  previous section, this bound is irrelevant for the $(C_V^L,\,C_S^L=-4C_T)$ and $(C_V^L,\,C_S^R)$ scenarios and also does not affect the complex $(C_S^L=+4C_T)$ scenario, unless the hypothetical future bound of 10\% is used. However, for the $(C_S^R,\,C_S^L)$ scenario it puts a stringent upper bound on $\mathcal{R}(\Lambda_c)$ depending on $\mathcal{R}(D)$.

\begin{figure*}[tp]
\begin{center}
	\subfigure{
		\includegraphics[width=0.45\textwidth, bb=0 0 360 363]{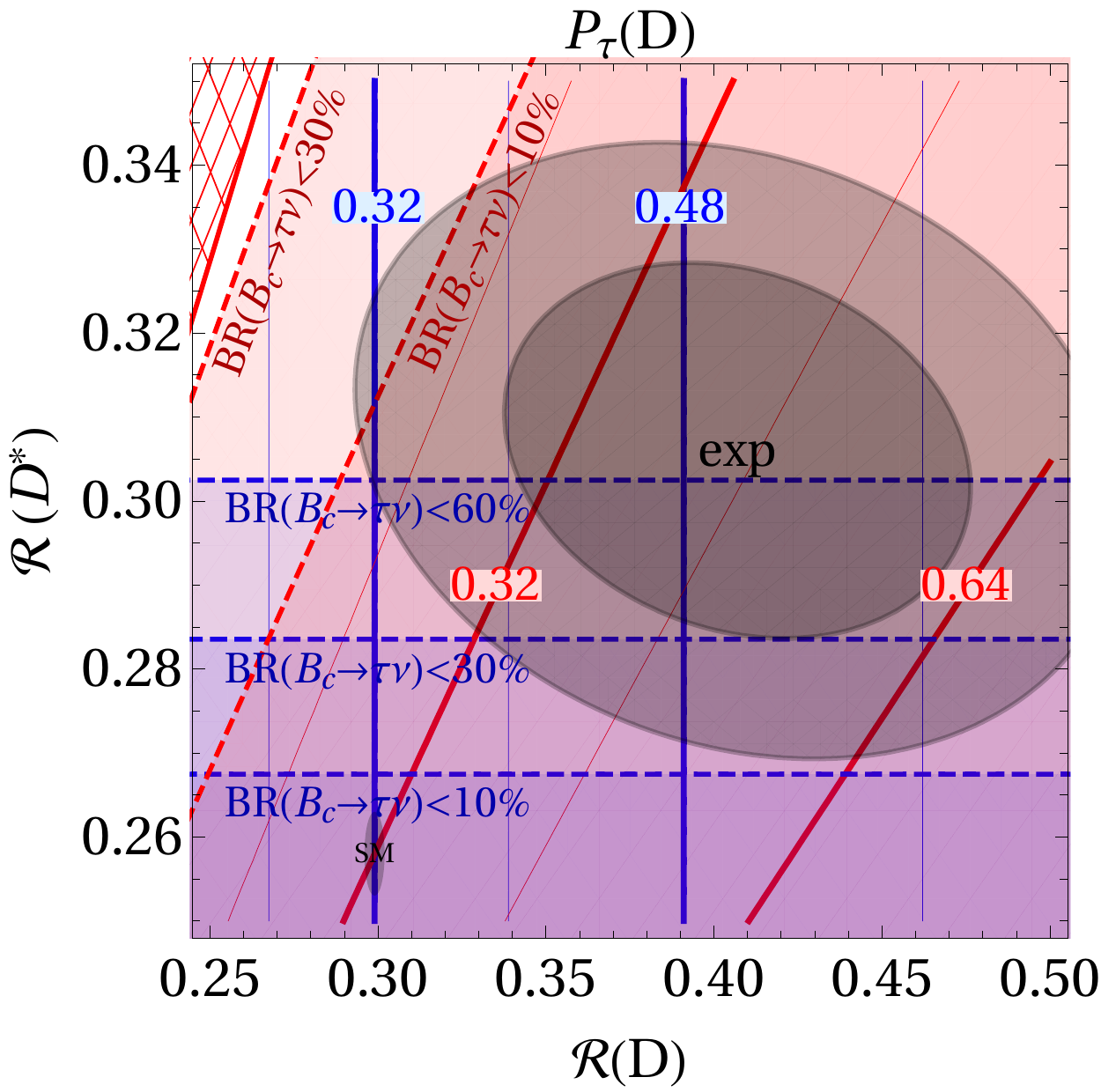}			}\qquad
	\subfigure{		
		\includegraphics[width=0.45\textwidth, bb =0 0  360 363]{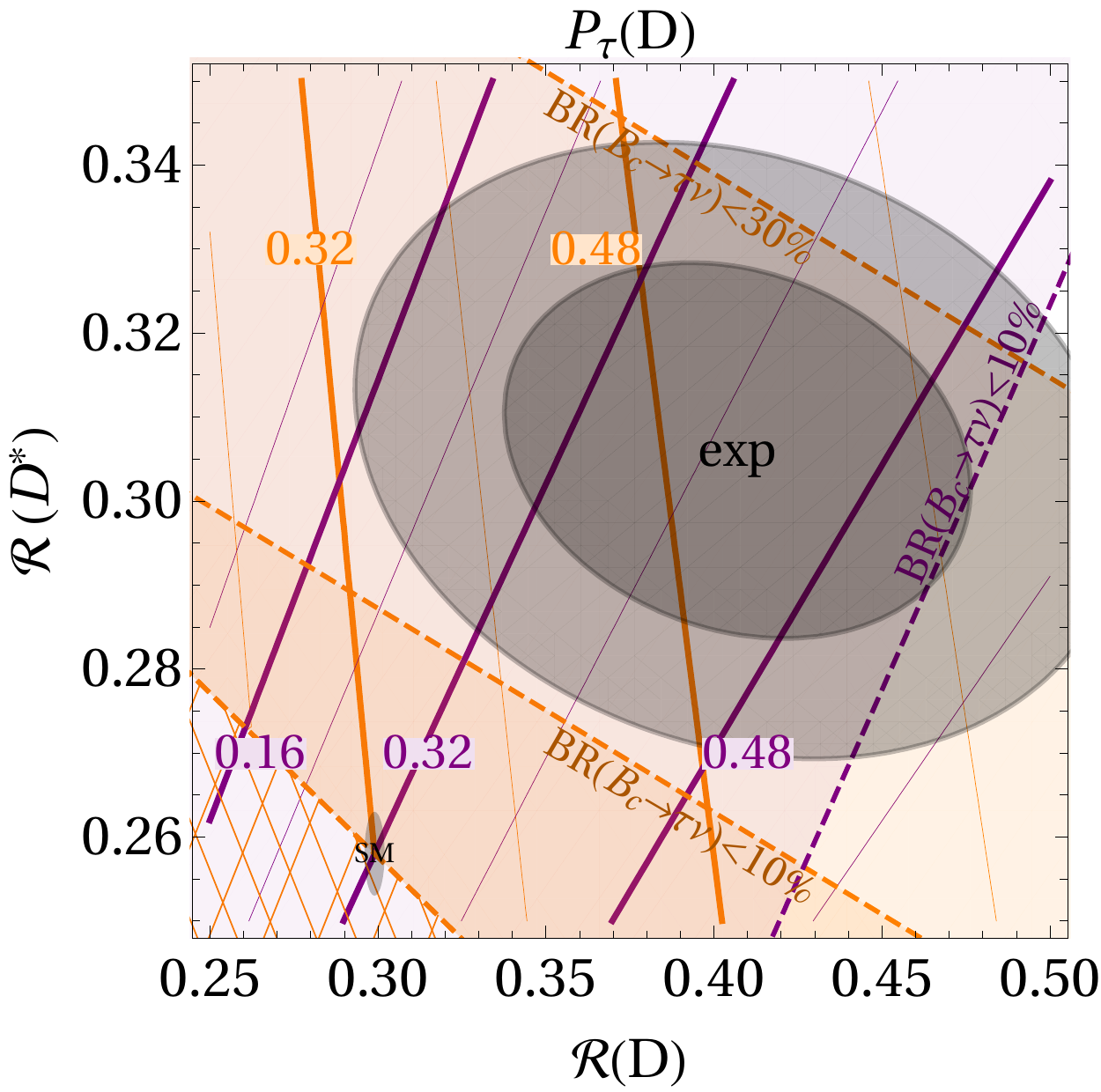}
		}\vspace{-5mm}\\
	\subfigure{	
		\includegraphics[width=0.45\textwidth, bb= 0 0 360 363]{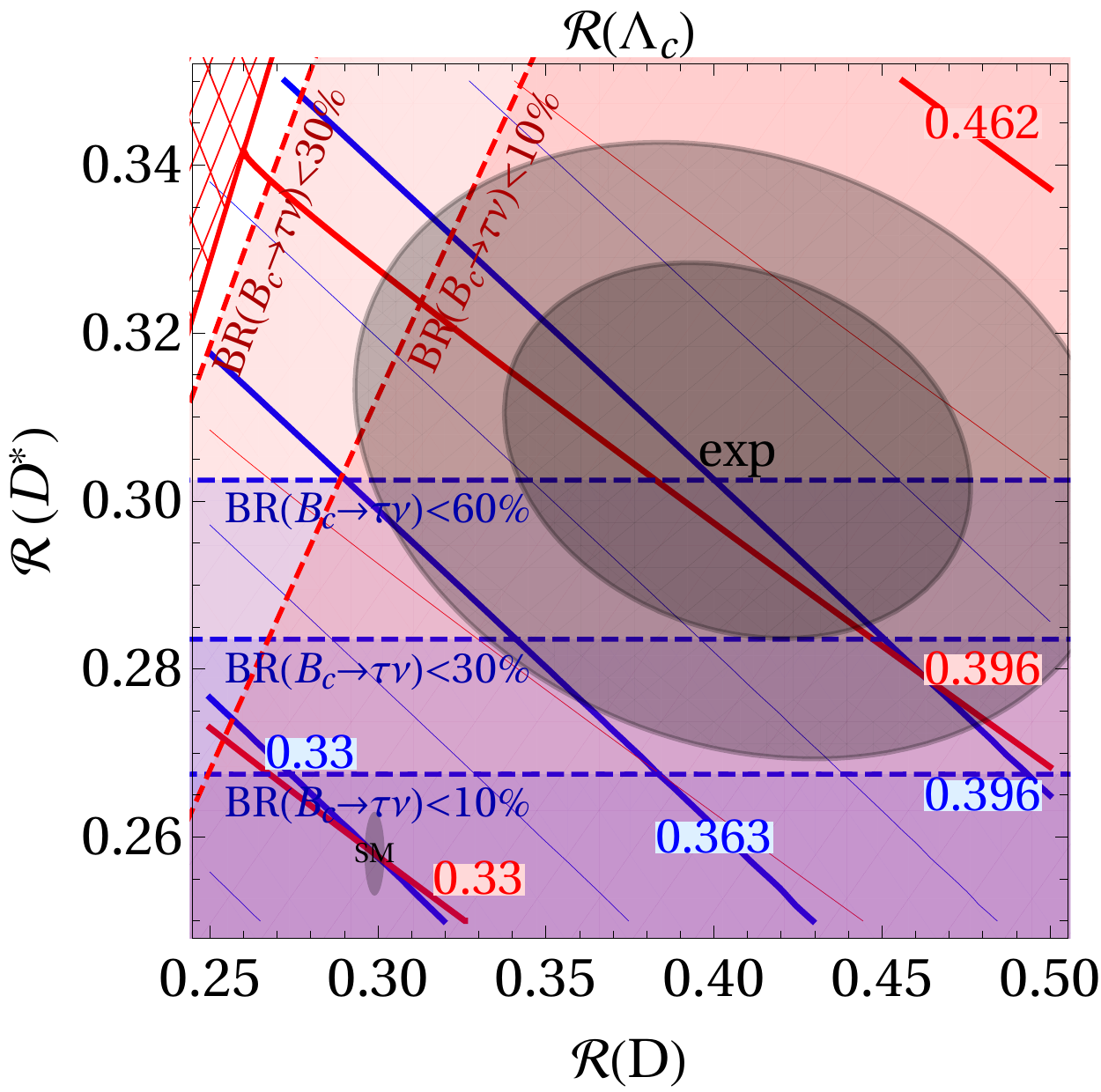}
		}\qquad
	\subfigure{
		\includegraphics[width=0.45\textwidth, bb= 0 0 360 363]{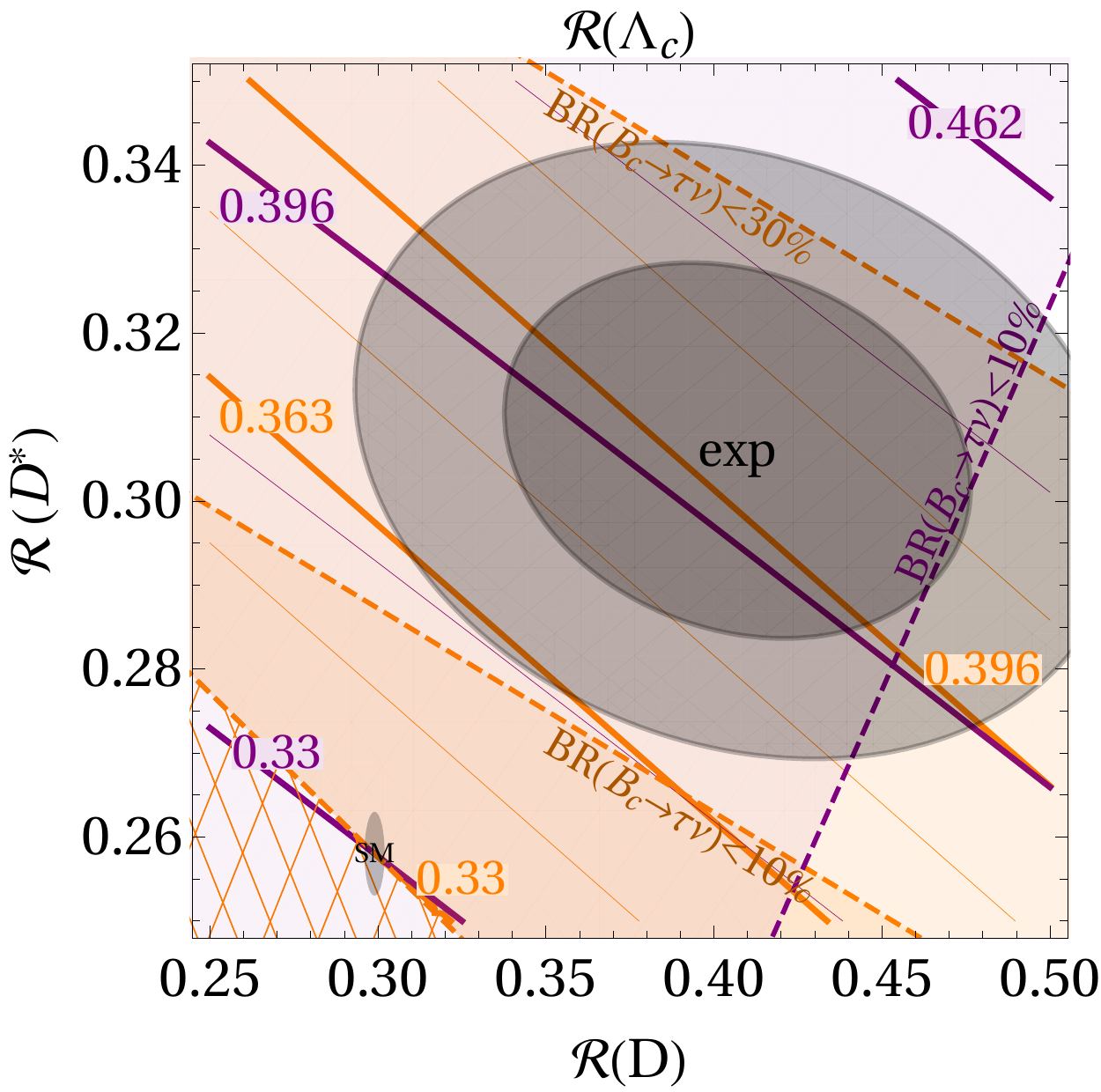}
		}\\	\vspace{1cm}
	\hspace{-5cm}	\includegraphics[width=0.2\textwidth, bb= -55 0 180 38]{legend1.pdf}\qquad\qquad\qquad\qquad\qquad\qquad\qquad\qquad	\includegraphics[width=0.1\textwidth,bb = -25 0 95 38]{legend2.pdf}
\end{center}
\vspace{-0.5cm}
\caption{Contour lines of $P_\tau(D)$ and ${\cal R}(\Lambda_c)$ for the two-dimensional scenarios in the ${\cal R}(D)$--${\cal R}(D^*)$ plane.	The colored regions (bounded by dashed lines) are allowed by the $10\%$, $30\%$ and $60\%$ limits on $\bbc$, where any area that would fill the entire plot is not shown for convenience. The contours show the predicted value for the various observables (neglecting the errors of the form factors). The thin lines carrying no labels depict the arithmetic means of the neighboring thick lines. The gray regions are currently preferred by data at the 1 and 2$\,\sigma$ level and the colored, hatched regions are excluded in the specific scenarios. Interestingly, the different scenarios display distinct correlations among the observables (except for $R(\Lambda_c)$), manifesting themselves in the different slopes of the contours and the different values associated with them. }
\label{RDRDstar2}
\end{figure*}

Interestingly, we find very similar patterns for $\mathcal{R}(\Lambda_c)$ in all scenarios and always predict
an enhancement of  $\mathcal{R}(\Lambda_c)$ over its SM value. 
We trace this behavior back to a \emph{sum rule}, which can be
derived from the expressions in Eqs.~(\ref{eq:rd}),  (\ref{eq:rds}) and (\ref{eq:rlc}): 
\begin{equation}
  \frac{\mathcal{R}(\Lambda_c)}{\mathcal{R}_{\rm SM}(\Lambda_c)}
  \,=\, 0.262 \frac{\mathcal{R}(D)}{\mathcal{R}_{\rm SM}(D)} + 0.738 \frac{\mathcal{R}(D^*)}{\mathcal{R}_{\rm
      SM}(D^*)}  - x \,,  \label{eq:sr}
\end{equation}
where the small remainder $x$ is well approximated by
\begin{align}
  x \simeq \, &  - \real \left[      (1+C_V^L) ( 0.32\, C_T^* \, +\, 0.03\, C_S^{L*}) \right] \\
  &+\,   1.76\, |C_T|^2\, 
  -\, 0.033 \, \real (C_S^L C_S^{R*}) \,, \nonumber
\end{align}
with all coefficients evaluated at $\mu=m_b$. The consequences of this sum rule are best visible in Fig.~\ref{RDRDstar2}, where the $\mathcal{R}(\Lambda_c)$ contours are essentially the same straight lines. Evolving the best-fit points of Table~\ref{tab:results2D} to $\mu=m_b$ with \eq{wcrun} (and using the exact formula for $x$) we find $x=6\cdot 10^{-4}$, $x=1\cdot 10^{-2}$, $x=-1\cdot 10^{-4}$, and $x=5\cdot 10^{-3}$ for the four scenarios. Even  beyond the considered scenarios and permitting more than two
coefficients to be nonzero one finds $|x|<0.05$ if the coefficients are chosen to explain $\mathcal{R}(D)$ and $\mathcal{R}(D^*)$. So $\mathcal{R}(\Lambda_c)$ must be enhanced over the SM value if $\mathcal{R}(D)$ and $\mathcal{R}(D^*)$ are. The existence of the sum rule in \eq{eq:sr}, which holds in \emph{any}\ model of new physics, implies that a future measurement of $\mathcal{R}(\Lambda_c)$ will serve as a check of the measurements of $\mathcal{R}(D)$ and $\mathcal{R}(D^*)$ and of the form-factor calculations. For all of our four two-dimensional scenarios we predict
\begin{eqnarray}
\begin{aligned}
 \mathcal{R}(\Lambda_c) \,=\,& \mathcal{R}_{\rm SM}(\Lambda_c) \left( 1.24 \pm 0.06 \right) 
  \label{eq:predlc1} \\
  \,=\,&  0.41 \pm 0.02 \pm 0.01      \label{eq:predlc2} ,
\end{aligned} 
\end{eqnarray}
where the first error stems from the experimental errors in
$\mathcal{R}(D^{(*)})$ in \eq{rdhflav} and the second error in \eq{eq:predlc2} reflects the present uncertainties of the form-factor ratios.

\begin{figure*}[tp]
	\subfigure{
		\includegraphics[width=0.40\textwidth, bb= 0 0 350 298]{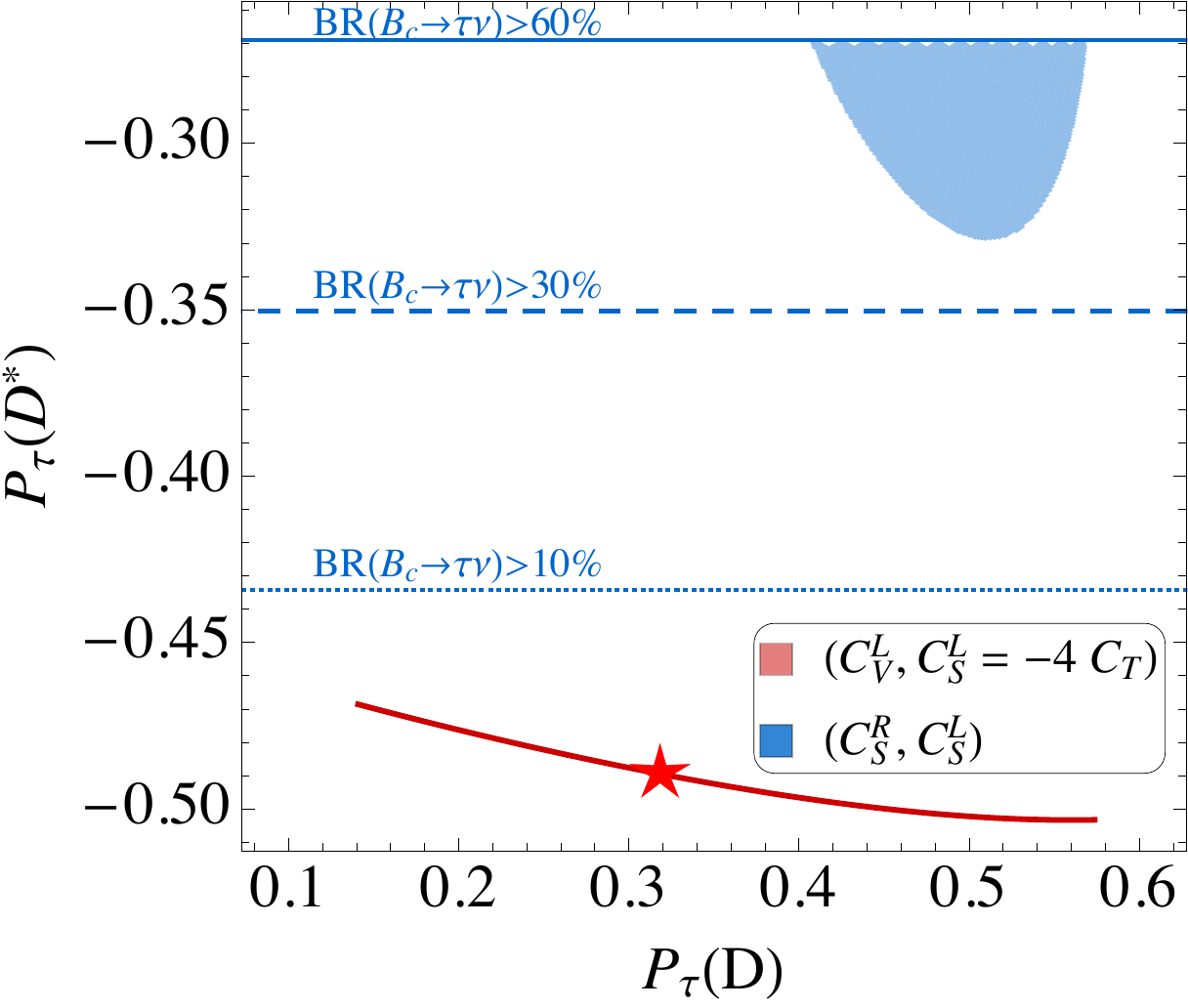}			}\quad
	\subfigure{	
		\includegraphics[width=0.40\textwidth, bb= 0 0 350 298]{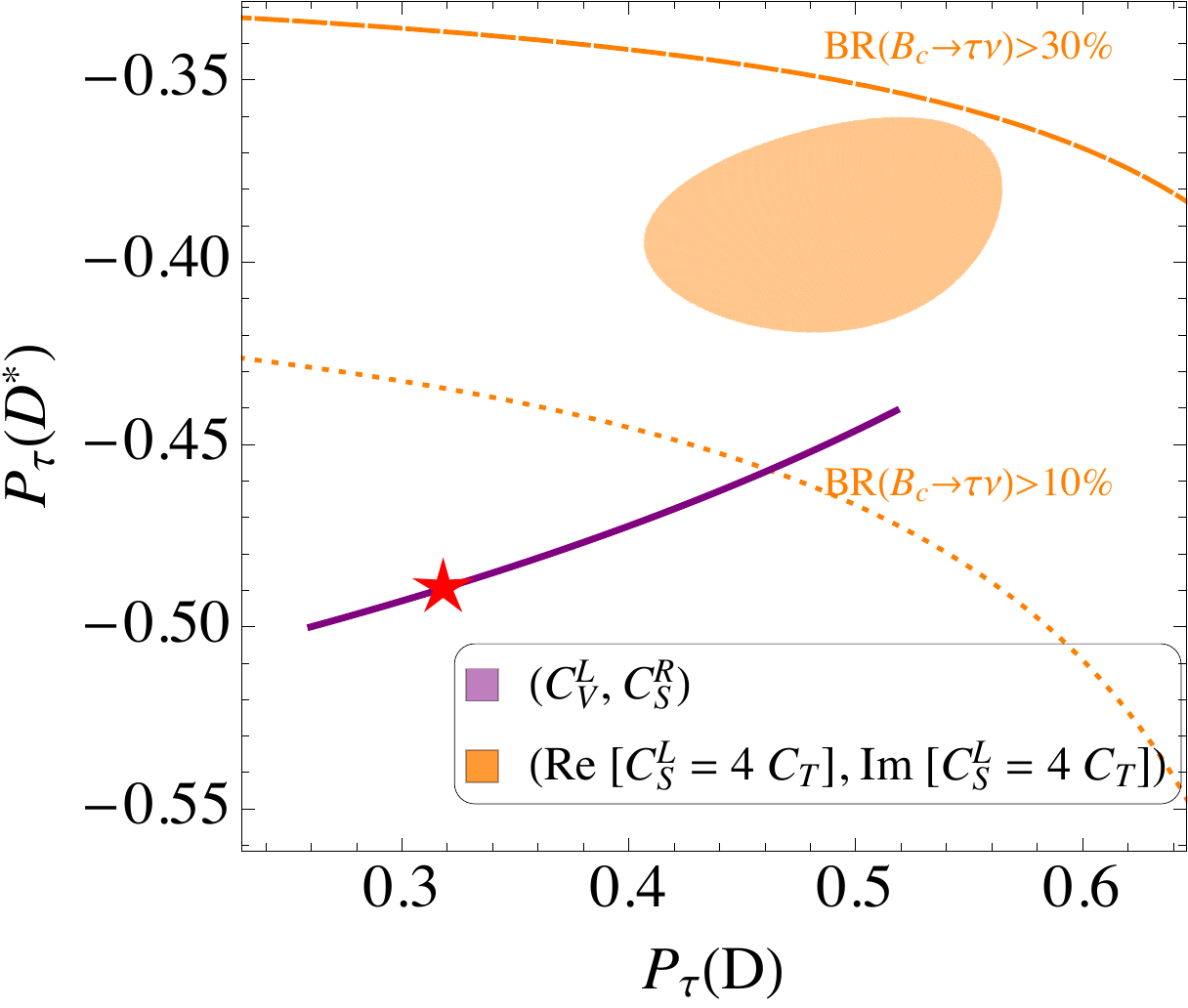}	        }\\
	\subfigure{
		\includegraphics[width=0.41\textwidth, bb=0 0  350 297]{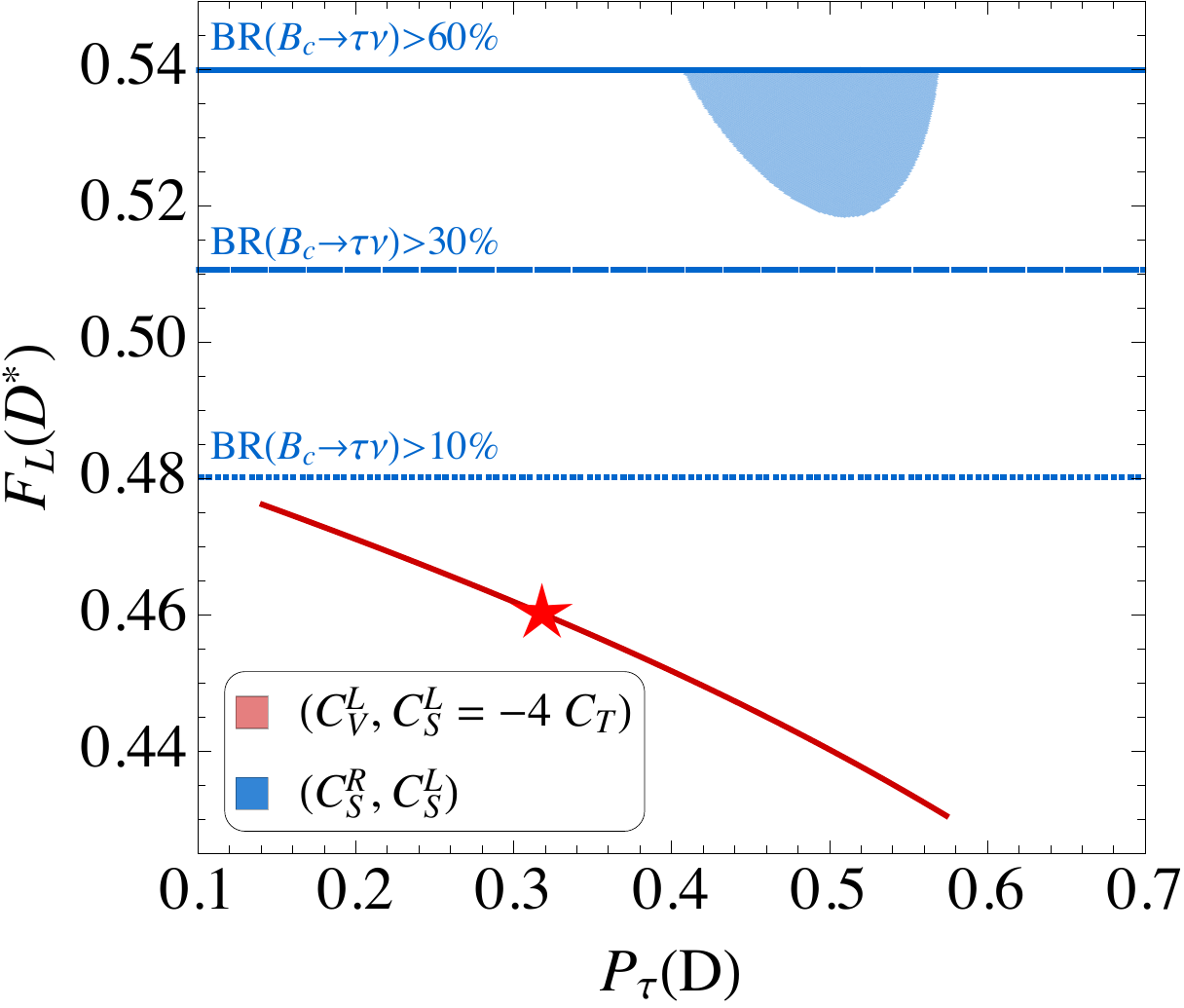}			}\quad
	\subfigure{	
		\includegraphics[width=0.39\textwidth, bb= 0 0 350 308 ]{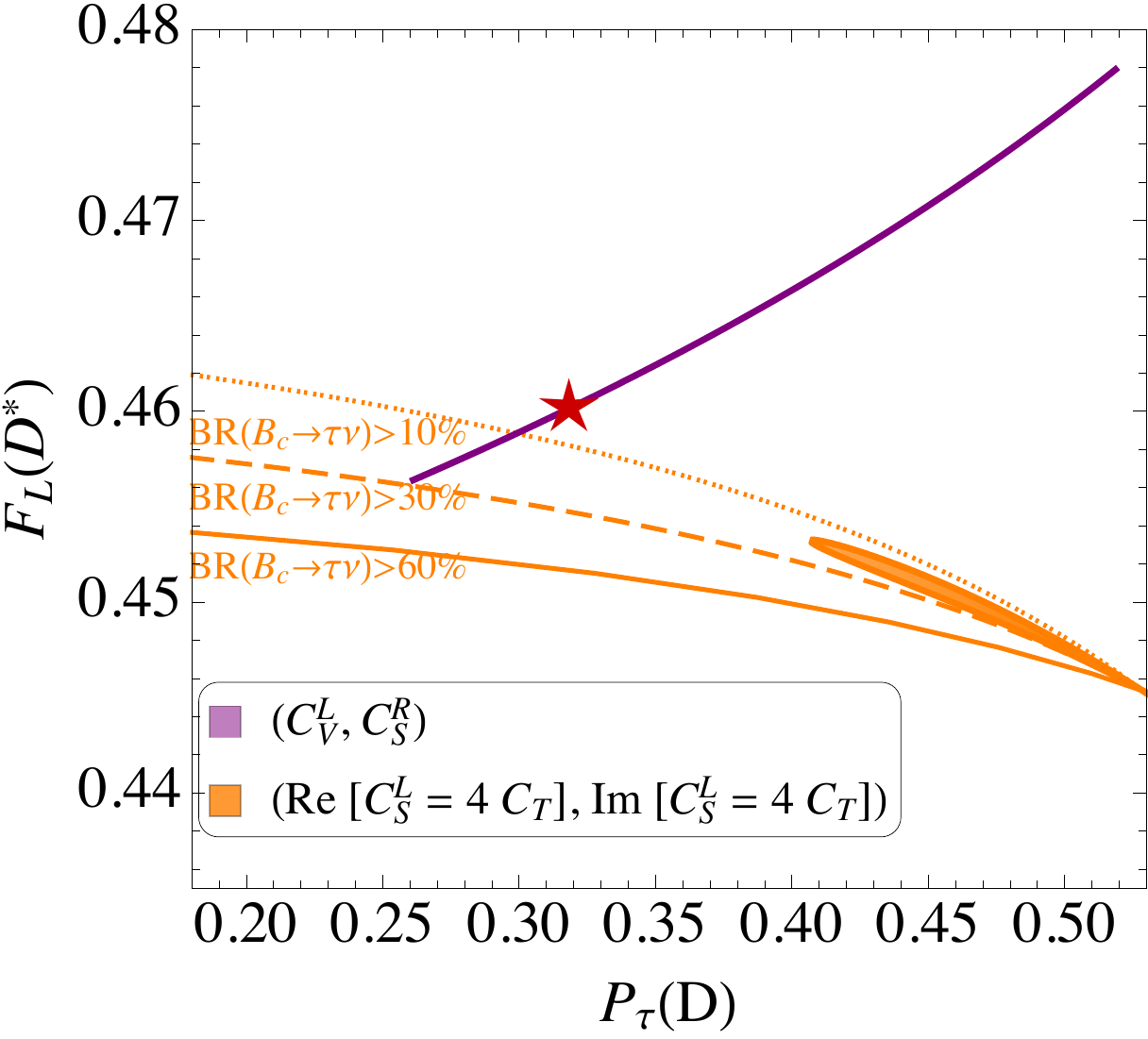}        }\\
	\subfigure{
		\includegraphics[width=0.41\textwidth, bb = 0 0 350 301]{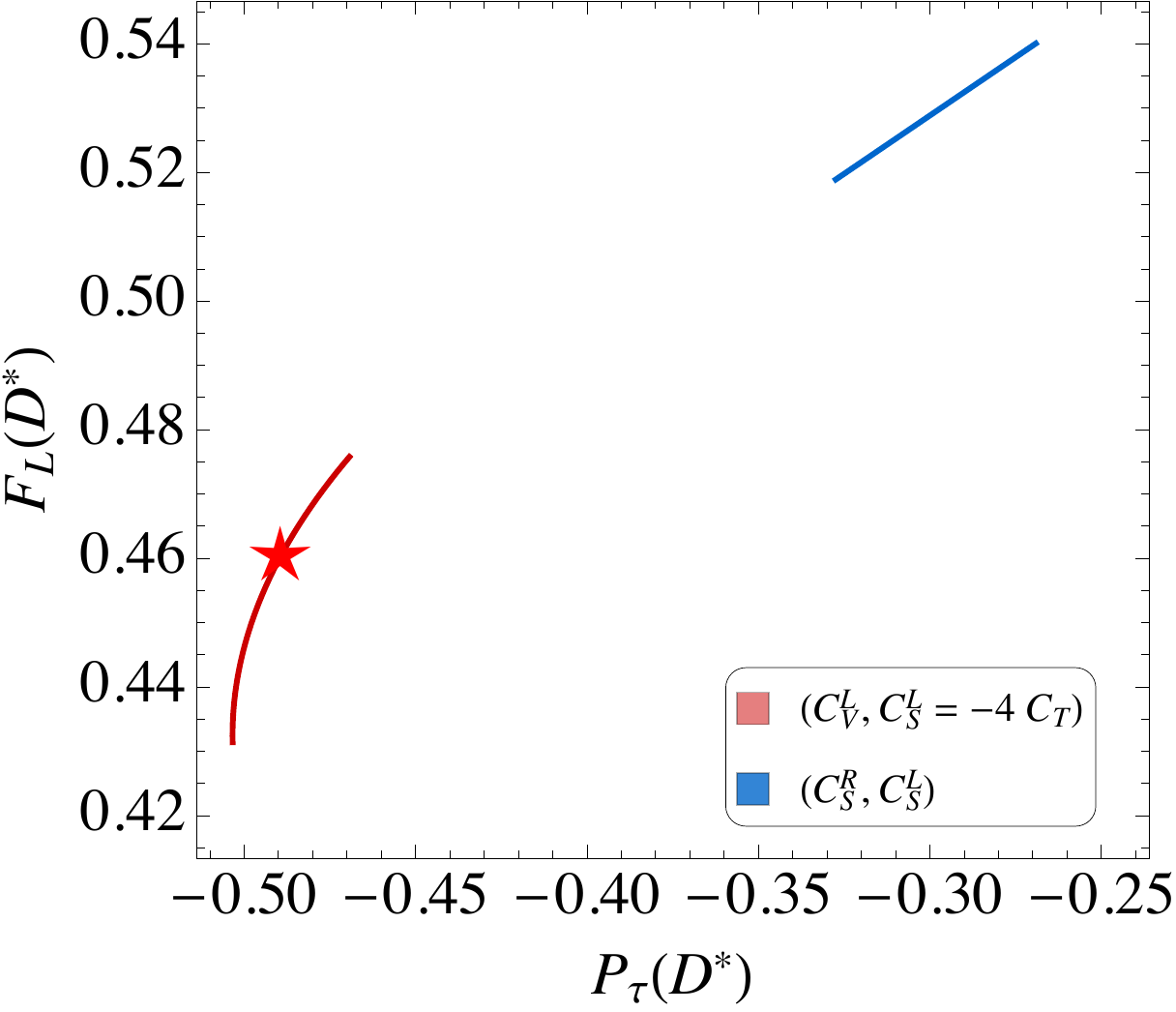}			}\quad
	\subfigure{	
		\includegraphics[width=0.40\textwidth, bb= 0 0 350 297]{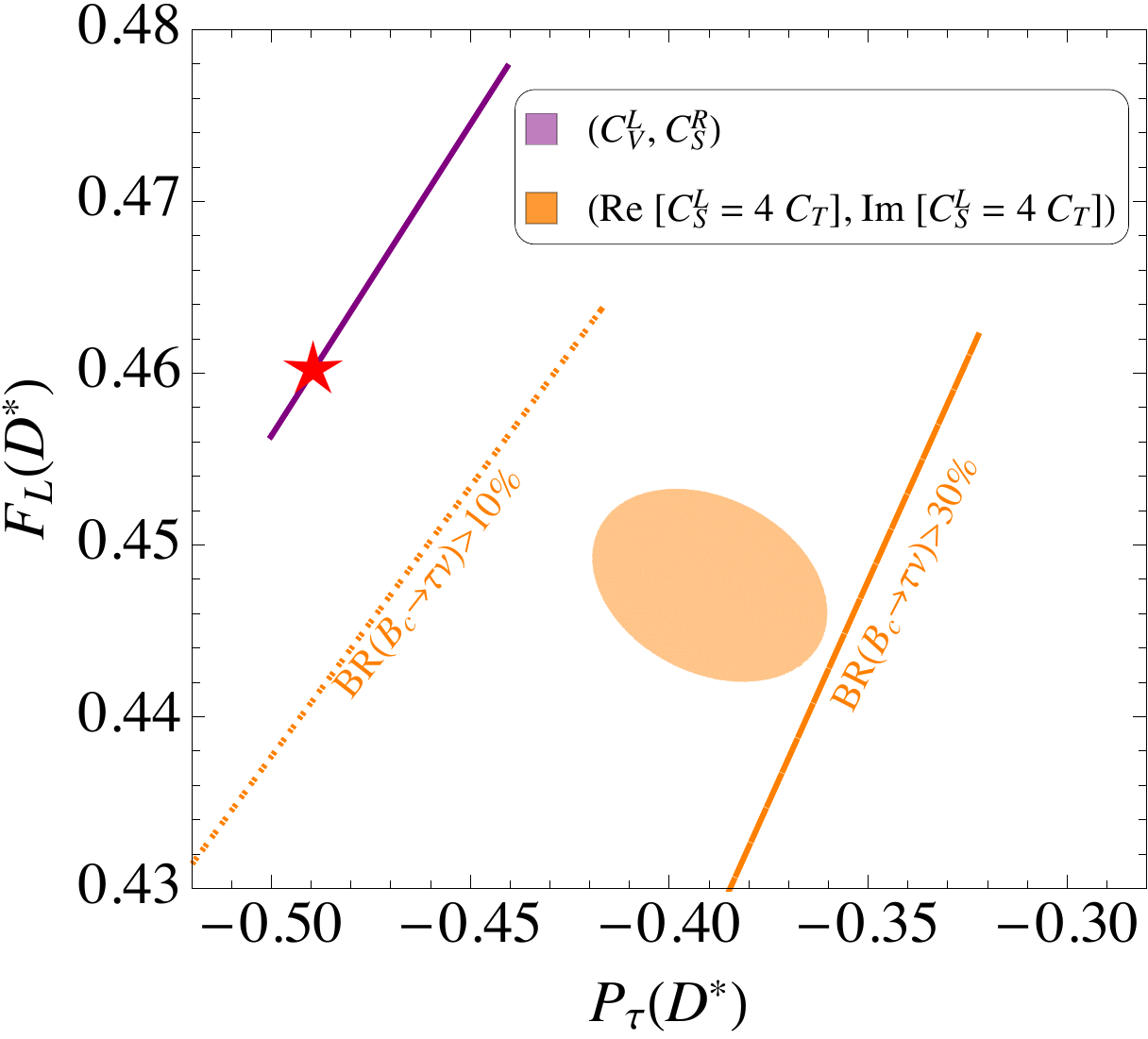}        }\\
\caption{Pairwise correlations between the observables
$P_\tau (D), P_\tau (D^\ast)$ and $F_L(D^\ast)$. The predicted $1\,\sigma$ regions in the four two-dimensional scenarios (assuming $\bbc<60\%$) are shown. The color coding is the same as in Fig.~\ref{Fig:Correl-First}. Note that in some cases the preferred regions shrink to lines, revealing the tight correlations between two observables in a given scenario. However, it is important to keep in mind that these correlations are obtained in the limit of vanishing form-factor uncertainties. Therefore, they represent the ideal correlations which can only in principle be obtained in a given scenario.}
\label{Fig:Correl-Second}
\end{figure*}

Figure~\ref{Fig:Correl-Second} reveals interesting correlations between polarization observables, including the yet unmeasured tau
polarization in the $B\to D\tau\nu$ decay mode. These correlations provide a
strong tool to discriminate between different NP scenarios, especially
in the cases in which the predicted regions shrink effectively to a line
(i.e., exhibit direct correlations). In the case of the correlation
between $P_\tau(D^\ast)$ and $F_L(D^\ast)$ considered within the $(C^R_S,
C^L_S)$ scenario (plot in the third row on the left) this follows trivially since both observables are affected by the pseudoscalar combination $C^R_S-C^L_S$ only. {The tight correlations in the other cases are, on the other hand, a result of the polarization observables being insensitive to the value of $C_V^L$.} However, it is very important to keep in mind that these correlations are obtained for vanishing uncertainties of the form factors. Therefore, they represent the correlations which can only in principle be obtained in a given scenario. Therefore, for exploiting such correlations future improvements on the theory predictions for form factors are imperative. 

\begin{figure*}[tp]
\begin{center}
	\subfigure{
		\includegraphics[width=0.45\textwidth, bb=0 0 360 363]{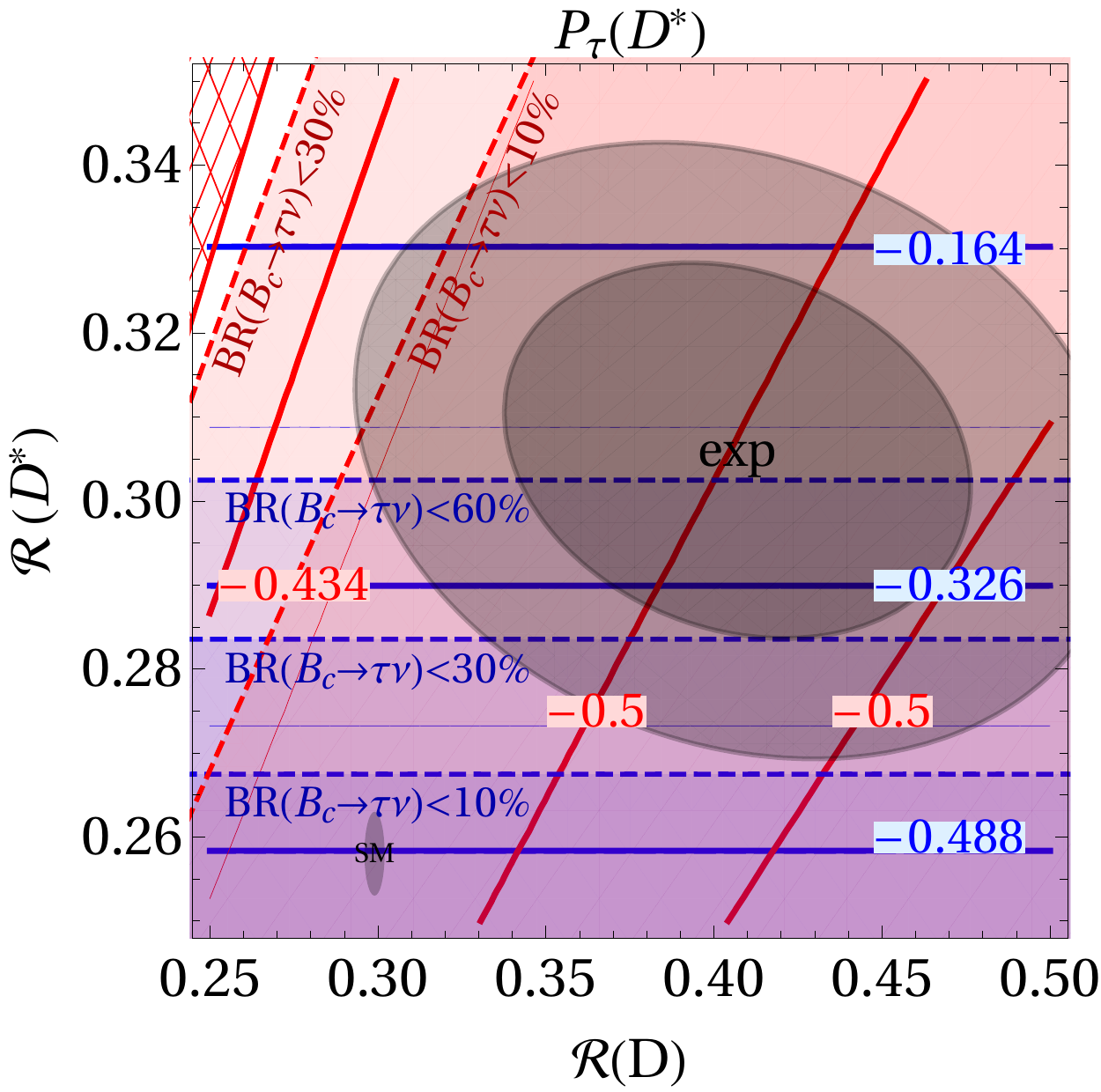}			}\qquad
	\subfigure{		
		\includegraphics[width=0.45\textwidth, bb =0 0  360 363]{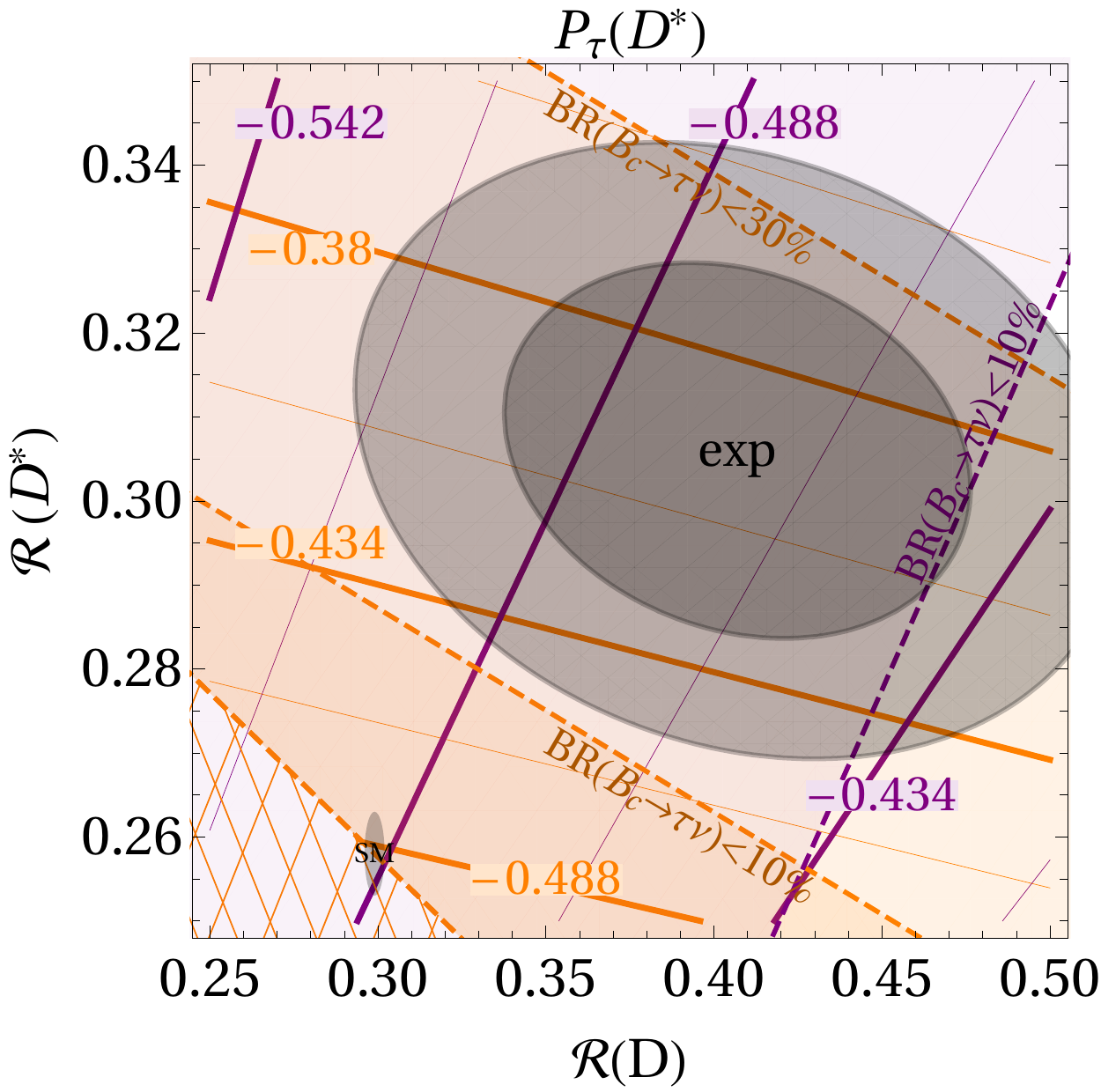}
		}\vspace{-5mm}\\
	\subfigure{	
		\includegraphics[width=0.45\textwidth, bb= 0 0 360 363]{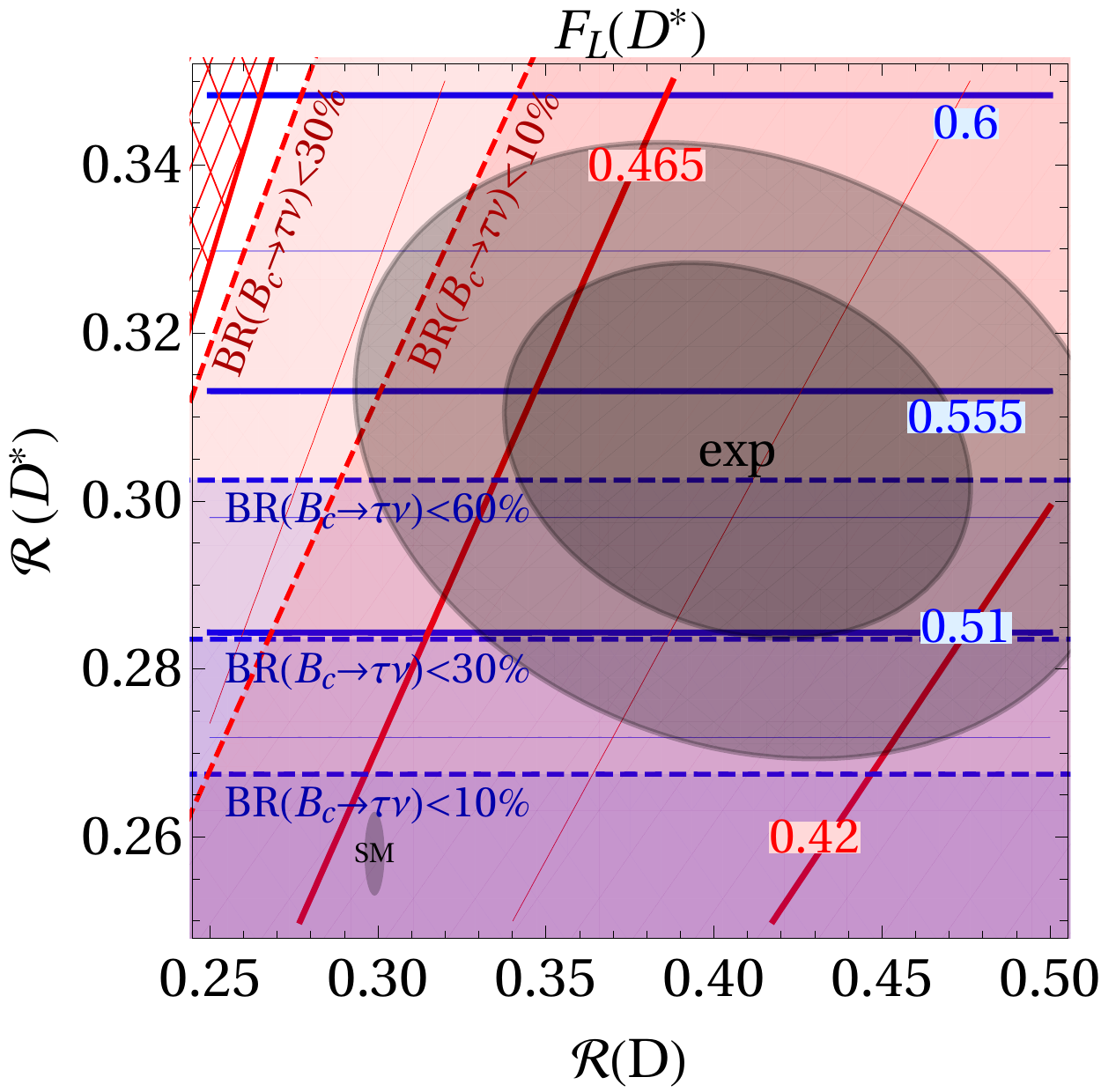}
		}\qquad
	\subfigure{
		\includegraphics[width=0.45\textwidth, bb= 0 0 360 363]{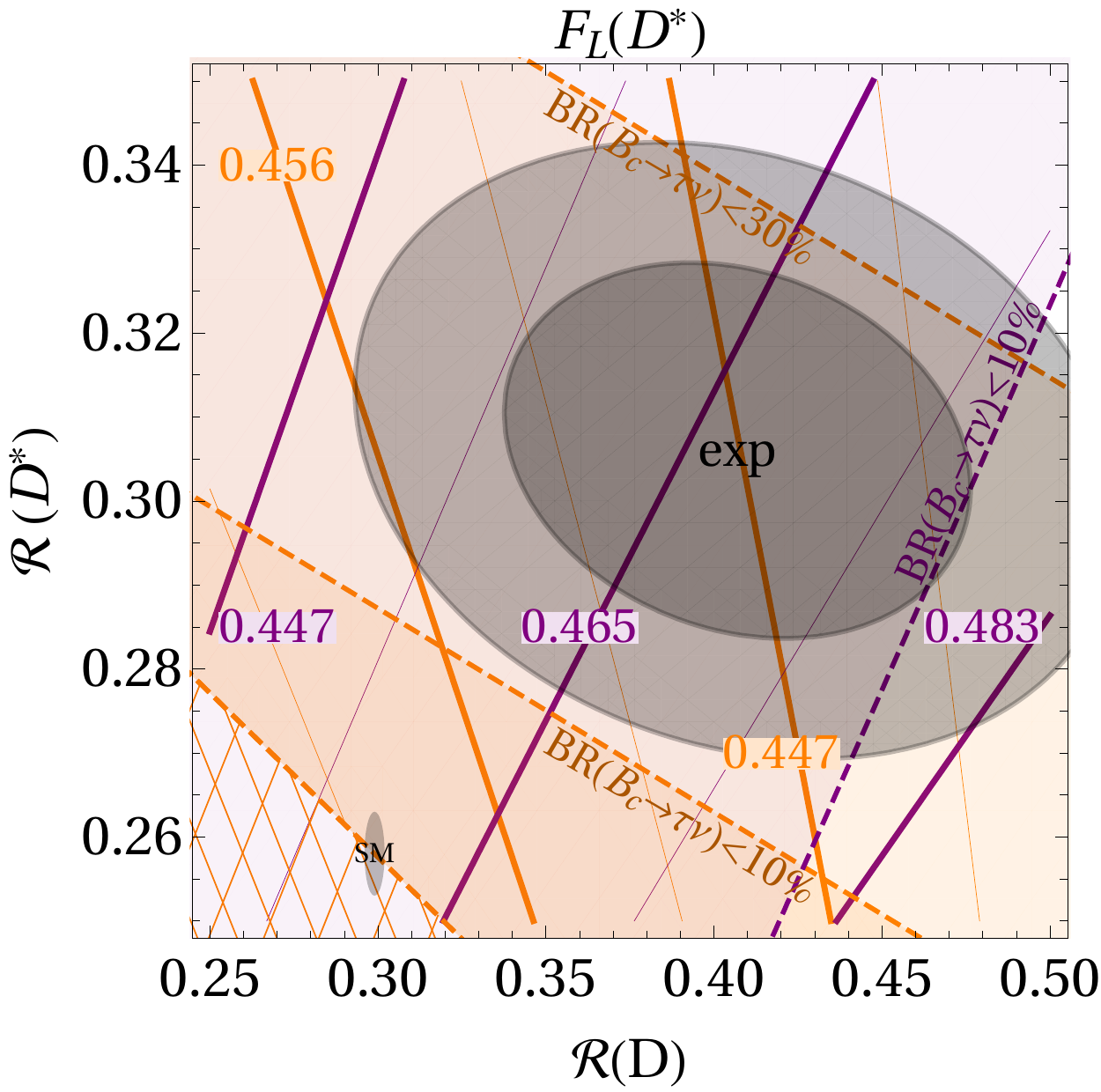}
		}\\	\vspace{1cm}
	\hspace{-5cm}	\includegraphics[width=0.2\textwidth, bb= -55 0 180 38]{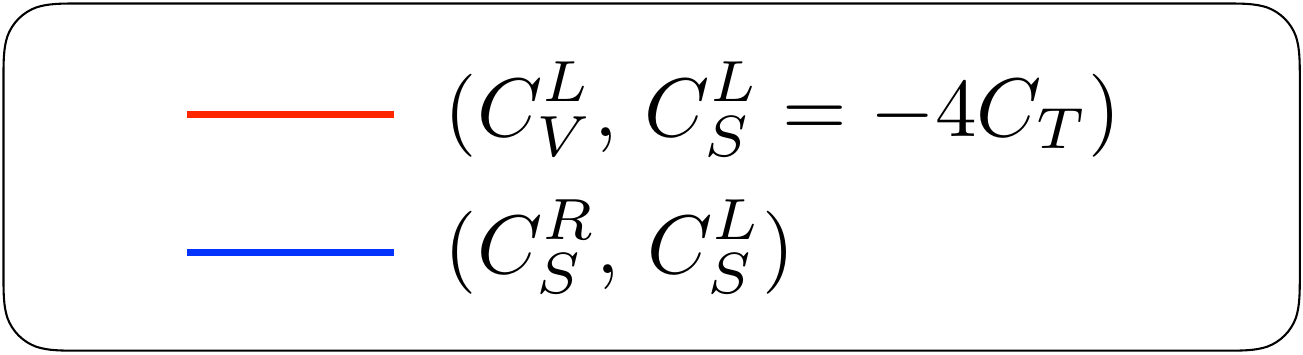}\qquad\qquad\qquad\qquad\qquad\qquad\qquad\qquad
\includegraphics[width=0.1\textwidth,bb = -25 0 95 38]{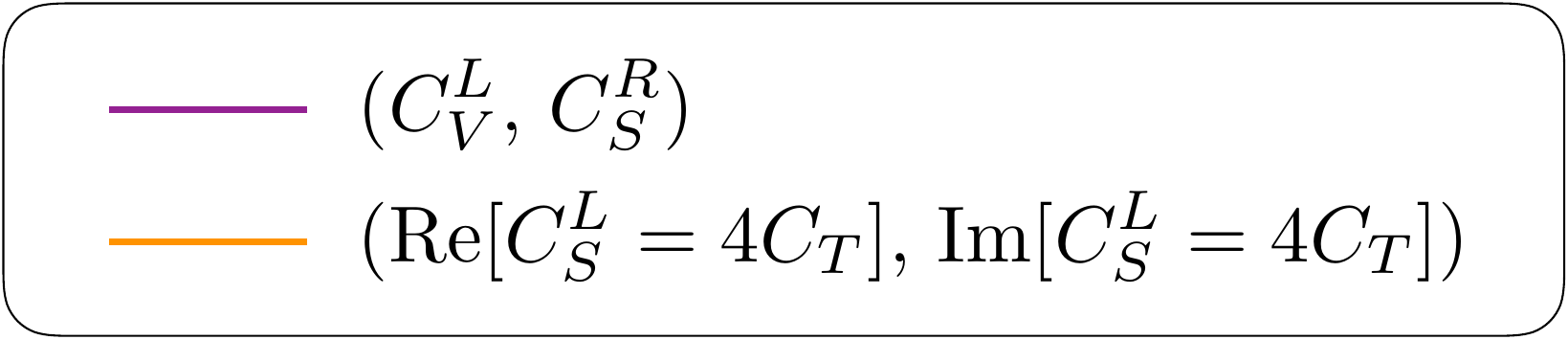}
\end{center}
\vspace{-0.5cm}
\caption{Contour lines of the $\tau$ polarization and the longitudinal
  $D^*$ polarization for the two-dimensional scenarios in the
  ${\cal R}(D)$--${\cal R}(D^*)$ plane. The colored regions (bounded by  dashed lines) are allowed by the $10\%$, $30\%$ and $60\%$ limits on  $\bbc$, where any area that would fill the entire plot is not shown  for convenience. The contours show the predicted values for the
  various observables (for vanishing form-factor uncertainties). The  thin lines carrying no labels depict the arithmetic means of the  neighboring thick lines. The gray regions are currently preferred by  data at the 1 and 2$\,\sigma$ levels. The colored, hatched regions  are excluded in the specific scenarios. Interestingly, the different  scenarios exhibit distinct correlations among the observables,
  manifesting themselves in the different slopes of the contours and the  different values associated with them. }
\label{RDRDstar1}
\end{figure*}

Let us now turn to future predictions and impact of the polarization observables. Here, we consider again the four two-dimensional scenarios of Sec.~\ref{sec:2D}. However, this time we do not use ${F_L(D^{\ast})}$, the tau polarizations as inputs for the fit, but rather predict them
as a function of ${\cal R}(D)$ and ${\cal R}(D^*)$. This is shown in Figs.~\ref{RDRDstar1} and \ref{RDRDstar2}. While the current experimental data for $P_\tau(D^*)$ do not significantly discriminate between the different scenarios, the preliminary $F_L(D^*)$ measurement shows a tension in the scenarios $(C_V^L,\,C_S^L=-4C_T)$ and complex $C_S^L=4C_T$. Furthermore, future measurements of $F_L(D^*)$ can be used to differentiate between different scenarios. This can be seen from the different slopes of the contour lines and the quite different values associated to them comparing the four scenarios.

\medskip
\section{Conclusions}\label{sec:conc}

Tauonic $B$ meson decays are excellent probes of physics beyond the SM (complementary to the direct searches at the LHC) since they are sensitive to lepton flavor universality violation in the tau sector, e.g., to Higgs bosons, $W^\prime$ bosons and leptoquarks. In order to distinguish different models, it is very important to be able to assess the presence of scalar and/or tensor operators: While Higgs bosons only generate scalar operators, LQs generate vector operators and possibly also scalar or tensor ones, while $W^\prime$ bosons only give  rise to vector operators. Thus, on the one hand, establishing the presence of scalar operators would rule out (pure) $W^\prime$ explanations while the presence of vector operators would exclude (pure) charged Higgs models. On the other hand, the combination of vector operators with scalar and/or tensor ones would be a strong indication for LQs. 

In this respect, the current Belle measurement of $F_L(D^{\ast})$ is very important and the limit on the NP contribution to $\bbc$ is crucial to establish or disprove scalar contributions.
Together with the measurements of the ratios $\mathcal{R}(D)$, $\mathcal{R}(D^*)$, $\mathcal{R}(J/\Psi)$, these observables can be used in the future to identify the Lorentz structure of NP. 

In this article we studied four one-dimensional scenarios (all with real Wilson coefficients) $C_V^L$,  $C_S^R$, $C_S^L$, and $C_S^L=4C_T$ and the four two-dimensional scenarios $(C_V^L,\,C_S^L=-4C_T)$, $(C_S^R,\,C_S^L)$, $(C_V^L,\,C_S^R)$ and $({\rm Re}[C_S^L=4C_T],\,{\rm Im}[C_S^L=4C_T])$. All these scenarios have in common that they can be generated by the exchange of a single new particle. The fit results are shown in Tables~\ref{tab:results1D} and~\ref{tab:results2D}.

For these scenarios we critically reconsidered the limits on the NP contribution to the decay $B_c\to\tau\nu$. Here we stress that the $10$\% limit~\cite{Akeroyd:2017mhr} 
on $\bbc$ from $Z\to b \bar b$ decays at LEP suffers from uncertainties related to the hadronization probability of a $b$ quark into a $B_c$ meson and should not be taken at face value. Furthermore, also the more conservative $30\%$ limit of Ref.~\cite{Alonso:2016oyd} is not strict since the error of the theory calculation of the $B_c$ lifetime has not fully been  taken into account. Therefore, a conservative limit of $60\%$ seems reasonable. Concerning the one-dimensional scenarios we found that the impact of the choice of the limit on $\bbc$ on the fit is very limited; only the $C_S^R$ scenario (which does not give a good fit to data anyway) is slightly affected if the hypothetical future bound of 10\% is chosen while the $C_V^L$ scenario always gives by far the best fit. However, on the two-dimensional scenarios the choice of the $\bbc$ limit has a significant impact. Using the conservative 60\% limit the $(C_S^R,\,C_S^L)$ scenario gives the best fit to data, while when enforcing the 30\% limit the agreement with data is significantly worse and for the 10\% limit this scenario is even strongly disfavored.

Next we studied the predictions for $\mathcal{R}(\Lambda_c)$ finding a \emph{sum rule} relating this ratio to $\mathcal{R}(D)$ and $\mathcal{R}(D^*)$, independent of any NP scenario up to small corrections. This implies that $\mathcal{R}(\Lambda_c)$ does not provide additional information on the Lorentz structure of NP but provides an important consistency check of the $\mathcal{R}(D)$ and $\mathcal{R}(D^*)$ measurements.

Finally, we considered the correlations among polarization observables and predicted them as functions of $\mathcal{R}(D)$ and $\mathcal{R}(D^*)$. Here we found strong correlations among the polarization observables, depending on the scenario chosen (see Fig.~\ref{Fig:Correl-Second}). Disregarding the form-factor uncertainties, even direct correlations are found. In the $(C^R_S, C^L_S)$ scenario this is due to the equal dependence of the observables on the Wilson coefficients while in the other cases the correlation is a result of the polarization observables being insensitive to the value of $C_V^L$.
 Furthermore, the polarization observables show a unique dependence on $\mathcal{R}(D)$ and $\mathcal{R}(D^*)$ for the different observables (see Figs.~\ref{RDRDstar1} and \ref{RDRDstar2}).

Therefore, future measurements of polarization observables together with $\mathcal{R}(D)$ and $\mathcal{R}(D^*)$ will be able to determine the Lorentz structure of NP while $\mathcal{R}(\Lambda_c)$ will serve as a consistency check. In this way different models (e.g., $W^\prime$, leptoquark and charged Higgs) can in principle be distinguished. However, for this exciting perspective also improved theory predictions for the form factors and ${
\rm BR}(B_c\to \tau \nu)$  are crucial.

\vspace{2mm} {\it Acknowledgments.}--- {\small The work of A.C. is
supported by an Ambizione Grant of the Swiss National Science
Foundation (PZ00P2\_154834).  M.M. acknowledges the support of the
DFG-funded Doctoral School ``Karlsruhe School of Elementary and
Astroparticle Physics: Science and Technology''. The work of S.dB.,
I.N., and U.N. is supported by BMBF under grant no. 05H18VKKB1. We thank
Syuhei Iguro, Yuji Omura, Ryoutaro Watanabe and Kei Yamamoto for sharing with us
their draft ``\emph{$D^\ast$ polarization vs. $R_{D^{(\ast)}}$ anomalies
  in the leptoquark models}'' \cite{Iguro:2018vqb} prior to its publication. We
acknowledge extensive discussions with Andrew Akeroyd  
{about Ref.~\cite{Akeroyd:2017mhr} and thank Admir Greljo for guiding us through 
Refs.~\cite{Greljo:2018tzh,greljotalk}.}}

\section*{Notes Added}

Recently Ref.~\cite{Greljo:2018tzh} appeared
which studies the constraints from the high-$p_T$ tails in mono-$\tau$
searches on the effective field theory (EFT) operators mediating $b\to c\tau\nu$ {and
  specific UV completions}. The authors {have found that the EFT
  analysis is valid for certain leptoquark models if the leptoquarks are
  sufficiently heavy, while UV completions with the exchange of a
  colorless particle in the s-channel require an explicit
  model-dependent study beyond the EFT framework. In
  Ref.~\cite{Greljo:2018tzh} a few scenarios are found for which already
  present high-$p_T$ data pose useful constraints on the Wilson
  coefficients challenging these scenarios as explanations of the
  $b\to c\tau\nu$ anomaly, but none of these scenarios is considered in
  this paper. However, we find that the study of
  Ref.~\cite{Greljo:2018tzh} constrains our two-dimensional scenario
  with complex $C_S^L=4C_T$, corresponding to the exchange of a
  (sufficiently heavy) leptoquark $S_2$. Inferring the allowed region
  from Ref.~\cite{greljotalk}, we realize that the best-fit point of the
  scenarios with $\bbc<30\%,60\%$ (see Table~\ref{tab:results2D}) and
  a large portion of the corresponding 2$\sigma$ area (in red) in the
  lower right plot of Fig.~\ref{WCdouble} is excluded by the
  2$\sigma$ bound $|C_S^L|\lesssim 0.35$ from high-$p_T$ data. 
 This latter bound  qualitatively mimics
  a stricter bound on $\bbc$, which would also push $|C_S^L|=4|C_T|$ to a
  smaller value. Thus if the $S_2$ scenario is realized in nature, we
  can expect effects in high-$p_T$ tails in mono-$\tau$ searches or
  eventually even the discovery of the $S_2$ leptoquark by ATLAS or CMS.} 
\vspace{2mm}

Updated results based on the HFLAV average for spring 2019 are available in
\cite{Blanke:2019qrx}.

\bibliography{BIB}

\begin{thebibliography}{138}%
\makeatletter
\providecommand \@ifxundefined [1]{%
 \@ifx{#1\undefined}
}%
\providecommand \@ifnum [1]{%
 \ifnum #1\expandafter \@firstoftwo
 \else \expandafter \@secondoftwo
 \fi
}%
\providecommand \@ifx [1]{%
 \ifx #1\expandafter \@firstoftwo
 \else \expandafter \@secondoftwo
 \fi
}%
\providecommand \natexlab [1]{#1}%
\providecommand \enquote  [1]{``#1''}%
\providecommand \bibnamefont  [1]{#1}%
\providecommand \bibfnamefont [1]{#1}%
\providecommand \citenamefont [1]{#1}%
\providecommand \href@noop [0]{\@secondoftwo}%
\providecommand \href [0]{\begingroup \@sanitize@url \@href}%
\providecommand \@href[1]{\@@startlink{#1}\@@href}%
\providecommand \@@href[1]{\endgroup#1\@@endlink}%
\providecommand \@sanitize@url [0]{\catcode `\\12\catcode `\$12\catcode
  `\&12\catcode `\#12\catcode `\^12\catcode `\_12\catcode `\%12\relax}%
\providecommand \@@startlink[1]{}%
\providecommand \@@endlink[0]{}%
\providecommand \url  [0]{\begingroup\@sanitize@url \@url }%
\providecommand \@url [1]{\endgroup\@href {#1}{\urlprefix }}%
\providecommand \urlprefix  [0]{URL }%
\providecommand \Eprint [0]{\href }%
\providecommand \doibase [0]{http://dx.doi.org/}%
\providecommand \selectlanguage [0]{\@gobble}%
\providecommand \bibinfo  [0]{\@secondoftwo}%
\providecommand \bibfield  [0]{\@secondoftwo}%
\providecommand \translation [1]{[#1]}%
\providecommand \BibitemOpen [0]{}%
\providecommand \bibitemStop [0]{}%
\providecommand \bibitemNoStop [0]{.\EOS\space}%
\providecommand \EOS [0]{\spacefactor3000\relax}%
\providecommand \BibitemShut  [1]{\csname bibitem#1\endcsname}%
\let\auto@bib@innerbib\@empty
\bibitem [{\citenamefont {Ricciardi}(2017)}]{Ricciardi:2016pmh}%
  \BibitemOpen
  \bibfield  {author} {\bibinfo {author} {\bibfnamefont {G.}~\bibnamefont
  {Ricciardi}},\ }\href {\doibase 10.1142/S0217732317300051} {\bibfield
  {journal} {\bibinfo  {journal} {Mod. Phys. Lett.}\ }\textbf {\bibinfo
  {volume} {A32}},\ \bibinfo {pages} {1730005} (\bibinfo {year} {2017})},\
  \Eprint {http://arxiv.org/abs/1610.04387} {arXiv:1610.04387 [hep-ph]}
  \BibitemShut {NoStop}%
\bibitem [{\citenamefont {Grinstein}(2017)}]{Grinstein:2016xpg}%
  \BibitemOpen
  \bibfield  {author} {\bibinfo {author} {\bibfnamefont {B.}~\bibnamefont
  {Grinstein}},\ }\bibfield  {booktitle} {\emph {\bibinfo {booktitle}
  {{Proceedings, 13th International Conference on Heavy Quarks and Leptons (HQL
  2016): Blacksburg, Virginia, USA, May 22-27, 2016}}},\ }\href {\doibase
  10.22323/1.274.0061} {\bibfield  {journal} {\bibinfo  {journal} {PoS}\
  }\textbf {\bibinfo {volume} {HQL2016}},\ \bibinfo {pages} {061} (\bibinfo
  {year} {2017})}\BibitemShut {NoStop}%
\bibitem [{\citenamefont {De~Fazio}(2017)}]{DeFazio:2017spv}%
  \BibitemOpen
  \bibfield  {author} {\bibinfo {author} {\bibfnamefont {F.}~\bibnamefont
  {De~Fazio}},\ }\bibfield  {booktitle} {\emph {\bibinfo {booktitle}
  {{Proceedings, 2017 European Physical Society Conference on High Energy
  Physics (EPS-HEP 2017): Venice, Italy, July 5-12, 2017}}},\ }\href {\doibase
  10.22323/1.314.0210} {\bibfield  {journal} {\bibinfo  {journal} {PoS}\
  }\textbf {\bibinfo {volume} {EPS-HEP2017}},\ \bibinfo {pages} {210} (\bibinfo
  {year} {2017})},\ \Eprint {http://arxiv.org/abs/1710.10017} {arXiv:1710.10017
  [hep-ph]} \BibitemShut {NoStop}%
\bibitem [{\citenamefont {Nierste}\ \emph {et~al.}(2008)\citenamefont
  {Nierste}, \citenamefont {Trine},\ and\ \citenamefont
  {Westhoff}}]{Nierste:2008qe}%
  \BibitemOpen
  \bibfield  {author} {\bibinfo {author} {\bibfnamefont {U.}~\bibnamefont
  {Nierste}}, \bibinfo {author} {\bibfnamefont {S.}~\bibnamefont {Trine}}, \
  and\ \bibinfo {author} {\bibfnamefont {S.}~\bibnamefont {Westhoff}},\ }\href
  {\doibase 10.1103/PhysRevD.78.015006} {\bibfield  {journal} {\bibinfo
  {journal} {Phys. Rev.}\ }\textbf {\bibinfo {volume} {D78}},\ \bibinfo {pages}
  {015006} (\bibinfo {year} {2008})},\ \Eprint {http://arxiv.org/abs/0801.4938}
  {arXiv:0801.4938 [hep-ph]} \BibitemShut {NoStop}%
\bibitem [{\citenamefont {Kamenik}\ and\ \citenamefont
  {Mescia}(2008)}]{Kamenik:2008tj}%
  \BibitemOpen
  \bibfield  {author} {\bibinfo {author} {\bibfnamefont {J.~F.}\ \bibnamefont
  {Kamenik}}\ and\ \bibinfo {author} {\bibfnamefont {F.}~\bibnamefont
  {Mescia}},\ }\href {\doibase 10.1103/PhysRevD.78.014003} {\bibfield
  {journal} {\bibinfo  {journal} {Phys. Rev.}\ }\textbf {\bibinfo {volume}
  {D78}},\ \bibinfo {pages} {014003} (\bibinfo {year} {2008})},\ \Eprint
  {http://arxiv.org/abs/0802.3790} {arXiv:0802.3790 [hep-ph]} \BibitemShut
  {NoStop}%
\bibitem [{\citenamefont {Fajfer}\ \emph
  {et~al.}(2012{\natexlab{a}})\citenamefont {Fajfer}, \citenamefont {Kamenik},\
  and\ \citenamefont {Nisandzic}}]{Fajfer:2012vx}%
  \BibitemOpen
  \bibfield  {author} {\bibinfo {author} {\bibfnamefont {S.}~\bibnamefont
  {Fajfer}}, \bibinfo {author} {\bibfnamefont {J.~F.}\ \bibnamefont {Kamenik}},
  \ and\ \bibinfo {author} {\bibfnamefont {I.}~\bibnamefont {Nisandzic}},\
  }\href {\doibase 10.1103/PhysRevD.85.094025} {\bibfield  {journal} {\bibinfo
  {journal} {Phys. Rev.}\ }\textbf {\bibinfo {volume} {D85}},\ \bibinfo {pages}
  {094025} (\bibinfo {year} {2012}{\natexlab{a}})},\ \Eprint
  {http://arxiv.org/abs/1203.2654} {arXiv:1203.2654 [hep-ph]} \BibitemShut
  {NoStop}%
\bibitem [{\citenamefont {Alonso}\ \emph
  {et~al.}(2017{\natexlab{a}})\citenamefont {Alonso}, \citenamefont
  {Martin~Camalich},\ and\ \citenamefont {Westhoff}}]{Alonso:2017ktd}%
  \BibitemOpen
  \bibfield  {author} {\bibinfo {author} {\bibfnamefont {R.}~\bibnamefont
  {Alonso}}, \bibinfo {author} {\bibfnamefont {J.}~\bibnamefont
  {Martin~Camalich}}, \ and\ \bibinfo {author} {\bibfnamefont {S.}~\bibnamefont
  {Westhoff}},\ }\href {\doibase 10.1103/PhysRevD.95.093006} {\bibfield
  {journal} {\bibinfo  {journal} {Phys. Rev.}\ }\textbf {\bibinfo {volume}
  {D95}},\ \bibinfo {pages} {093006} (\bibinfo {year} {2017}{\natexlab{a}})},\
  \Eprint {http://arxiv.org/abs/1702.02773} {arXiv:1702.02773 [hep-ph]}
  \BibitemShut {NoStop}%
\bibitem [{\citenamefont {Lees}\ \emph {et~al.}(2012)\citenamefont {Lees} \emph
  {et~al.}}]{Lees:2012xj}%
  \BibitemOpen
  \bibfield  {author} {\bibinfo {author} {\bibfnamefont {J.~P.}\ \bibnamefont
  {Lees}} \emph {et~al.} (\bibinfo {collaboration} {BaBar}),\ }\href {\doibase
  10.1103/PhysRevLett.109.101802} {\bibfield  {journal} {\bibinfo  {journal}
  {Phys. Rev. Lett.}\ }\textbf {\bibinfo {volume} {109}},\ \bibinfo {pages}
  {101802} (\bibinfo {year} {2012})},\ \Eprint {http://arxiv.org/abs/1205.5442}
  {arXiv:1205.5442 [hep-ex]} \BibitemShut {NoStop}%
\bibitem [{\citenamefont {Lees}\ \emph {et~al.}(2013)\citenamefont {Lees} \emph
  {et~al.}}]{Lees:2013uzd}%
  \BibitemOpen
  \bibfield  {author} {\bibinfo {author} {\bibfnamefont {J.~P.}\ \bibnamefont
  {Lees}} \emph {et~al.} (\bibinfo {collaboration} {BaBar}),\ }\href {\doibase
  10.1103/PhysRevD.88.072012} {\bibfield  {journal} {\bibinfo  {journal} {Phys.
  Rev.}\ }\textbf {\bibinfo {volume} {D88}},\ \bibinfo {pages} {072012}
  (\bibinfo {year} {2013})},\ \Eprint {http://arxiv.org/abs/1303.0571}
  {arXiv:1303.0571 [hep-ex]} \BibitemShut {NoStop}%
\bibitem [{\citenamefont {Huschle}\ \emph {et~al.}(2015)\citenamefont {Huschle}
  \emph {et~al.}}]{Huschle:2015rga}%
  \BibitemOpen
  \bibfield  {author} {\bibinfo {author} {\bibfnamefont {M.}~\bibnamefont
  {Huschle}} \emph {et~al.} (\bibinfo {collaboration} {Belle}),\ }\href
  {\doibase 10.1103/PhysRevD.92.072014} {\bibfield  {journal} {\bibinfo
  {journal} {Phys. Rev.}\ }\textbf {\bibinfo {volume} {D92}},\ \bibinfo {pages}
  {072014} (\bibinfo {year} {2015})},\ \Eprint
  {http://arxiv.org/abs/1507.03233} {arXiv:1507.03233 [hep-ex]} \BibitemShut
  {NoStop}%
\bibitem [{\citenamefont {Sato}\ \emph {et~al.}(2016)\citenamefont {Sato} \emph
  {et~al.}}]{Sato:2016svk}%
  \BibitemOpen
  \bibfield  {author} {\bibinfo {author} {\bibfnamefont {Y.}~\bibnamefont
  {Sato}} \emph {et~al.} (\bibinfo {collaboration} {Belle}),\ }\href {\doibase
  10.1103/PhysRevD.94.072007} {\bibfield  {journal} {\bibinfo  {journal} {Phys.
  Rev.}\ }\textbf {\bibinfo {volume} {D94}},\ \bibinfo {pages} {072007}
  (\bibinfo {year} {2016})},\ \Eprint {http://arxiv.org/abs/1607.07923}
  {arXiv:1607.07923 [hep-ex]} \BibitemShut {NoStop}%
\bibitem [{\citenamefont {Hirose}\ \emph {et~al.}(2017)\citenamefont {Hirose}
  \emph {et~al.}}]{Hirose:2016wfn}%
  \BibitemOpen
  \bibfield  {author} {\bibinfo {author} {\bibfnamefont {S.}~\bibnamefont
  {Hirose}} \emph {et~al.} (\bibinfo {collaboration} {Belle}),\ }\href
  {\doibase 10.1103/PhysRevLett.118.211801} {\bibfield  {journal} {\bibinfo
  {journal} {Phys. Rev. Lett.}\ }\textbf {\bibinfo {volume} {118}},\ \bibinfo
  {pages} {211801} (\bibinfo {year} {2017})},\ \Eprint
  {http://arxiv.org/abs/1612.00529} {arXiv:1612.00529 [hep-ex]} \BibitemShut
  {NoStop}%
\bibitem [{\citenamefont {Hirose}\ \emph {et~al.}(2018)\citenamefont {Hirose}
  \emph {et~al.}}]{Hirose:2017dxl}%
  \BibitemOpen
  \bibfield  {author} {\bibinfo {author} {\bibfnamefont {S.}~\bibnamefont
  {Hirose}} \emph {et~al.} (\bibinfo {collaboration} {Belle}),\ }\href
  {\doibase 10.1103/PhysRevD.97.012004} {\bibfield  {journal} {\bibinfo
  {journal} {Phys. Rev.}\ }\textbf {\bibinfo {volume} {D97}},\ \bibinfo {pages}
  {012004} (\bibinfo {year} {2018})},\ \Eprint
  {http://arxiv.org/abs/1709.00129} {arXiv:1709.00129 [hep-ex]} \BibitemShut
  {NoStop}%
\bibitem [{\citenamefont {Aaij}\ \emph
  {et~al.}(2015{\natexlab{a}})\citenamefont {Aaij} \emph
  {et~al.}}]{Aaij:2015yra}%
  \BibitemOpen
  \bibfield  {author} {\bibinfo {author} {\bibfnamefont {R.}~\bibnamefont
  {Aaij}} \emph {et~al.} (\bibinfo {collaboration} {LHCb}),\ }\href {\doibase
  10.1103/PhysRevLett.115.159901, 10.1103/PhysRevLett.115.111803} {\bibfield
  {journal} {\bibinfo  {journal} {Phys. Rev. Lett.}\ }\textbf {\bibinfo
  {volume} {115}},\ \bibinfo {pages} {111803} (\bibinfo {year}
  {2015}{\natexlab{a}})},\ \bibinfo {note} {[Erratum: Phys. Rev.
  Lett.115,no.15,159901(2015)]},\ \Eprint {http://arxiv.org/abs/1506.08614}
  {arXiv:1506.08614 [hep-ex]} \BibitemShut {NoStop}%
\bibitem [{\citenamefont {Aaij}\ \emph
  {et~al.}(2018{\natexlab{a}})\citenamefont {Aaij} \emph
  {et~al.}}]{Aaij:2017uff}%
  \BibitemOpen
  \bibfield  {author} {\bibinfo {author} {\bibfnamefont {R.}~\bibnamefont
  {Aaij}} \emph {et~al.} (\bibinfo {collaboration} {LHCb}),\ }\href {\doibase
  10.1103/PhysRevLett.120.171802} {\bibfield  {journal} {\bibinfo  {journal}
  {Phys. Rev. Lett.}\ }\textbf {\bibinfo {volume} {120}},\ \bibinfo {pages}
  {171802} (\bibinfo {year} {2018}{\natexlab{a}})},\ \Eprint
  {http://arxiv.org/abs/1708.08856} {arXiv:1708.08856 [hep-ex]} \BibitemShut
  {NoStop}%
\bibitem [{\citenamefont {Aaij}\ \emph
  {et~al.}(2018{\natexlab{b}})\citenamefont {Aaij} \emph
  {et~al.}}]{Aaij:2017deq}%
  \BibitemOpen
  \bibfield  {author} {\bibinfo {author} {\bibfnamefont {R.}~\bibnamefont
  {Aaij}} \emph {et~al.} (\bibinfo {collaboration} {LHCb}),\ }\href {\doibase
  10.1103/PhysRevD.97.072013} {\bibfield  {journal} {\bibinfo  {journal} {Phys.
  Rev.}\ }\textbf {\bibinfo {volume} {D97}},\ \bibinfo {pages} {072013}
  (\bibinfo {year} {2018}{\natexlab{b}})},\ \Eprint
  {http://arxiv.org/abs/1711.02505} {arXiv:1711.02505 [hep-ex]} \BibitemShut
  {NoStop}%
\bibitem [{\citenamefont {Amhis}\ \emph {et~al.}(2017)\citenamefont {Amhis}
  \emph {et~al.}}]{Amhis:2016xyh}%
  \BibitemOpen
  \bibfield  {author} {\bibinfo {author} {\bibfnamefont {Y.}~\bibnamefont
  {Amhis}} \emph {et~al.} (\bibinfo {collaboration} {HFLAV}),\ }\href {\doibase
  10.1140/epjc/s10052-017-5058-4} {\bibfield  {journal} {\bibinfo  {journal}
  {Eur. Phys. J.}\ }\textbf {\bibinfo {volume} {C77}},\ \bibinfo {pages} {895}
  (\bibinfo {year} {2017})},\ \bibinfo {note} {{U}pdated average of $R(D)$ and
  $R(D^{\ast})$ for Summer 2018 at
  \url{https://hflav-eos.web.cern.ch/hflav-eos/semi/summer18/RDRDs.html}},\
  \Eprint {http://arxiv.org/abs/1612.07233} {arXiv:1612.07233 [hep-ex]}
  \BibitemShut {NoStop}%
\bibitem [{\citenamefont {Bigi}\ and\ \citenamefont
  {Gambino}(2016)}]{Bigi:2016mdz}%
  \BibitemOpen
  \bibfield  {author} {\bibinfo {author} {\bibfnamefont {D.}~\bibnamefont
  {Bigi}}\ and\ \bibinfo {author} {\bibfnamefont {P.}~\bibnamefont {Gambino}},\
  }\href {\doibase 10.1103/PhysRevD.94.094008} {\bibfield  {journal} {\bibinfo
  {journal} {Phys. Rev.}\ }\textbf {\bibinfo {volume} {D94}},\ \bibinfo {pages}
  {094008} (\bibinfo {year} {2016})},\ \Eprint
  {http://arxiv.org/abs/1606.08030} {arXiv:1606.08030 [hep-ph]} \BibitemShut
  {NoStop}%
\bibitem [{\citenamefont {Bernlochner}\ \emph {et~al.}(2017)\citenamefont
  {Bernlochner}, \citenamefont {Ligeti}, \citenamefont {Papucci},\ and\
  \citenamefont {Robinson}}]{Bernlochner:2017jka}%
  \BibitemOpen
  \bibfield  {author} {\bibinfo {author} {\bibfnamefont {F.~U.}\ \bibnamefont
  {Bernlochner}}, \bibinfo {author} {\bibfnamefont {Z.}~\bibnamefont {Ligeti}},
  \bibinfo {author} {\bibfnamefont {M.}~\bibnamefont {Papucci}}, \ and\
  \bibinfo {author} {\bibfnamefont {D.~J.}\ \bibnamefont {Robinson}},\ }\href
  {\doibase 10.1103/PhysRevD.95.115008, 10.1103/PhysRevD.97.059902} {\bibfield
  {journal} {\bibinfo  {journal} {Phys. Rev.}\ }\textbf {\bibinfo {volume}
  {D95}},\ \bibinfo {pages} {115008} (\bibinfo {year} {2017})},\ \bibinfo
  {note} {[Erratum: Phys. Rev.D97,no.5,059902(2018)]},\ \Eprint
  {http://arxiv.org/abs/1703.05330} {arXiv:1703.05330 [hep-ph]} \BibitemShut
  {NoStop}%
\bibitem [{\citenamefont {Bigi}\ \emph {et~al.}(2017)\citenamefont {Bigi},
  \citenamefont {Gambino},\ and\ \citenamefont {Schacht}}]{Bigi:2017jbd}%
  \BibitemOpen
  \bibfield  {author} {\bibinfo {author} {\bibfnamefont {D.}~\bibnamefont
  {Bigi}}, \bibinfo {author} {\bibfnamefont {P.}~\bibnamefont {Gambino}}, \
  and\ \bibinfo {author} {\bibfnamefont {S.}~\bibnamefont {Schacht}},\ }\href
  {\doibase 10.1007/JHEP11(2017)061} {\bibfield  {journal} {\bibinfo  {journal}
  {JHEP}\ }\textbf {\bibinfo {volume} {11}},\ \bibinfo {pages} {061} (\bibinfo
  {year} {2017})},\ \Eprint {http://arxiv.org/abs/1707.09509} {arXiv:1707.09509
  [hep-ph]} \BibitemShut {NoStop}%
\bibitem [{\citenamefont {Jaiswal}\ \emph {et~al.}(2017)\citenamefont
  {Jaiswal}, \citenamefont {Nandi},\ and\ \citenamefont
  {Patra}}]{Jaiswal:2017rve}%
  \BibitemOpen
  \bibfield  {author} {\bibinfo {author} {\bibfnamefont {S.}~\bibnamefont
  {Jaiswal}}, \bibinfo {author} {\bibfnamefont {S.}~\bibnamefont {Nandi}}, \
  and\ \bibinfo {author} {\bibfnamefont {S.~K.}\ \bibnamefont {Patra}},\ }\href
  {\doibase 10.1007/JHEP12(2017)060} {\bibfield  {journal} {\bibinfo  {journal}
  {JHEP}\ }\textbf {\bibinfo {volume} {12}},\ \bibinfo {pages} {060} (\bibinfo
  {year} {2017})},\ \Eprint {http://arxiv.org/abs/1707.09977} {arXiv:1707.09977
  [hep-ph]} \BibitemShut {NoStop}%
\bibitem [{\citenamefont {Becirevic}\ and\ \citenamefont
  {Kosnik}(2010)}]{Becirevic:2009fy}%
  \BibitemOpen
  \bibfield  {author} {\bibinfo {author} {\bibfnamefont {D.}~\bibnamefont
  {Becirevic}}\ and\ \bibinfo {author} {\bibfnamefont {N.}~\bibnamefont
  {Kosnik}},\ }\bibfield  {booktitle} {\emph {\bibinfo {booktitle}
  {{Primosten09: Progress and Challenges in Flavour Physics Primosten, Croatia,
  September 29-October 3, 2009}}},\ }\href@noop {} {\bibfield  {journal}
  {\bibinfo  {journal} {Acta Phys. Polon. Supp.}\ }\textbf {\bibinfo {volume}
  {3}},\ \bibinfo {pages} {207} (\bibinfo {year} {2010})},\ \Eprint
  {http://arxiv.org/abs/0910.5031} {arXiv:0910.5031 [hep-ph]} \BibitemShut
  {NoStop}%
\bibitem [{\citenamefont {de~Boer}\ \emph {et~al.}(2018)\citenamefont
  {de~Boer}, \citenamefont {Kitahara},\ and\ \citenamefont
  {Nisandzic}}]{deBoer:2018ipi}%
  \BibitemOpen
  \bibfield  {author} {\bibinfo {author} {\bibfnamefont {S.}~\bibnamefont
  {de~Boer}}, \bibinfo {author} {\bibfnamefont {T.}~\bibnamefont {Kitahara}}, \
  and\ \bibinfo {author} {\bibfnamefont {I.}~\bibnamefont {Nisandzic}},\ }\href
  {\doibase 10.1103/PhysRevLett.120.261804} {\bibfield  {journal} {\bibinfo
  {journal} {Phys. Rev. Lett.}\ }\textbf {\bibinfo {volume} {120}},\ \bibinfo
  {pages} {261804} (\bibinfo {year} {2018})},\ \Eprint
  {http://arxiv.org/abs/1803.05881} {arXiv:1803.05881 [hep-ph]} \BibitemShut
  {NoStop}%
\bibitem [{\citenamefont {Bailey}\ \emph {et~al.}(2015)\citenamefont {Bailey}
  \emph {et~al.}}]{Lattice:2015rga}%
  \BibitemOpen
  \bibfield  {author} {\bibinfo {author} {\bibfnamefont {J.~A.}\ \bibnamefont
  {Bailey}} \emph {et~al.} (\bibinfo {collaboration} {MILC}),\ }\href {\doibase
  10.1103/PhysRevD.92.034506} {\bibfield  {journal} {\bibinfo  {journal} {Phys.
  Rev.}\ }\textbf {\bibinfo {volume} {D92}},\ \bibinfo {pages} {034506}
  (\bibinfo {year} {2015})},\ \Eprint {http://arxiv.org/abs/1503.07237}
  {arXiv:1503.07237 [hep-lat]} \BibitemShut {NoStop}%
\bibitem [{\citenamefont {Na}\ \emph {et~al.}(2015)\citenamefont {Na},
  \citenamefont {Bouchard}, \citenamefont {Lepage}, \citenamefont {Monahan},\
  and\ \citenamefont {Shigemitsu}}]{Na:2015kha}%
  \BibitemOpen
  \bibfield  {author} {\bibinfo {author} {\bibfnamefont {H.}~\bibnamefont
  {Na}}, \bibinfo {author} {\bibfnamefont {C.~M.}\ \bibnamefont {Bouchard}},
  \bibinfo {author} {\bibfnamefont {G.~P.}\ \bibnamefont {Lepage}}, \bibinfo
  {author} {\bibfnamefont {C.}~\bibnamefont {Monahan}}, \ and\ \bibinfo
  {author} {\bibfnamefont {J.}~\bibnamefont {Shigemitsu}} (\bibinfo
  {collaboration} {HPQCD}),\ }\href {\doibase 10.1103/PhysRevD.93.119906,
  10.1103/PhysRevD.92.054510} {\bibfield  {journal} {\bibinfo  {journal} {Phys.
  Rev.}\ }\textbf {\bibinfo {volume} {D92}},\ \bibinfo {pages} {054510}
  (\bibinfo {year} {2015})},\ \bibinfo {note} {[Erratum: Phys.
  Rev.D93,no.11,119906(2016)]},\ \Eprint {http://arxiv.org/abs/1505.03925}
  {arXiv:1505.03925 [hep-lat]} \BibitemShut {NoStop}%
\bibitem [{\citenamefont {Aaij}\ \emph
  {et~al.}(2018{\natexlab{c}})\citenamefont {Aaij} \emph
  {et~al.}}]{Aaij:2017tyk}%
  \BibitemOpen
  \bibfield  {author} {\bibinfo {author} {\bibfnamefont {R.}~\bibnamefont
  {Aaij}} \emph {et~al.} (\bibinfo {collaboration} {LHCb}),\ }\href {\doibase
  10.1103/PhysRevLett.120.121801} {\bibfield  {journal} {\bibinfo  {journal}
  {Phys. Rev. Lett.}\ }\textbf {\bibinfo {volume} {120}},\ \bibinfo {pages}
  {121801} (\bibinfo {year} {2018}{\natexlab{c}})},\ \Eprint
  {http://arxiv.org/abs/1711.05623} {arXiv:1711.05623 [hep-ex]} \BibitemShut
  {NoStop}%
\bibitem [{\citenamefont {Murphy}\ and\ \citenamefont
  {Soni}(2018)}]{Murphy:2018sqg}%
  \BibitemOpen
  \bibfield  {author} {\bibinfo {author} {\bibfnamefont {C.~W.}\ \bibnamefont
  {Murphy}}\ and\ \bibinfo {author} {\bibfnamefont {A.}~\bibnamefont {Soni}},\
  }\href@noop {} {\  (\bibinfo {year} {2018})},\ \Eprint
  {http://arxiv.org/abs/1808.05932} {arXiv:1808.05932 [hep-ph]} \BibitemShut
  {NoStop}%
\bibitem [{\citenamefont {Cohen}\ \emph {et~al.}(2018)\citenamefont {Cohen},
  \citenamefont {Lamm},\ and\ \citenamefont {Lebed}}]{Cohen:2018dgz}%
  \BibitemOpen
  \bibfield  {author} {\bibinfo {author} {\bibfnamefont {T.~D.}\ \bibnamefont
  {Cohen}}, \bibinfo {author} {\bibfnamefont {H.}~\bibnamefont {Lamm}}, \ and\
  \bibinfo {author} {\bibfnamefont {R.~F.}\ \bibnamefont {Lebed}},\ }\href
  {\doibase 10.1007/JHEP09(2018)168} {\bibfield  {journal} {\bibinfo  {journal}
  {JHEP}\ }\textbf {\bibinfo {volume} {09}},\ \bibinfo {pages} {168} (\bibinfo
  {year} {2018})},\ \Eprint {http://arxiv.org/abs/1807.02730} {arXiv:1807.02730
  [hep-ph]} \BibitemShut {NoStop}%
\bibitem [{\citenamefont {Tran}\ \emph {et~al.}(2018)\citenamefont {Tran},
  \citenamefont {Ivanov}, \citenamefont {K{\"o}rner},\ and\ \citenamefont
  {Santorelli}}]{Tran:2018kuv}%
  \BibitemOpen
  \bibfield  {author} {\bibinfo {author} {\bibfnamefont {C.-T.}\ \bibnamefont
  {Tran}}, \bibinfo {author} {\bibfnamefont {M.~A.}\ \bibnamefont {Ivanov}},
  \bibinfo {author} {\bibfnamefont {J.~G.}\ \bibnamefont {K{\"o}rner}}, \ and\
  \bibinfo {author} {\bibfnamefont {P.}~\bibnamefont {Santorelli}},\ }\href
  {\doibase 10.1103/PhysRevD.97.054014} {\bibfield  {journal} {\bibinfo
  {journal} {Phys. Rev.}\ }\textbf {\bibinfo {volume} {D97}},\ \bibinfo {pages}
  {054014} (\bibinfo {year} {2018})},\ \Eprint
  {http://arxiv.org/abs/1801.06927} {arXiv:1801.06927 [hep-ph]} \BibitemShut
  {NoStop}%
\bibitem [{\citenamefont {Watanabe}(2018)}]{Watanabe:2017mip}%
  \BibitemOpen
  \bibfield  {author} {\bibinfo {author} {\bibfnamefont {R.}~\bibnamefont
  {Watanabe}},\ }\href {\doibase 10.1016/j.physletb.2017.11.016} {\bibfield
  {journal} {\bibinfo  {journal} {Phys. Lett.}\ }\textbf {\bibinfo {volume}
  {B776}},\ \bibinfo {pages} {5} (\bibinfo {year} {2018})},\ \Eprint
  {http://arxiv.org/abs/1709.08644} {arXiv:1709.08644 [hep-ph]} \BibitemShut
  {NoStop}%
\bibitem [{\citenamefont {Chauhan}\ and\ \citenamefont
  {Kindra}(2017)}]{Chauhan:2017uil}%
  \BibitemOpen
  \bibfield  {author} {\bibinfo {author} {\bibfnamefont {B.}~\bibnamefont
  {Chauhan}}\ and\ \bibinfo {author} {\bibfnamefont {B.}~\bibnamefont
  {Kindra}},\ }\href@noop {} {\  (\bibinfo {year} {2017})},\ \Eprint
  {http://arxiv.org/abs/1709.09989} {arXiv:1709.09989 [hep-ph]} \BibitemShut
  {NoStop}%
\bibitem [{\citenamefont {Detmold}\ \emph {et~al.}(2015)\citenamefont
  {Detmold}, \citenamefont {Lehner},\ and\ \citenamefont
  {Meinel}}]{Detmold:2015aaa}%
  \BibitemOpen
  \bibfield  {author} {\bibinfo {author} {\bibfnamefont {W.}~\bibnamefont
  {Detmold}}, \bibinfo {author} {\bibfnamefont {C.}~\bibnamefont {Lehner}}, \
  and\ \bibinfo {author} {\bibfnamefont {S.}~\bibnamefont {Meinel}},\ }\href
  {\doibase 10.1103/PhysRevD.92.034503} {\bibfield  {journal} {\bibinfo
  {journal} {Phys. Rev.}\ }\textbf {\bibinfo {volume} {D92}},\ \bibinfo {pages}
  {034503} (\bibinfo {year} {2015})},\ \Eprint
  {http://arxiv.org/abs/1503.01421} {arXiv:1503.01421 [hep-lat]} \BibitemShut
  {NoStop}%
\bibitem [{\citenamefont {Adamczyk}(2018)}]{Adamczyk}%
  \BibitemOpen
  \bibfield  {author} {\bibinfo {author} {\bibfnamefont {K.}~\bibnamefont
  {Adamczyk}},\ }\href@noop {} {\enquote {\bibinfo {title} {{B to semitauonic
  decays at Belle/Belle II}},}\ } (\bibinfo {year} {2018}),\ \bibinfo {note}
  {talk at \emph{10th International Workshop on the CKM Unitarity Triangle},
  Heidelberg, 17-21 Sep 2018}\BibitemShut {NoStop}%
\bibitem [{\citenamefont {Alok}\ \emph {et~al.}(2017)\citenamefont {Alok},
  \citenamefont {Kumar}, \citenamefont {Kumbhakar},\ and\ \citenamefont
  {Sankar}}]{Alok:2016qyh}%
  \BibitemOpen
  \bibfield  {author} {\bibinfo {author} {\bibfnamefont {A.~K.}\ \bibnamefont
  {Alok}}, \bibinfo {author} {\bibfnamefont {D.}~\bibnamefont {Kumar}},
  \bibinfo {author} {\bibfnamefont {S.}~\bibnamefont {Kumbhakar}}, \ and\
  \bibinfo {author} {\bibfnamefont {S.~U.}\ \bibnamefont {Sankar}},\ }\href
  {\doibase 10.1103/PhysRevD.95.115038} {\bibfield  {journal} {\bibinfo
  {journal} {Phys. Rev.}\ }\textbf {\bibinfo {volume} {D95}},\ \bibinfo {pages}
  {115038} (\bibinfo {year} {2017})},\ \Eprint
  {http://arxiv.org/abs/1606.03164} {arXiv:1606.03164 [hep-ph]} \BibitemShut
  {NoStop}%
\bibitem [{\citenamefont {Celis}\ \emph {et~al.}(2017)\citenamefont {Celis},
  \citenamefont {Jung}, \citenamefont {Li},\ and\ \citenamefont
  {Pich}}]{Celis:2016azn}%
  \BibitemOpen
  \bibfield  {author} {\bibinfo {author} {\bibfnamefont {A.}~\bibnamefont
  {Celis}}, \bibinfo {author} {\bibfnamefont {M.}~\bibnamefont {Jung}},
  \bibinfo {author} {\bibfnamefont {X.-Q.}\ \bibnamefont {Li}}, \ and\ \bibinfo
  {author} {\bibfnamefont {A.}~\bibnamefont {Pich}},\ }\href {\doibase
  10.1016/j.physletb.2017.05.037} {\bibfield  {journal} {\bibinfo  {journal}
  {Phys. Lett.}\ }\textbf {\bibinfo {volume} {B771}},\ \bibinfo {pages} {168}
  (\bibinfo {year} {2017})},\ \Eprint {http://arxiv.org/abs/1612.07757}
  {arXiv:1612.07757 [hep-ph]} \BibitemShut {NoStop}%
\bibitem [{\citenamefont {Alonso}\ \emph
  {et~al.}(2017{\natexlab{b}})\citenamefont {Alonso}, \citenamefont
  {Grinstein},\ and\ \citenamefont {Martin~Camalich}}]{Alonso:2016oyd}%
  \BibitemOpen
  \bibfield  {author} {\bibinfo {author} {\bibfnamefont {R.}~\bibnamefont
  {Alonso}}, \bibinfo {author} {\bibfnamefont {B.}~\bibnamefont {Grinstein}}, \
  and\ \bibinfo {author} {\bibfnamefont {J.}~\bibnamefont {Martin~Camalich}},\
  }\href {\doibase 10.1103/PhysRevLett.118.081802} {\bibfield  {journal}
  {\bibinfo  {journal} {Phys. Rev. Lett.}\ }\textbf {\bibinfo {volume} {118}},\
  \bibinfo {pages} {081802} (\bibinfo {year} {2017}{\natexlab{b}})},\ \Eprint
  {http://arxiv.org/abs/1611.06676} {arXiv:1611.06676 [hep-ph]} \BibitemShut
  {NoStop}%
\bibitem [{\citenamefont {Tanabashi}\ \emph {et~al.}(2018)\citenamefont
  {Tanabashi} \emph {et~al.}}]{Tanabashi:2018oca}%
  \BibitemOpen
  \bibfield  {author} {\bibinfo {author} {\bibfnamefont {M.}~\bibnamefont
  {Tanabashi}} \emph {et~al.} (\bibinfo {collaboration} {Particle Data
  Group}),\ }\href {\doibase 10.1103/PhysRevD.98.030001} {\bibfield  {journal}
  {\bibinfo  {journal} {Phys. Rev.}\ }\textbf {\bibinfo {volume} {D98}},\
  \bibinfo {pages} {030001} (\bibinfo {year} {2018})}\BibitemShut {NoStop}%
\bibitem [{\citenamefont {Fajfer}\ \emph
  {et~al.}(2012{\natexlab{b}})\citenamefont {Fajfer}, \citenamefont {Kamenik},
  \citenamefont {Nisandzic},\ and\ \citenamefont {Zupan}}]{Fajfer:2012jt}%
  \BibitemOpen
  \bibfield  {author} {\bibinfo {author} {\bibfnamefont {S.}~\bibnamefont
  {Fajfer}}, \bibinfo {author} {\bibfnamefont {J.~F.}\ \bibnamefont {Kamenik}},
  \bibinfo {author} {\bibfnamefont {I.}~\bibnamefont {Nisandzic}}, \ and\
  \bibinfo {author} {\bibfnamefont {J.}~\bibnamefont {Zupan}},\ }\href
  {\doibase 10.1103/PhysRevLett.109.161801} {\bibfield  {journal} {\bibinfo
  {journal} {Phys. Rev. Lett.}\ }\textbf {\bibinfo {volume} {109}},\ \bibinfo
  {pages} {161801} (\bibinfo {year} {2012}{\natexlab{b}})},\ \Eprint
  {http://arxiv.org/abs/1206.1872} {arXiv:1206.1872 [hep-ph]} \BibitemShut
  {NoStop}%
\bibitem [{\citenamefont {Sakaki}\ and\ \citenamefont
  {Tanaka}(2013)}]{Sakaki:2012ft}%
  \BibitemOpen
  \bibfield  {author} {\bibinfo {author} {\bibfnamefont {Y.}~\bibnamefont
  {Sakaki}}\ and\ \bibinfo {author} {\bibfnamefont {H.}~\bibnamefont
  {Tanaka}},\ }\href {\doibase 10.1103/PhysRevD.87.054002} {\bibfield
  {journal} {\bibinfo  {journal} {Phys. Rev.}\ }\textbf {\bibinfo {volume}
  {D87}},\ \bibinfo {pages} {054002} (\bibinfo {year} {2013})},\ \Eprint
  {http://arxiv.org/abs/1205.4908} {arXiv:1205.4908 [hep-ph]} \BibitemShut
  {NoStop}%
\bibitem [{\citenamefont {Tanaka}\ and\ \citenamefont
  {Watanabe}(2013)}]{Tanaka:2012nw}%
  \BibitemOpen
  \bibfield  {author} {\bibinfo {author} {\bibfnamefont {M.}~\bibnamefont
  {Tanaka}}\ and\ \bibinfo {author} {\bibfnamefont {R.}~\bibnamefont
  {Watanabe}},\ }\href {\doibase 10.1103/PhysRevD.87.034028} {\bibfield
  {journal} {\bibinfo  {journal} {Phys. Rev.}\ }\textbf {\bibinfo {volume}
  {D87}},\ \bibinfo {pages} {034028} (\bibinfo {year} {2013})},\ \Eprint
  {http://arxiv.org/abs/1212.1878} {arXiv:1212.1878 [hep-ph]} \BibitemShut
  {NoStop}%
\bibitem [{\citenamefont {Be\v{c}irevi\'{c}}\ \emph {et~al.}(2012)\citenamefont
  {Be\v{c}irevi\'{c}}, \citenamefont {Ko\v{s}nik},\ and\ \citenamefont
  {Tayduganov}}]{Becirevic:2012jf}%
  \BibitemOpen
  \bibfield  {author} {\bibinfo {author} {\bibfnamefont {D.}~\bibnamefont
  {Be\v{c}irevi\'{c}}}, \bibinfo {author} {\bibfnamefont {N.}~\bibnamefont
  {Ko\v{s}nik}}, \ and\ \bibinfo {author} {\bibfnamefont {A.}~\bibnamefont
  {Tayduganov}},\ }\href {\doibase 10.1016/j.physletb.2012.08.016} {\bibfield
  {journal} {\bibinfo  {journal} {Phys. Lett.}\ }\textbf {\bibinfo {volume}
  {B716}},\ \bibinfo {pages} {208} (\bibinfo {year} {2012})},\ \Eprint
  {http://arxiv.org/abs/1206.4977} {arXiv:1206.4977 [hep-ph]} \BibitemShut
  {NoStop}%
\bibitem [{\citenamefont {Datta}\ \emph {et~al.}(2012)\citenamefont {Datta},
  \citenamefont {Duraisamy},\ and\ \citenamefont {Ghosh}}]{Datta:2012qk}%
  \BibitemOpen
  \bibfield  {author} {\bibinfo {author} {\bibfnamefont {A.}~\bibnamefont
  {Datta}}, \bibinfo {author} {\bibfnamefont {M.}~\bibnamefont {Duraisamy}}, \
  and\ \bibinfo {author} {\bibfnamefont {D.}~\bibnamefont {Ghosh}},\ }\href
  {\doibase 10.1103/PhysRevD.86.034027} {\bibfield  {journal} {\bibinfo
  {journal} {Phys. Rev.}\ }\textbf {\bibinfo {volume} {D86}},\ \bibinfo {pages}
  {034027} (\bibinfo {year} {2012})},\ \Eprint {http://arxiv.org/abs/1206.3760}
  {arXiv:1206.3760 [hep-ph]} \BibitemShut {NoStop}%
\bibitem [{\citenamefont {Duraisamy}\ and\ \citenamefont
  {Datta}(2013)}]{Duraisamy:2013kcw}%
  \BibitemOpen
  \bibfield  {author} {\bibinfo {author} {\bibfnamefont {M.}~\bibnamefont
  {Duraisamy}}\ and\ \bibinfo {author} {\bibfnamefont {A.}~\bibnamefont
  {Datta}},\ }\href {\doibase 10.1007/JHEP09(2013)059} {\bibfield  {journal}
  {\bibinfo  {journal} {JHEP}\ }\textbf {\bibinfo {volume} {09}},\ \bibinfo
  {pages} {059} (\bibinfo {year} {2013})},\ \Eprint
  {http://arxiv.org/abs/1302.7031} {arXiv:1302.7031 [hep-ph]} \BibitemShut
  {NoStop}%
\bibitem [{\citenamefont {Dutta}\ \emph {et~al.}(2013)\citenamefont {Dutta},
  \citenamefont {Bhol},\ and\ \citenamefont {Giri}}]{Dutta:2013qaa}%
  \BibitemOpen
  \bibfield  {author} {\bibinfo {author} {\bibfnamefont {R.}~\bibnamefont
  {Dutta}}, \bibinfo {author} {\bibfnamefont {A.}~\bibnamefont {Bhol}}, \ and\
  \bibinfo {author} {\bibfnamefont {A.~K.}\ \bibnamefont {Giri}},\ }\href
  {\doibase 10.1103/PhysRevD.88.114023} {\bibfield  {journal} {\bibinfo
  {journal} {Phys. Rev.}\ }\textbf {\bibinfo {volume} {D88}},\ \bibinfo {pages}
  {114023} (\bibinfo {year} {2013})},\ \Eprint {http://arxiv.org/abs/1307.6653}
  {arXiv:1307.6653 [hep-ph]} \BibitemShut {NoStop}%
\bibitem [{\citenamefont {Duraisamy}\ \emph {et~al.}(2014)\citenamefont
  {Duraisamy}, \citenamefont {Sharma},\ and\ \citenamefont
  {Datta}}]{Duraisamy:2014sna}%
  \BibitemOpen
  \bibfield  {author} {\bibinfo {author} {\bibfnamefont {M.}~\bibnamefont
  {Duraisamy}}, \bibinfo {author} {\bibfnamefont {P.}~\bibnamefont {Sharma}}, \
  and\ \bibinfo {author} {\bibfnamefont {A.}~\bibnamefont {Datta}},\ }\href
  {\doibase 10.1103/PhysRevD.90.074013} {\bibfield  {journal} {\bibinfo
  {journal} {Phys. Rev.}\ }\textbf {\bibinfo {volume} {D90}},\ \bibinfo {pages}
  {074013} (\bibinfo {year} {2014})},\ \Eprint {http://arxiv.org/abs/1405.3719}
  {arXiv:1405.3719 [hep-ph]} \BibitemShut {NoStop}%
\bibitem [{\citenamefont {Sakaki}\ \emph {et~al.}(2015)\citenamefont {Sakaki},
  \citenamefont {Tanaka}, \citenamefont {Tayduganov},\ and\ \citenamefont
  {Watanabe}}]{Sakaki:2014sea}%
  \BibitemOpen
  \bibfield  {author} {\bibinfo {author} {\bibfnamefont {Y.}~\bibnamefont
  {Sakaki}}, \bibinfo {author} {\bibfnamefont {M.}~\bibnamefont {Tanaka}},
  \bibinfo {author} {\bibfnamefont {A.}~\bibnamefont {Tayduganov}}, \ and\
  \bibinfo {author} {\bibfnamefont {R.}~\bibnamefont {Watanabe}},\ }\href
  {\doibase 10.1103/PhysRevD.91.114028} {\bibfield  {journal} {\bibinfo
  {journal} {Phys. Rev.}\ }\textbf {\bibinfo {volume} {D91}},\ \bibinfo {pages}
  {114028} (\bibinfo {year} {2015})},\ \Eprint {http://arxiv.org/abs/1412.3761}
  {arXiv:1412.3761 [hep-ph]} \BibitemShut {NoStop}%
\bibitem [{\citenamefont {Freytsis}\ \emph {et~al.}(2015)\citenamefont
  {Freytsis}, \citenamefont {Ligeti},\ and\ \citenamefont
  {Ruderman}}]{Freytsis:2015qca}%
  \BibitemOpen
  \bibfield  {author} {\bibinfo {author} {\bibfnamefont {M.}~\bibnamefont
  {Freytsis}}, \bibinfo {author} {\bibfnamefont {Z.}~\bibnamefont {Ligeti}}, \
  and\ \bibinfo {author} {\bibfnamefont {J.~T.}\ \bibnamefont {Ruderman}},\
  }\href {\doibase 10.1103/PhysRevD.92.054018} {\bibfield  {journal} {\bibinfo
  {journal} {Phys. Rev.}\ }\textbf {\bibinfo {volume} {D92}},\ \bibinfo {pages}
  {054018} (\bibinfo {year} {2015})},\ \Eprint
  {http://arxiv.org/abs/1506.08896} {arXiv:1506.08896 [hep-ph]} \BibitemShut
  {NoStop}%
\bibitem [{\citenamefont {Alonso}\ \emph {et~al.}(2016)\citenamefont {Alonso},
  \citenamefont {Kobach},\ and\ \citenamefont
  {Martin~Camalich}}]{Alonso:2016gym}%
  \BibitemOpen
  \bibfield  {author} {\bibinfo {author} {\bibfnamefont {R.}~\bibnamefont
  {Alonso}}, \bibinfo {author} {\bibfnamefont {A.}~\bibnamefont {Kobach}}, \
  and\ \bibinfo {author} {\bibfnamefont {J.}~\bibnamefont {Martin~Camalich}},\
  }\href {\doibase 10.1103/PhysRevD.94.094021} {\bibfield  {journal} {\bibinfo
  {journal} {Phys. Rev.}\ }\textbf {\bibinfo {volume} {D94}},\ \bibinfo {pages}
  {094021} (\bibinfo {year} {2016})},\ \Eprint
  {http://arxiv.org/abs/1602.07671} {arXiv:1602.07671 [hep-ph]} \BibitemShut
  {NoStop}%
\bibitem [{\citenamefont {Bardhan}\ \emph {et~al.}(2017)\citenamefont
  {Bardhan}, \citenamefont {Byakti},\ and\ \citenamefont
  {Ghosh}}]{Bardhan:2016uhr}%
  \BibitemOpen
  \bibfield  {author} {\bibinfo {author} {\bibfnamefont {D.}~\bibnamefont
  {Bardhan}}, \bibinfo {author} {\bibfnamefont {P.}~\bibnamefont {Byakti}}, \
  and\ \bibinfo {author} {\bibfnamefont {D.}~\bibnamefont {Ghosh}},\ }\href
  {\doibase 10.1007/JHEP01(2017)125} {\bibfield  {journal} {\bibinfo  {journal}
  {JHEP}\ }\textbf {\bibinfo {volume} {01}},\ \bibinfo {pages} {125} (\bibinfo
  {year} {2017})},\ \Eprint {http://arxiv.org/abs/1610.03038} {arXiv:1610.03038
  [hep-ph]} \BibitemShut {NoStop}%
\bibitem [{\citenamefont {Bhattacharya}\ \emph
  {et~al.}(2017{\natexlab{a}})\citenamefont {Bhattacharya}, \citenamefont
  {Nandi},\ and\ \citenamefont {Patra}}]{Bhattacharya:2016zcw}%
  \BibitemOpen
  \bibfield  {author} {\bibinfo {author} {\bibfnamefont {S.}~\bibnamefont
  {Bhattacharya}}, \bibinfo {author} {\bibfnamefont {S.}~\bibnamefont {Nandi}},
  \ and\ \bibinfo {author} {\bibfnamefont {S.~K.}\ \bibnamefont {Patra}},\
  }\href {\doibase 10.1103/PhysRevD.95.075012} {\bibfield  {journal} {\bibinfo
  {journal} {Phys. Rev.}\ }\textbf {\bibinfo {volume} {D95}},\ \bibinfo {pages}
  {075012} (\bibinfo {year} {2017}{\natexlab{a}})},\ \Eprint
  {http://arxiv.org/abs/1611.04605} {arXiv:1611.04605 [hep-ph]} \BibitemShut
  {NoStop}%
\bibitem [{\citenamefont {Alok}\ \emph {et~al.}(2018)\citenamefont {Alok},
  \citenamefont {Kumar}, \citenamefont {Kumar}, \citenamefont {Kumbhakar},\
  and\ \citenamefont {Sankar}}]{Alok:2017qsi}%
  \BibitemOpen
  \bibfield  {author} {\bibinfo {author} {\bibfnamefont {A.~K.}\ \bibnamefont
  {Alok}}, \bibinfo {author} {\bibfnamefont {D.}~\bibnamefont {Kumar}},
  \bibinfo {author} {\bibfnamefont {J.}~\bibnamefont {Kumar}}, \bibinfo
  {author} {\bibfnamefont {S.}~\bibnamefont {Kumbhakar}}, \ and\ \bibinfo
  {author} {\bibfnamefont {S.~U.}\ \bibnamefont {Sankar}},\ }\href {\doibase
  10.1007/JHEP09(2018)152} {\bibfield  {journal} {\bibinfo  {journal} {JHEP}\
  }\textbf {\bibinfo {volume} {09}},\ \bibinfo {pages} {152} (\bibinfo {year}
  {2018})},\ \Eprint {http://arxiv.org/abs/1710.04127} {arXiv:1710.04127
  [hep-ph]} \BibitemShut {NoStop}%
\bibitem [{\citenamefont {Dutta}(2017)}]{Dutta:2017wpq}%
  \BibitemOpen
  \bibfield  {author} {\bibinfo {author} {\bibfnamefont {R.}~\bibnamefont
  {Dutta}},\ }\href@noop {} {\  (\bibinfo {year} {2017})},\ \Eprint
  {http://arxiv.org/abs/1710.00351} {arXiv:1710.00351 [hep-ph]} \BibitemShut
  {NoStop}%
\bibitem [{\citenamefont {Azatov}\ \emph
  {et~al.}(2018{\natexlab{a}})\citenamefont {Azatov}, \citenamefont {Bardhan},
  \citenamefont {Ghosh}, \citenamefont {Sgarlata},\ and\ \citenamefont
  {Venturini}}]{Azatov:2018knx}%
  \BibitemOpen
  \bibfield  {author} {\bibinfo {author} {\bibfnamefont {A.}~\bibnamefont
  {Azatov}}, \bibinfo {author} {\bibfnamefont {D.}~\bibnamefont {Bardhan}},
  \bibinfo {author} {\bibfnamefont {D.}~\bibnamefont {Ghosh}}, \bibinfo
  {author} {\bibfnamefont {F.}~\bibnamefont {Sgarlata}}, \ and\ \bibinfo
  {author} {\bibfnamefont {E.}~\bibnamefont {Venturini}},\ }\href@noop {} {\
  (\bibinfo {year} {2018}{\natexlab{a}})},\ \Eprint
  {http://arxiv.org/abs/1805.03209} {arXiv:1805.03209 [hep-ph]} \BibitemShut
  {NoStop}%
\bibitem [{\citenamefont {Bifani}\ \emph {et~al.}(2018)\citenamefont {Bifani},
  \citenamefont {Descotes-Genon}, \citenamefont {Romero~Vidal},\ and\
  \citenamefont {Schune}}]{Bifani:2018zmi}%
  \BibitemOpen
  \bibfield  {author} {\bibinfo {author} {\bibfnamefont {S.}~\bibnamefont
  {Bifani}}, \bibinfo {author} {\bibfnamefont {S.}~\bibnamefont
  {Descotes-Genon}}, \bibinfo {author} {\bibfnamefont {A.}~\bibnamefont
  {Romero~Vidal}}, \ and\ \bibinfo {author} {\bibfnamefont {M.-H.}\
  \bibnamefont {Schune}},\ }\href@noop {} {\  (\bibinfo {year} {2018})},\
  \Eprint {http://arxiv.org/abs/1809.06229} {arXiv:1809.06229 [hep-ex]}
  \BibitemShut {NoStop}%
\bibitem [{\citenamefont {Huang}\ \emph {et~al.}(2018)\citenamefont {Huang},
  \citenamefont {Li}, \citenamefont {Lu}, \citenamefont {Paracha},\ and\
  \citenamefont {Wang}}]{Huang:2018nnq}%
  \BibitemOpen
  \bibfield  {author} {\bibinfo {author} {\bibfnamefont {Z.-R.}\ \bibnamefont
  {Huang}}, \bibinfo {author} {\bibfnamefont {Y.}~\bibnamefont {Li}}, \bibinfo
  {author} {\bibfnamefont {C.-D.}\ \bibnamefont {Lu}}, \bibinfo {author}
  {\bibfnamefont {M.~A.}\ \bibnamefont {Paracha}}, \ and\ \bibinfo {author}
  {\bibfnamefont {C.}~\bibnamefont {Wang}},\ }\href@noop {} {\  (\bibinfo
  {year} {2018})},\ \Eprint {http://arxiv.org/abs/1808.03565} {arXiv:1808.03565
  [hep-ph]} \BibitemShut {NoStop}%
\bibitem [{\citenamefont {Asadi}\ \emph {et~al.}(2018)\citenamefont {Asadi},
  \citenamefont {Buckley},\ and\ \citenamefont {Shih}}]{Asadi:2018sym}%
  \BibitemOpen
  \bibfield  {author} {\bibinfo {author} {\bibfnamefont {P.}~\bibnamefont
  {Asadi}}, \bibinfo {author} {\bibfnamefont {M.~R.}\ \bibnamefont {Buckley}},
  \ and\ \bibinfo {author} {\bibfnamefont {D.}~\bibnamefont {Shih}},\
  }\href@noop {} {\  (\bibinfo {year} {2018})},\ \Eprint
  {http://arxiv.org/abs/1810.06597} {arXiv:1810.06597 [hep-ph]} \BibitemShut
  {NoStop}%
\bibitem [{\citenamefont {Hu}\ \emph {et~al.}(2018)\citenamefont {Hu},
  \citenamefont {Li},\ and\ \citenamefont {Yang}}]{Hu:2018veh}%
  \BibitemOpen
  \bibfield  {author} {\bibinfo {author} {\bibfnamefont {Q.-Y.}\ \bibnamefont
  {Hu}}, \bibinfo {author} {\bibfnamefont {X.-Q.}\ \bibnamefont {Li}}, \ and\
  \bibinfo {author} {\bibfnamefont {Y.-D.}\ \bibnamefont {Yang}},\ }\href@noop
  {} {\  (\bibinfo {year} {2018})},\ \Eprint {http://arxiv.org/abs/1810.04939}
  {arXiv:1810.04939 [hep-ph]} \BibitemShut {NoStop}%
\bibitem [{\citenamefont {Feruglio}\ \emph {et~al.}(2018)\citenamefont
  {Feruglio}, \citenamefont {Paradisi},\ and\ \citenamefont
  {Sumensari}}]{Feruglio:2018fxo}%
  \BibitemOpen
  \bibfield  {author} {\bibinfo {author} {\bibfnamefont {F.}~\bibnamefont
  {Feruglio}}, \bibinfo {author} {\bibfnamefont {P.}~\bibnamefont {Paradisi}},
  \ and\ \bibinfo {author} {\bibfnamefont {O.}~\bibnamefont {Sumensari}},\
  }\href@noop {} {\  (\bibinfo {year} {2018})},\ \Eprint
  {http://arxiv.org/abs/1806.10155} {arXiv:1806.10155 [hep-ph]} \BibitemShut
  {NoStop}%
\bibitem [{\citenamefont {Angelescu}\ \emph {et~al.}(2018)\citenamefont
  {Angelescu}, \citenamefont {Be\v{c}irevi\'{c}}, \citenamefont {Faroughy},\
  and\ \citenamefont {Sumensari}}]{Angelescu:2018tyl}%
  \BibitemOpen
  \bibfield  {author} {\bibinfo {author} {\bibfnamefont {A.}~\bibnamefont
  {Angelescu}}, \bibinfo {author} {\bibfnamefont {D.}~\bibnamefont
  {Be\v{c}irevi\'{c}}}, \bibinfo {author} {\bibfnamefont {D.~A.}\ \bibnamefont
  {Faroughy}}, \ and\ \bibinfo {author} {\bibfnamefont {O.}~\bibnamefont
  {Sumensari}},\ }\href@noop {} {\  (\bibinfo {year} {2018})},\ \Eprint
  {http://arxiv.org/abs/1808.08179} {arXiv:1808.08179 [hep-ph]} \BibitemShut
  {NoStop}%
\bibitem [{\citenamefont {Iguro}\ \emph {et~al.}(2018)\citenamefont {Iguro},
  \citenamefont {Omura},\ and\ \citenamefont {Takeuchi}}]{Iguro:2018fni}%
  \BibitemOpen
  \bibfield  {author} {\bibinfo {author} {\bibfnamefont {S.}~\bibnamefont
  {Iguro}}, \bibinfo {author} {\bibfnamefont {Y.}~\bibnamefont {Omura}}, \ and\
  \bibinfo {author} {\bibfnamefont {M.}~\bibnamefont {Takeuchi}},\ }\href@noop
  {} {\  (\bibinfo {year} {2018})},\ \Eprint {http://arxiv.org/abs/1810.05843}
  {arXiv:1810.05843 [hep-ph]} \BibitemShut {NoStop}%
\bibitem [{\citenamefont {Bhattacharya}\ \emph {et~al.}(2018)\citenamefont
  {Bhattacharya}, \citenamefont {Nandi},\ and\ \citenamefont
  {Kumar~Patra}}]{Bhattacharya:2018kig}%
  \BibitemOpen
  \bibfield  {author} {\bibinfo {author} {\bibfnamefont {S.}~\bibnamefont
  {Bhattacharya}}, \bibinfo {author} {\bibfnamefont {S.}~\bibnamefont {Nandi}},
  \ and\ \bibinfo {author} {\bibfnamefont {S.}~\bibnamefont {Kumar~Patra}},\
  }\href@noop {} {\  (\bibinfo {year} {2018})},\ \Eprint
  {http://arxiv.org/abs/1805.08222} {arXiv:1805.08222 [hep-ph]} \BibitemShut
  {NoStop}%
\bibitem [{\citenamefont {Aebischer}\ \emph {et~al.}(2018)\citenamefont
  {Aebischer}, \citenamefont {Kumar}, \citenamefont {Stangl},\ and\
  \citenamefont {Straub}}]{Aebischer:2018iyb}%
  \BibitemOpen
  \bibfield  {author} {\bibinfo {author} {\bibfnamefont {J.}~\bibnamefont
  {Aebischer}}, \bibinfo {author} {\bibfnamefont {J.}~\bibnamefont {Kumar}},
  \bibinfo {author} {\bibfnamefont {P.}~\bibnamefont {Stangl}}, \ and\ \bibinfo
  {author} {\bibfnamefont {D.~M.}\ \bibnamefont {Straub}},\ }\href@noop {} {\
  (\bibinfo {year} {2018})},\ \Eprint {http://arxiv.org/abs/1810.07698}
  {arXiv:1810.07698 [hep-ph]} \BibitemShut {NoStop}%
\bibitem [{\citenamefont {Iguro}\ and\ \citenamefont
  {Omura}(2018)}]{Iguro:2018qzf}%
  \BibitemOpen
  \bibfield  {author} {\bibinfo {author} {\bibfnamefont {S.}~\bibnamefont
  {Iguro}}\ and\ \bibinfo {author} {\bibfnamefont {Y.}~\bibnamefont {Omura}},\
  }\href {\doibase 10.1007/JHEP05(2018)173} {\bibfield  {journal} {\bibinfo
  {journal} {JHEP}\ }\textbf {\bibinfo {volume} {05}},\ \bibinfo {pages} {173}
  (\bibinfo {year} {2018})},\ \Eprint {http://arxiv.org/abs/1802.01732}
  {arXiv:1802.01732 [hep-ph]} \BibitemShut {NoStop}%
\bibitem [{\citenamefont {Greljo}\ \emph
  {et~al.}(2018{\natexlab{a}})\citenamefont {Greljo}, \citenamefont {Robinson},
  \citenamefont {Shakya},\ and\ \citenamefont {Zupan}}]{Greljo:2018ogz}%
  \BibitemOpen
  \bibfield  {author} {\bibinfo {author} {\bibfnamefont {A.}~\bibnamefont
  {Greljo}}, \bibinfo {author} {\bibfnamefont {D.~J.}\ \bibnamefont
  {Robinson}}, \bibinfo {author} {\bibfnamefont {B.}~\bibnamefont {Shakya}}, \
  and\ \bibinfo {author} {\bibfnamefont {J.}~\bibnamefont {Zupan}},\ }\href
  {\doibase 10.1007/JHEP09(2018)169} {\bibfield  {journal} {\bibinfo  {journal}
  {JHEP}\ }\textbf {\bibinfo {volume} {09}},\ \bibinfo {pages} {169} (\bibinfo
  {year} {2018}{\natexlab{a}})},\ \Eprint {http://arxiv.org/abs/1804.04642}
  {arXiv:1804.04642 [hep-ph]} \BibitemShut {NoStop}%
\bibitem [{\citenamefont {Robinson}\ \emph {et~al.}(2018)\citenamefont
  {Robinson}, \citenamefont {Shakya},\ and\ \citenamefont
  {Zupan}}]{Robinson:2018gza}%
  \BibitemOpen
  \bibfield  {author} {\bibinfo {author} {\bibfnamefont {D.~J.}\ \bibnamefont
  {Robinson}}, \bibinfo {author} {\bibfnamefont {B.}~\bibnamefont {Shakya}}, \
  and\ \bibinfo {author} {\bibfnamefont {J.}~\bibnamefont {Zupan}},\
  }\href@noop {} {\  (\bibinfo {year} {2018})},\ \Eprint
  {http://arxiv.org/abs/1807.04753} {arXiv:1807.04753 [hep-ph]} \BibitemShut
  {NoStop}%
\bibitem [{\citenamefont {Azatov}\ \emph
  {et~al.}(2018{\natexlab{b}})\citenamefont {Azatov}, \citenamefont {Barducci},
  \citenamefont {Ghosh}, \citenamefont {Marzocca},\ and\ \citenamefont
  {Ubaldi}}]{Azatov:2018kzb}%
  \BibitemOpen
  \bibfield  {author} {\bibinfo {author} {\bibfnamefont {A.}~\bibnamefont
  {Azatov}}, \bibinfo {author} {\bibfnamefont {D.}~\bibnamefont {Barducci}},
  \bibinfo {author} {\bibfnamefont {D.}~\bibnamefont {Ghosh}}, \bibinfo
  {author} {\bibfnamefont {D.}~\bibnamefont {Marzocca}}, \ and\ \bibinfo
  {author} {\bibfnamefont {L.}~\bibnamefont {Ubaldi}},\ }\href {\doibase
  10.1007/JHEP10(2018)092} {\bibfield  {journal} {\bibinfo  {journal} {JHEP}\
  }\textbf {\bibinfo {volume} {10}},\ \bibinfo {pages} {092} (\bibinfo {year}
  {2018}{\natexlab{b}})},\ \Eprint {http://arxiv.org/abs/1807.10745}
  {arXiv:1807.10745 [hep-ph]} \BibitemShut {NoStop}%
\bibitem [{\citenamefont {Jung}\ and\ \citenamefont
  {Straub}(2018)}]{Jung:2018lfu}%
  \BibitemOpen
  \bibfield  {author} {\bibinfo {author} {\bibfnamefont {M.}~\bibnamefont
  {Jung}}\ and\ \bibinfo {author} {\bibfnamefont {D.~M.}\ \bibnamefont
  {Straub}},\ }\href@noop {} {\  (\bibinfo {year} {2018})},\ \Eprint
  {http://arxiv.org/abs/1801.01112} {arXiv:1801.01112 [hep-ph]} \BibitemShut
  {NoStop}%
\bibitem [{\citenamefont {Buchmuller}\ and\ \citenamefont
  {Wyler}(1986)}]{Buchmuller:1985jz}%
  \BibitemOpen
  \bibfield  {author} {\bibinfo {author} {\bibfnamefont {W.}~\bibnamefont
  {Buchmuller}}\ and\ \bibinfo {author} {\bibfnamefont {D.}~\bibnamefont
  {Wyler}},\ }\href {\doibase 10.1016/0550-3213(86)90262-2} {\bibfield
  {journal} {\bibinfo  {journal} {Nucl. Phys.}\ }\textbf {\bibinfo {volume}
  {B268}},\ \bibinfo {pages} {621} (\bibinfo {year} {1986})}\BibitemShut
  {NoStop}%
\bibitem [{\citenamefont {Grzadkowski}\ \emph {et~al.}(2010)\citenamefont
  {Grzadkowski}, \citenamefont {Iskrzynski}, \citenamefont {Misiak},\ and\
  \citenamefont {Rosiek}}]{Grzadkowski:2010es}%
  \BibitemOpen
  \bibfield  {author} {\bibinfo {author} {\bibfnamefont {B.}~\bibnamefont
  {Grzadkowski}}, \bibinfo {author} {\bibfnamefont {M.}~\bibnamefont
  {Iskrzynski}}, \bibinfo {author} {\bibfnamefont {M.}~\bibnamefont {Misiak}},
  \ and\ \bibinfo {author} {\bibfnamefont {J.}~\bibnamefont {Rosiek}},\ }\href
  {\doibase 10.1007/JHEP10(2010)085} {\bibfield  {journal} {\bibinfo  {journal}
  {JHEP}\ }\textbf {\bibinfo {volume} {10}},\ \bibinfo {pages} {085} (\bibinfo
  {year} {2010})},\ \Eprint {http://arxiv.org/abs/1008.4884} {arXiv:1008.4884
  [hep-ph]} \BibitemShut {NoStop}%
\bibitem [{\citenamefont {Aebischer}\ \emph {et~al.}(2016)\citenamefont
  {Aebischer}, \citenamefont {Crivellin}, \citenamefont {Fael},\ and\
  \citenamefont {Greub}}]{Aebischer:2015fzz}%
  \BibitemOpen
  \bibfield  {author} {\bibinfo {author} {\bibfnamefont {J.}~\bibnamefont
  {Aebischer}}, \bibinfo {author} {\bibfnamefont {A.}~\bibnamefont
  {Crivellin}}, \bibinfo {author} {\bibfnamefont {M.}~\bibnamefont {Fael}}, \
  and\ \bibinfo {author} {\bibfnamefont {C.}~\bibnamefont {Greub}},\ }\href
  {\doibase 10.1007/JHEP05(2016)037} {\bibfield  {journal} {\bibinfo  {journal}
  {JHEP}\ }\textbf {\bibinfo {volume} {05}},\ \bibinfo {pages} {037} (\bibinfo
  {year} {2016})},\ \Eprint {http://arxiv.org/abs/1512.02830} {arXiv:1512.02830
  [hep-ph]} \BibitemShut {NoStop}%
\bibitem [{\citenamefont {Gonz\'{a}lez-Alonso}\ \emph
  {et~al.}(2017)\citenamefont {Gonz\'{a}lez-Alonso}, \citenamefont
  {Martin~Camalich},\ and\ \citenamefont {Mimouni}}]{Gonzalez-Alonso:2017iyc}%
  \BibitemOpen
  \bibfield  {author} {\bibinfo {author} {\bibfnamefont {M.}~\bibnamefont
  {Gonz\'{a}lez-Alonso}}, \bibinfo {author} {\bibfnamefont {J.}~\bibnamefont
  {Martin~Camalich}}, \ and\ \bibinfo {author} {\bibfnamefont {K.}~\bibnamefont
  {Mimouni}},\ }\href {\doibase 10.1016/j.physletb.2017.07.003} {\bibfield
  {journal} {\bibinfo  {journal} {Phys. Lett.}\ }\textbf {\bibinfo {volume}
  {B772}},\ \bibinfo {pages} {777} (\bibinfo {year} {2017})},\ \Eprint
  {http://arxiv.org/abs/1706.00410} {arXiv:1706.00410 [hep-ph]} \BibitemShut
  {NoStop}%
\bibitem [{\citenamefont {Gershtein}\ \emph {et~al.}(1995)\citenamefont
  {Gershtein}, \citenamefont {Kiselev}, \citenamefont {Likhoded},\ and\
  \citenamefont {Tkabladze}}]{Gershtein:1994jw}%
  \BibitemOpen
  \bibfield  {author} {\bibinfo {author} {\bibfnamefont {S.~S.}\ \bibnamefont
  {Gershtein}}, \bibinfo {author} {\bibfnamefont {V.~V.}\ \bibnamefont
  {Kiselev}}, \bibinfo {author} {\bibfnamefont {A.~K.}\ \bibnamefont
  {Likhoded}}, \ and\ \bibinfo {author} {\bibfnamefont {A.~V.}\ \bibnamefont
  {Tkabladze}},\ }\href {\doibase 10.1070/PU1995v038n01ABEH000063} {\bibfield
  {journal} {\bibinfo  {journal} {Phys. Usp.}\ }\textbf {\bibinfo {volume}
  {38}},\ \bibinfo {pages} {1} (\bibinfo {year} {1995})},\ \bibinfo {note}
  {[Usp. Fiz. Nauk165,3(1995)]},\ \Eprint {http://arxiv.org/abs/hep-ph/9504319}
  {arXiv:hep-ph/9504319 [hep-ph]} \BibitemShut {NoStop}%
\bibitem [{\citenamefont {Bigi}(1996)}]{Bigi:1995fs}%
  \BibitemOpen
  \bibfield  {author} {\bibinfo {author} {\bibfnamefont {I.~I.~Y.}\
  \bibnamefont {Bigi}},\ }\href {\doibase 10.1016/0370-2693(95)01574-4}
  {\bibfield  {journal} {\bibinfo  {journal} {Phys. Lett.}\ }\textbf {\bibinfo
  {volume} {B371}},\ \bibinfo {pages} {105} (\bibinfo {year} {1996})},\ \Eprint
  {http://arxiv.org/abs/hep-ph/9510325} {arXiv:hep-ph/9510325 [hep-ph]}
  \BibitemShut {NoStop}%
\bibitem [{\citenamefont {Beneke}\ and\ \citenamefont
  {Buchalla}(1996)}]{Beneke:1996xe}%
  \BibitemOpen
  \bibfield  {author} {\bibinfo {author} {\bibfnamefont {M.}~\bibnamefont
  {Beneke}}\ and\ \bibinfo {author} {\bibfnamefont {G.}~\bibnamefont
  {Buchalla}},\ }\href {\doibase 10.1103/PhysRevD.53.4991} {\bibfield
  {journal} {\bibinfo  {journal} {Phys. Rev.}\ }\textbf {\bibinfo {volume}
  {D53}},\ \bibinfo {pages} {4991} (\bibinfo {year} {1996})},\ \Eprint
  {http://arxiv.org/abs/hep-ph/9601249} {arXiv:hep-ph/9601249 [hep-ph]}
  \BibitemShut {NoStop}%
\bibitem [{\citenamefont {Chang}\ \emph {et~al.}(2001)\citenamefont {Chang},
  \citenamefont {Chen}, \citenamefont {Feng},\ and\ \citenamefont
  {Li}}]{Chang:2000ac}%
  \BibitemOpen
  \bibfield  {author} {\bibinfo {author} {\bibfnamefont {C.-H.}\ \bibnamefont
  {Chang}}, \bibinfo {author} {\bibfnamefont {S.-L.}\ \bibnamefont {Chen}},
  \bibinfo {author} {\bibfnamefont {T.-F.}\ \bibnamefont {Feng}}, \ and\
  \bibinfo {author} {\bibfnamefont {X.-Q.}\ \bibnamefont {Li}},\ }\href
  {\doibase 10.1103/PhysRevD.64.014003} {\bibfield  {journal} {\bibinfo
  {journal} {Phys. Rev.}\ }\textbf {\bibinfo {volume} {D64}},\ \bibinfo {pages}
  {014003} (\bibinfo {year} {2001})},\ \Eprint
  {http://arxiv.org/abs/hep-ph/0007162} {arXiv:hep-ph/0007162 [hep-ph]}
  \BibitemShut {NoStop}%
\bibitem [{\citenamefont {Kiselev}\ \emph {et~al.}(2000)\citenamefont
  {Kiselev}, \citenamefont {Kovalsky},\ and\ \citenamefont
  {Likhoded}}]{Kiselev:2000pp}%
  \BibitemOpen
  \bibfield  {author} {\bibinfo {author} {\bibfnamefont {V.~V.}\ \bibnamefont
  {Kiselev}}, \bibinfo {author} {\bibfnamefont {A.~E.}\ \bibnamefont
  {Kovalsky}}, \ and\ \bibinfo {author} {\bibfnamefont {A.~K.}\ \bibnamefont
  {Likhoded}},\ }\href {\doibase 10.1016/S0550-3213(00)00386-2} {\bibfield
  {journal} {\bibinfo  {journal} {Nucl. Phys.}\ }\textbf {\bibinfo {volume}
  {B585}},\ \bibinfo {pages} {353} (\bibinfo {year} {2000})},\ \Eprint
  {http://arxiv.org/abs/hep-ph/0002127} {arXiv:hep-ph/0002127 [hep-ph]}
  \BibitemShut {NoStop}%
\bibitem [{\citenamefont {Akeroyd}\ and\ \citenamefont
  {Chen}(2017)}]{Akeroyd:2017mhr}%
  \BibitemOpen
  \bibfield  {author} {\bibinfo {author} {\bibfnamefont {A.~G.}\ \bibnamefont
  {Akeroyd}}\ and\ \bibinfo {author} {\bibfnamefont {C.-H.}\ \bibnamefont
  {Chen}},\ }\href {\doibase 10.1103/PhysRevD.96.075011} {\bibfield  {journal}
  {\bibinfo  {journal} {Phys. Rev.}\ }\textbf {\bibinfo {volume} {D96}},\
  \bibinfo {pages} {075011} (\bibinfo {year} {2017})},\ \Eprint
  {http://arxiv.org/abs/1708.04072} {arXiv:1708.04072 [hep-ph]} \BibitemShut
  {NoStop}%
\bibitem [{\citenamefont {Aaij}\ \emph {et~al.}(2013)\citenamefont {Aaij} \emph
  {et~al.}}]{Aaij:2013qqa}%
  \BibitemOpen
  \bibfield  {author} {\bibinfo {author} {\bibfnamefont {R.}~\bibnamefont
  {Aaij}} \emph {et~al.} (\bibinfo {collaboration} {LHCb}),\ }\href {\doibase
  10.1007/JHEP04(2013)001} {\bibfield  {journal} {\bibinfo  {journal} {JHEP}\
  }\textbf {\bibinfo {volume} {04}},\ \bibinfo {pages} {001} (\bibinfo {year}
  {2013})},\ \Eprint {http://arxiv.org/abs/1301.5286} {arXiv:1301.5286
  [hep-ex]} \BibitemShut {NoStop}%
\bibitem [{\citenamefont {Khachatryan}\ \emph {et~al.}(2015)\citenamefont
  {Khachatryan} \emph {et~al.}}]{Khachatryan:2014nfa}%
  \BibitemOpen
  \bibfield  {author} {\bibinfo {author} {\bibfnamefont {V.}~\bibnamefont
  {Khachatryan}} \emph {et~al.} (\bibinfo {collaboration} {CMS}),\ }\href
  {\doibase 10.1007/JHEP01(2015)063} {\bibfield  {journal} {\bibinfo  {journal}
  {JHEP}\ }\textbf {\bibinfo {volume} {01}},\ \bibinfo {pages} {063} (\bibinfo
  {year} {2015})},\ \Eprint {http://arxiv.org/abs/1410.5729} {arXiv:1410.5729
  [hep-ex]} \BibitemShut {NoStop}%
\bibitem [{\citenamefont {Aaij}\ \emph
  {et~al.}(2015{\natexlab{b}})\citenamefont {Aaij} \emph
  {et~al.}}]{Aaij:2014ija}%
  \BibitemOpen
  \bibfield  {author} {\bibinfo {author} {\bibfnamefont {R.}~\bibnamefont
  {Aaij}} \emph {et~al.} (\bibinfo {collaboration} {LHCb}),\ }\href {\doibase
  10.1103/PhysRevLett.114.132001} {\bibfield  {journal} {\bibinfo  {journal}
  {Phys. Rev. Lett.}\ }\textbf {\bibinfo {volume} {114}},\ \bibinfo {pages}
  {132001} (\bibinfo {year} {2015}{\natexlab{b}})},\ \Eprint
  {http://arxiv.org/abs/1411.2943} {arXiv:1411.2943 [hep-ex]} \BibitemShut
  {NoStop}%
\bibitem [{\citenamefont {Aoki}\ \emph {et~al.}(2017)\citenamefont {Aoki} \emph
  {et~al.}}]{Aoki:2016frl}%
  \BibitemOpen
  \bibfield  {author} {\bibinfo {author} {\bibfnamefont {S.}~\bibnamefont
  {Aoki}} \emph {et~al.},\ }\href {\doibase 10.1140/epjc/s10052-016-4509-7}
  {\bibfield  {journal} {\bibinfo  {journal} {Eur. Phys. J.}\ }\textbf
  {\bibinfo {volume} {C77}},\ \bibinfo {pages} {112} (\bibinfo {year}
  {2017})},\ \Eprint {http://arxiv.org/abs/1607.00299} {arXiv:1607.00299
  [hep-lat]} \BibitemShut {NoStop}%
\bibitem [{\citenamefont {McNeile}\ \emph {et~al.}(2012)\citenamefont
  {McNeile}, \citenamefont {Davies}, \citenamefont {Follana}, \citenamefont
  {Hornbostel},\ and\ \citenamefont {Lepage}}]{McNeile:2012qf}%
  \BibitemOpen
  \bibfield  {author} {\bibinfo {author} {\bibfnamefont {C.}~\bibnamefont
  {McNeile}}, \bibinfo {author} {\bibfnamefont {C.~T.~H.}\ \bibnamefont
  {Davies}}, \bibinfo {author} {\bibfnamefont {E.}~\bibnamefont {Follana}},
  \bibinfo {author} {\bibfnamefont {K.}~\bibnamefont {Hornbostel}}, \ and\
  \bibinfo {author} {\bibfnamefont {G.~P.}\ \bibnamefont {Lepage}},\ }\href
  {\doibase 10.1103/PhysRevD.86.074503} {\bibfield  {journal} {\bibinfo
  {journal} {Phys. Rev.}\ }\textbf {\bibinfo {volume} {D86}},\ \bibinfo {pages}
  {074503} (\bibinfo {year} {2012})},\ \Eprint {http://arxiv.org/abs/1207.0994}
  {arXiv:1207.0994 [hep-lat]} \BibitemShut {NoStop}%
\bibitem [{\citenamefont {Datta}\ \emph {et~al.}(2017)\citenamefont {Datta},
  \citenamefont {Kamali}, \citenamefont {Meinel},\ and\ \citenamefont
  {Rashed}}]{Datta:2017aue}%
  \BibitemOpen
  \bibfield  {author} {\bibinfo {author} {\bibfnamefont {A.}~\bibnamefont
  {Datta}}, \bibinfo {author} {\bibfnamefont {S.}~\bibnamefont {Kamali}},
  \bibinfo {author} {\bibfnamefont {S.}~\bibnamefont {Meinel}}, \ and\ \bibinfo
  {author} {\bibfnamefont {A.}~\bibnamefont {Rashed}},\ }\href {\doibase
  10.1007/JHEP08(2017)131} {\bibfield  {journal} {\bibinfo  {journal} {JHEP}\
  }\textbf {\bibinfo {volume} {08}},\ \bibinfo {pages} {131} (\bibinfo {year}
  {2017})},\ \Eprint {http://arxiv.org/abs/1702.02243} {arXiv:1702.02243
  [hep-ph]} \BibitemShut {NoStop}%
\bibitem [{\citenamefont {Bernlochner}\ \emph {et~al.}(2018)\citenamefont
  {Bernlochner}, \citenamefont {Ligeti}, \citenamefont {Robinson},\ and\
  \citenamefont {Sutcliffe}}]{Bernlochner:2018kxh}%
  \BibitemOpen
  \bibfield  {author} {\bibinfo {author} {\bibfnamefont {F.~U.}\ \bibnamefont
  {Bernlochner}}, \bibinfo {author} {\bibfnamefont {Z.}~\bibnamefont {Ligeti}},
  \bibinfo {author} {\bibfnamefont {D.~J.}\ \bibnamefont {Robinson}}, \ and\
  \bibinfo {author} {\bibfnamefont {W.~L.}\ \bibnamefont {Sutcliffe}},\
  }\href@noop {} {\  (\bibinfo {year} {2018})},\ \Eprint
  {http://arxiv.org/abs/1808.09464} {arXiv:1808.09464 [hep-ph]} \BibitemShut
  {NoStop}%
\bibitem [{\citenamefont {Descotes-Genon}\ \emph {et~al.}(2016)\citenamefont
  {Descotes-Genon}, \citenamefont {Hofer}, \citenamefont {Matias},\ and\
  \citenamefont {Virto}}]{Descotes-Genon:2015uva}%
  \BibitemOpen
  \bibfield  {author} {\bibinfo {author} {\bibfnamefont {S.}~\bibnamefont
  {Descotes-Genon}}, \bibinfo {author} {\bibfnamefont {L.}~\bibnamefont
  {Hofer}}, \bibinfo {author} {\bibfnamefont {J.}~\bibnamefont {Matias}}, \
  and\ \bibinfo {author} {\bibfnamefont {J.}~\bibnamefont {Virto}},\ }\href
  {\doibase 10.1007/JHEP06(2016)092} {\bibfield  {journal} {\bibinfo  {journal}
  {JHEP}\ }\textbf {\bibinfo {volume} {06}},\ \bibinfo {pages} {092} (\bibinfo
  {year} {2016})},\ \Eprint {http://arxiv.org/abs/1510.04239} {arXiv:1510.04239
  [hep-ph]} \BibitemShut {NoStop}%
\bibitem [{\citenamefont {Alonso}\ \emph {et~al.}(2015)\citenamefont {Alonso},
  \citenamefont {Grinstein},\ and\ \citenamefont
  {Martin~Camalich}}]{Alonso:2015sja}%
  \BibitemOpen
  \bibfield  {author} {\bibinfo {author} {\bibfnamefont {R.}~\bibnamefont
  {Alonso}}, \bibinfo {author} {\bibfnamefont {B.}~\bibnamefont {Grinstein}}, \
  and\ \bibinfo {author} {\bibfnamefont {J.}~\bibnamefont {Martin~Camalich}},\
  }\href {\doibase 10.1007/JHEP10(2015)184} {\bibfield  {journal} {\bibinfo
  {journal} {JHEP}\ }\textbf {\bibinfo {volume} {10}},\ \bibinfo {pages} {184}
  (\bibinfo {year} {2015})},\ \Eprint {http://arxiv.org/abs/1505.05164}
  {arXiv:1505.05164 [hep-ph]} \BibitemShut {NoStop}%
\bibitem [{\citenamefont {Calibbi}\ \emph {et~al.}(2015)\citenamefont
  {Calibbi}, \citenamefont {Crivellin},\ and\ \citenamefont
  {Ota}}]{Calibbi:2015kma}%
  \BibitemOpen
  \bibfield  {author} {\bibinfo {author} {\bibfnamefont {L.}~\bibnamefont
  {Calibbi}}, \bibinfo {author} {\bibfnamefont {A.}~\bibnamefont {Crivellin}},
  \ and\ \bibinfo {author} {\bibfnamefont {T.}~\bibnamefont {Ota}},\ }\href
  {\doibase 10.1103/PhysRevLett.115.181801} {\bibfield  {journal} {\bibinfo
  {journal} {Phys. Rev. Lett.}\ }\textbf {\bibinfo {volume} {115}},\ \bibinfo
  {pages} {181801} (\bibinfo {year} {2015})},\ \Eprint
  {http://arxiv.org/abs/1506.02661} {arXiv:1506.02661 [hep-ph]} \BibitemShut
  {NoStop}%
\bibitem [{\citenamefont {Fajfer}\ and\ \citenamefont
  {Ko\v{s}nik}(2016)}]{Fajfer:2015ycq}%
  \BibitemOpen
  \bibfield  {author} {\bibinfo {author} {\bibfnamefont {S.}~\bibnamefont
  {Fajfer}}\ and\ \bibinfo {author} {\bibfnamefont {N.}~\bibnamefont
  {Ko\v{s}nik}},\ }\href {\doibase 10.1016/j.physletb.2016.02.018} {\bibfield
  {journal} {\bibinfo  {journal} {Phys. Lett.}\ }\textbf {\bibinfo {volume}
  {B755}},\ \bibinfo {pages} {270} (\bibinfo {year} {2016})},\ \Eprint
  {http://arxiv.org/abs/1511.06024} {arXiv:1511.06024 [hep-ph]} \BibitemShut
  {NoStop}%
\bibitem [{\citenamefont {Barbieri}\ \emph {et~al.}(2016)\citenamefont
  {Barbieri}, \citenamefont {Isidori}, \citenamefont {Pattori},\ and\
  \citenamefont {Senia}}]{Barbieri:2015yvd}%
  \BibitemOpen
  \bibfield  {author} {\bibinfo {author} {\bibfnamefont {R.}~\bibnamefont
  {Barbieri}}, \bibinfo {author} {\bibfnamefont {G.}~\bibnamefont {Isidori}},
  \bibinfo {author} {\bibfnamefont {A.}~\bibnamefont {Pattori}}, \ and\
  \bibinfo {author} {\bibfnamefont {F.}~\bibnamefont {Senia}},\ }\href
  {\doibase 10.1140/epjc/s10052-016-3905-3} {\bibfield  {journal} {\bibinfo
  {journal} {Eur. Phys. J.}\ }\textbf {\bibinfo {volume} {C76}},\ \bibinfo
  {pages} {67} (\bibinfo {year} {2016})},\ \Eprint
  {http://arxiv.org/abs/1512.01560} {arXiv:1512.01560 [hep-ph]} \BibitemShut
  {NoStop}%
\bibitem [{\citenamefont {Barbieri}\ \emph {et~al.}(2017)\citenamefont
  {Barbieri}, \citenamefont {Murphy},\ and\ \citenamefont
  {Senia}}]{Barbieri:2016las}%
  \BibitemOpen
  \bibfield  {author} {\bibinfo {author} {\bibfnamefont {R.}~\bibnamefont
  {Barbieri}}, \bibinfo {author} {\bibfnamefont {C.~W.}\ \bibnamefont
  {Murphy}}, \ and\ \bibinfo {author} {\bibfnamefont {F.}~\bibnamefont
  {Senia}},\ }\href {\doibase 10.1140/epjc/s10052-016-4578-7} {\bibfield
  {journal} {\bibinfo  {journal} {Eur. Phys. J.}\ }\textbf {\bibinfo {volume}
  {C77}},\ \bibinfo {pages} {8} (\bibinfo {year} {2017})},\ \Eprint
  {http://arxiv.org/abs/1611.04930} {arXiv:1611.04930 [hep-ph]} \BibitemShut
  {NoStop}%
\bibitem [{\citenamefont {Hiller}\ \emph {et~al.}(2016)\citenamefont {Hiller},
  \citenamefont {Loose},\ and\ \citenamefont {Schonwald}}]{Hiller:2016kry}%
  \BibitemOpen
  \bibfield  {author} {\bibinfo {author} {\bibfnamefont {G.}~\bibnamefont
  {Hiller}}, \bibinfo {author} {\bibfnamefont {D.}~\bibnamefont {Loose}}, \
  and\ \bibinfo {author} {\bibfnamefont {K.}~\bibnamefont {Schonwald}},\ }\href
  {\doibase 10.1007/JHEP12(2016)027} {\bibfield  {journal} {\bibinfo  {journal}
  {JHEP}\ }\textbf {\bibinfo {volume} {12}},\ \bibinfo {pages} {027} (\bibinfo
  {year} {2016})},\ \Eprint {http://arxiv.org/abs/1609.08895} {arXiv:1609.08895
  [hep-ph]} \BibitemShut {NoStop}%
\bibitem [{\citenamefont {Bhattacharya}\ \emph
  {et~al.}(2017{\natexlab{b}})\citenamefont {Bhattacharya}, \citenamefont
  {Datta}, \citenamefont {Gu\'{e}vin}, \citenamefont {London},\ and\
  \citenamefont {Watanabe}}]{Bhattacharya:2016mcc}%
  \BibitemOpen
  \bibfield  {author} {\bibinfo {author} {\bibfnamefont {B.}~\bibnamefont
  {Bhattacharya}}, \bibinfo {author} {\bibfnamefont {A.}~\bibnamefont {Datta}},
  \bibinfo {author} {\bibfnamefont {J.-P.}\ \bibnamefont {Gu\'{e}vin}},
  \bibinfo {author} {\bibfnamefont {D.}~\bibnamefont {London}}, \ and\ \bibinfo
  {author} {\bibfnamefont {R.}~\bibnamefont {Watanabe}},\ }\href {\doibase
  10.1007/JHEP01(2017)015} {\bibfield  {journal} {\bibinfo  {journal} {JHEP}\
  }\textbf {\bibinfo {volume} {01}},\ \bibinfo {pages} {015} (\bibinfo {year}
  {2017}{\natexlab{b}})},\ \Eprint {http://arxiv.org/abs/1609.09078}
  {arXiv:1609.09078 [hep-ph]} \BibitemShut {NoStop}%
\bibitem [{\citenamefont {Buttazzo}\ \emph {et~al.}(2017)\citenamefont
  {Buttazzo}, \citenamefont {Greljo}, \citenamefont {Isidori},\ and\
  \citenamefont {Marzocca}}]{Buttazzo:2017ixm}%
  \BibitemOpen
  \bibfield  {author} {\bibinfo {author} {\bibfnamefont {D.}~\bibnamefont
  {Buttazzo}}, \bibinfo {author} {\bibfnamefont {A.}~\bibnamefont {Greljo}},
  \bibinfo {author} {\bibfnamefont {G.}~\bibnamefont {Isidori}}, \ and\
  \bibinfo {author} {\bibfnamefont {D.}~\bibnamefont {Marzocca}},\ }\href
  {\doibase 10.1007/JHEP11(2017)044} {\bibfield  {journal} {\bibinfo  {journal}
  {JHEP}\ }\textbf {\bibinfo {volume} {11}},\ \bibinfo {pages} {044} (\bibinfo
  {year} {2017})},\ \Eprint {http://arxiv.org/abs/1706.07808} {arXiv:1706.07808
  [hep-ph]} \BibitemShut {NoStop}%
\bibitem [{\citenamefont {Kumar}\ \emph {et~al.}(2018)\citenamefont {Kumar},
  \citenamefont {London},\ and\ \citenamefont {Watanabe}}]{Kumar:2018kmr}%
  \BibitemOpen
  \bibfield  {author} {\bibinfo {author} {\bibfnamefont {J.}~\bibnamefont
  {Kumar}}, \bibinfo {author} {\bibfnamefont {D.}~\bibnamefont {London}}, \
  and\ \bibinfo {author} {\bibfnamefont {R.}~\bibnamefont {Watanabe}},\
  }\href@noop {} {\  (\bibinfo {year} {2018})},\ \Eprint
  {http://arxiv.org/abs/1806.07403} {arXiv:1806.07403 [hep-ph]} \BibitemShut
  {NoStop}%
\bibitem [{\citenamefont {Assad}\ \emph {et~al.}(2018)\citenamefont {Assad},
  \citenamefont {Fornal},\ and\ \citenamefont {Grinstein}}]{Assad:2017iib}%
  \BibitemOpen
  \bibfield  {author} {\bibinfo {author} {\bibfnamefont {N.}~\bibnamefont
  {Assad}}, \bibinfo {author} {\bibfnamefont {B.}~\bibnamefont {Fornal}}, \
  and\ \bibinfo {author} {\bibfnamefont {B.}~\bibnamefont {Grinstein}},\ }\href
  {\doibase 10.1016/j.physletb.2017.12.042} {\bibfield  {journal} {\bibinfo
  {journal} {Phys. Lett.}\ }\textbf {\bibinfo {volume} {B777}},\ \bibinfo
  {pages} {324} (\bibinfo {year} {2018})},\ \Eprint
  {http://arxiv.org/abs/1708.06350} {arXiv:1708.06350 [hep-ph]} \BibitemShut
  {NoStop}%
\bibitem [{\citenamefont {Di~Luzio}\ \emph {et~al.}(2017)\citenamefont
  {Di~Luzio}, \citenamefont {Greljo},\ and\ \citenamefont
  {Nardecchia}}]{DiLuzio:2017vat}%
  \BibitemOpen
  \bibfield  {author} {\bibinfo {author} {\bibfnamefont {L.}~\bibnamefont
  {Di~Luzio}}, \bibinfo {author} {\bibfnamefont {A.}~\bibnamefont {Greljo}}, \
  and\ \bibinfo {author} {\bibfnamefont {M.}~\bibnamefont {Nardecchia}},\
  }\href {\doibase 10.1103/PhysRevD.96.115011} {\bibfield  {journal} {\bibinfo
  {journal} {Phys. Rev.}\ }\textbf {\bibinfo {volume} {D96}},\ \bibinfo {pages}
  {115011} (\bibinfo {year} {2017})},\ \Eprint
  {http://arxiv.org/abs/1708.08450} {arXiv:1708.08450 [hep-ph]} \BibitemShut
  {NoStop}%
\bibitem [{\citenamefont {Calibbi}\ \emph {et~al.}(2017)\citenamefont
  {Calibbi}, \citenamefont {Crivellin},\ and\ \citenamefont
  {Li}}]{Calibbi:2017qbu}%
  \BibitemOpen
  \bibfield  {author} {\bibinfo {author} {\bibfnamefont {L.}~\bibnamefont
  {Calibbi}}, \bibinfo {author} {\bibfnamefont {A.}~\bibnamefont {Crivellin}},
  \ and\ \bibinfo {author} {\bibfnamefont {T.}~\bibnamefont {Li}},\ }\href@noop
  {} {\  (\bibinfo {year} {2017})},\ \Eprint {http://arxiv.org/abs/1709.00692}
  {arXiv:1709.00692 [hep-ph]} \BibitemShut {NoStop}%
\bibitem [{\citenamefont {Bordone}\ \emph
  {et~al.}(2018{\natexlab{a}})\citenamefont {Bordone}, \citenamefont
  {Cornella}, \citenamefont {Fuentes-Martin},\ and\ \citenamefont
  {Isidori}}]{Bordone:2017bld}%
  \BibitemOpen
  \bibfield  {author} {\bibinfo {author} {\bibfnamefont {M.}~\bibnamefont
  {Bordone}}, \bibinfo {author} {\bibfnamefont {C.}~\bibnamefont {Cornella}},
  \bibinfo {author} {\bibfnamefont {J.}~\bibnamefont {Fuentes-Martin}}, \ and\
  \bibinfo {author} {\bibfnamefont {G.}~\bibnamefont {Isidori}},\ }\href
  {\doibase 10.1016/j.physletb.2018.02.011} {\bibfield  {journal} {\bibinfo
  {journal} {Phys. Lett.}\ }\textbf {\bibinfo {volume} {B779}},\ \bibinfo
  {pages} {317} (\bibinfo {year} {2018}{\natexlab{a}})},\ \Eprint
  {http://arxiv.org/abs/1712.01368} {arXiv:1712.01368 [hep-ph]} \BibitemShut
  {NoStop}%
\bibitem [{\citenamefont {Barbieri}\ and\ \citenamefont
  {Tesi}(2018)}]{Barbieri:2017tuq}%
  \BibitemOpen
  \bibfield  {author} {\bibinfo {author} {\bibfnamefont {R.}~\bibnamefont
  {Barbieri}}\ and\ \bibinfo {author} {\bibfnamefont {A.}~\bibnamefont
  {Tesi}},\ }\href {\doibase 10.1140/epjc/s10052-018-5680-9} {\bibfield
  {journal} {\bibinfo  {journal} {Eur. Phys. J.}\ }\textbf {\bibinfo {volume}
  {C78}},\ \bibinfo {pages} {193} (\bibinfo {year} {2018})},\ \Eprint
  {http://arxiv.org/abs/1712.06844} {arXiv:1712.06844 [hep-ph]} \BibitemShut
  {NoStop}%
\bibitem [{\citenamefont {Blanke}\ and\ \citenamefont
  {Crivellin}(2018)}]{Blanke:2018sro}%
  \BibitemOpen
  \bibfield  {author} {\bibinfo {author} {\bibfnamefont {M.}~\bibnamefont
  {Blanke}}\ and\ \bibinfo {author} {\bibfnamefont {A.}~\bibnamefont
  {Crivellin}},\ }\href {\doibase 10.1103/PhysRevLett.121.011801} {\bibfield
  {journal} {\bibinfo  {journal} {Phys. Rev. Lett.}\ }\textbf {\bibinfo
  {volume} {121}},\ \bibinfo {pages} {011801} (\bibinfo {year} {2018})},\
  \Eprint {http://arxiv.org/abs/1801.07256} {arXiv:1801.07256 [hep-ph]}
  \BibitemShut {NoStop}%
\bibitem [{\citenamefont {Greljo}\ and\ \citenamefont
  {Stefanek}(2018)}]{Greljo:2018tuh}%
  \BibitemOpen
  \bibfield  {author} {\bibinfo {author} {\bibfnamefont {A.}~\bibnamefont
  {Greljo}}\ and\ \bibinfo {author} {\bibfnamefont {B.~A.}\ \bibnamefont
  {Stefanek}},\ }\href {\doibase 10.1016/j.physletb.2018.05.033} {\bibfield
  {journal} {\bibinfo  {journal} {Phys. Lett.}\ }\textbf {\bibinfo {volume}
  {B782}},\ \bibinfo {pages} {131} (\bibinfo {year} {2018})},\ \Eprint
  {http://arxiv.org/abs/1802.04274} {arXiv:1802.04274 [hep-ph]} \BibitemShut
  {NoStop}%
\bibitem [{\citenamefont {Bordone}\ \emph
  {et~al.}(2018{\natexlab{b}})\citenamefont {Bordone}, \citenamefont
  {Cornella}, \citenamefont {Fuentes-Mart\'{i}n},\ and\ \citenamefont
  {Isidori}}]{Bordone:2018nbg}%
  \BibitemOpen
  \bibfield  {author} {\bibinfo {author} {\bibfnamefont {M.}~\bibnamefont
  {Bordone}}, \bibinfo {author} {\bibfnamefont {C.}~\bibnamefont {Cornella}},
  \bibinfo {author} {\bibfnamefont {J.}~\bibnamefont {Fuentes-Mart\'{i}n}}, \
  and\ \bibinfo {author} {\bibfnamefont {G.}~\bibnamefont {Isidori}},\
  }\href@noop {} {\  (\bibinfo {year} {2018}{\natexlab{b}})},\ \Eprint
  {http://arxiv.org/abs/1805.09328} {arXiv:1805.09328 [hep-ph]} \BibitemShut
  {NoStop}%
\bibitem [{\citenamefont {Matsuzaki}\ \emph {et~al.}(2018)\citenamefont
  {Matsuzaki}, \citenamefont {Nishiwaki},\ and\ \citenamefont
  {Yamamoto}}]{Matsuzaki:2018jui}%
  \BibitemOpen
  \bibfield  {author} {\bibinfo {author} {\bibfnamefont {S.}~\bibnamefont
  {Matsuzaki}}, \bibinfo {author} {\bibfnamefont {K.}~\bibnamefont
  {Nishiwaki}}, \ and\ \bibinfo {author} {\bibfnamefont {K.}~\bibnamefont
  {Yamamoto}},\ }\href@noop {} {\  (\bibinfo {year} {2018})},\ \Eprint
  {http://arxiv.org/abs/1806.02312} {arXiv:1806.02312 [hep-ph]} \BibitemShut
  {NoStop}%
\bibitem [{\citenamefont {Crivellin}\ \emph {et~al.}(2018)\citenamefont
  {Crivellin}, \citenamefont {Greub}, \citenamefont {Saturnino},\ and\
  \citenamefont {Muller}}]{Crivellin:2018yvo}%
  \BibitemOpen
  \bibfield  {author} {\bibinfo {author} {\bibfnamefont {A.}~\bibnamefont
  {Crivellin}}, \bibinfo {author} {\bibfnamefont {C.}~\bibnamefont {Greub}},
  \bibinfo {author} {\bibfnamefont {F.}~\bibnamefont {Saturnino}}, \ and\
  \bibinfo {author} {\bibfnamefont {D.}~\bibnamefont {Muller}},\ }\href@noop {}
  {\  (\bibinfo {year} {2018})},\ \Eprint {http://arxiv.org/abs/1807.02068}
  {arXiv:1807.02068 [hep-ph]} \BibitemShut {NoStop}%
\bibitem [{\citenamefont {Di~Luzio}\ \emph {et~al.}(2018)\citenamefont
  {Di~Luzio}, \citenamefont {Fuentes-Martin}, \citenamefont {Greljo},
  \citenamefont {Nardecchia},\ and\ \citenamefont {Renner}}]{DiLuzio:2018zxy}%
  \BibitemOpen
  \bibfield  {author} {\bibinfo {author} {\bibfnamefont {L.}~\bibnamefont
  {Di~Luzio}}, \bibinfo {author} {\bibfnamefont {J.}~\bibnamefont
  {Fuentes-Martin}}, \bibinfo {author} {\bibfnamefont {A.}~\bibnamefont
  {Greljo}}, \bibinfo {author} {\bibfnamefont {M.}~\bibnamefont {Nardecchia}},
  \ and\ \bibinfo {author} {\bibfnamefont {S.}~\bibnamefont {Renner}},\
  }\href@noop {} {\  (\bibinfo {year} {2018})},\ \Eprint
  {http://arxiv.org/abs/1808.00942} {arXiv:1808.00942 [hep-ph]} \BibitemShut
  {NoStop}%
\bibitem [{\citenamefont {Biswas}\ \emph
  {et~al.}(2018{\natexlab{a}})\citenamefont {Biswas}, \citenamefont
  {Kumar~Ghosh}, \citenamefont {Ghosh}, \citenamefont {Shaw},\ and\
  \citenamefont {Swain}}]{Biswas:2018snp}%
  \BibitemOpen
  \bibfield  {author} {\bibinfo {author} {\bibfnamefont {A.}~\bibnamefont
  {Biswas}}, \bibinfo {author} {\bibfnamefont {D.}~\bibnamefont {Kumar~Ghosh}},
  \bibinfo {author} {\bibfnamefont {N.}~\bibnamefont {Ghosh}}, \bibinfo
  {author} {\bibfnamefont {A.}~\bibnamefont {Shaw}}, \ and\ \bibinfo {author}
  {\bibfnamefont {A.~K.}\ \bibnamefont {Swain}},\ }\href@noop {} {\  (\bibinfo
  {year} {2018}{\natexlab{a}})},\ \Eprint {http://arxiv.org/abs/1808.04169}
  {arXiv:1808.04169 [hep-ph]} \BibitemShut {NoStop}%
\bibitem [{\citenamefont {Deshpande}\ and\ \citenamefont
  {Menon}(2013)}]{Deshpande:2012rr}%
  \BibitemOpen
  \bibfield  {author} {\bibinfo {author} {\bibfnamefont {N.~G.}\ \bibnamefont
  {Deshpande}}\ and\ \bibinfo {author} {\bibfnamefont {A.}~\bibnamefont
  {Menon}},\ }\href {\doibase 10.1007/JHEP01(2013)025} {\bibfield  {journal}
  {\bibinfo  {journal} {JHEP}\ }\textbf {\bibinfo {volume} {01}},\ \bibinfo
  {pages} {025} (\bibinfo {year} {2013})},\ \Eprint
  {http://arxiv.org/abs/1208.4134} {arXiv:1208.4134 [hep-ph]} \BibitemShut
  {NoStop}%
\bibitem [{\citenamefont {Sakaki}\ \emph {et~al.}(2013)\citenamefont {Sakaki},
  \citenamefont {Tanaka}, \citenamefont {Tayduganov},\ and\ \citenamefont
  {Watanabe}}]{Sakaki:2013bfa}%
  \BibitemOpen
  \bibfield  {author} {\bibinfo {author} {\bibfnamefont {Y.}~\bibnamefont
  {Sakaki}}, \bibinfo {author} {\bibfnamefont {M.}~\bibnamefont {Tanaka}},
  \bibinfo {author} {\bibfnamefont {A.}~\bibnamefont {Tayduganov}}, \ and\
  \bibinfo {author} {\bibfnamefont {R.}~\bibnamefont {Watanabe}},\ }\href
  {\doibase 10.1103/PhysRevD.88.094012} {\bibfield  {journal} {\bibinfo
  {journal} {Phys. Rev.}\ }\textbf {\bibinfo {volume} {D88}},\ \bibinfo {pages}
  {094012} (\bibinfo {year} {2013})},\ \Eprint {http://arxiv.org/abs/1309.0301}
  {arXiv:1309.0301 [hep-ph]} \BibitemShut {NoStop}%
\bibitem [{\citenamefont {Bauer}\ and\ \citenamefont
  {Neubert}(2016)}]{Bauer:2015knc}%
  \BibitemOpen
  \bibfield  {author} {\bibinfo {author} {\bibfnamefont {M.}~\bibnamefont
  {Bauer}}\ and\ \bibinfo {author} {\bibfnamefont {M.}~\bibnamefont
  {Neubert}},\ }\href {\doibase 10.1103/PhysRevLett.116.141802} {\bibfield
  {journal} {\bibinfo  {journal} {Phys. Rev. Lett.}\ }\textbf {\bibinfo
  {volume} {116}},\ \bibinfo {pages} {141802} (\bibinfo {year} {2016})},\
  \Eprint {http://arxiv.org/abs/1511.01900} {arXiv:1511.01900 [hep-ph]}
  \BibitemShut {NoStop}%
\bibitem [{\citenamefont {Cai}\ \emph {et~al.}(2017)\citenamefont {Cai},
  \citenamefont {Gargalionis}, \citenamefont {Schmidt},\ and\ \citenamefont
  {Volkas}}]{Cai:2017wry}%
  \BibitemOpen
  \bibfield  {author} {\bibinfo {author} {\bibfnamefont {Y.}~\bibnamefont
  {Cai}}, \bibinfo {author} {\bibfnamefont {J.}~\bibnamefont {Gargalionis}},
  \bibinfo {author} {\bibfnamefont {M.~A.}\ \bibnamefont {Schmidt}}, \ and\
  \bibinfo {author} {\bibfnamefont {R.~R.}\ \bibnamefont {Volkas}},\ }\href
  {\doibase 10.1007/JHEP10(2017)047} {\bibfield  {journal} {\bibinfo  {journal}
  {JHEP}\ }\textbf {\bibinfo {volume} {10}},\ \bibinfo {pages} {047} (\bibinfo
  {year} {2017})},\ \Eprint {http://arxiv.org/abs/1704.05849} {arXiv:1704.05849
  [hep-ph]} \BibitemShut {NoStop}%
\bibitem [{\citenamefont {Crivellin}\ \emph {et~al.}(2017)\citenamefont
  {Crivellin}, \citenamefont {Muller},\ and\ \citenamefont
  {Ota}}]{Crivellin:2017zlb}%
  \BibitemOpen
  \bibfield  {author} {\bibinfo {author} {\bibfnamefont {A.}~\bibnamefont
  {Crivellin}}, \bibinfo {author} {\bibfnamefont {D.}~\bibnamefont {Muller}}, \
  and\ \bibinfo {author} {\bibfnamefont {T.}~\bibnamefont {Ota}},\ }\href
  {\doibase 10.1007/JHEP09(2017)040} {\bibfield  {journal} {\bibinfo  {journal}
  {JHEP}\ }\textbf {\bibinfo {volume} {09}},\ \bibinfo {pages} {040} (\bibinfo
  {year} {2017})},\ \Eprint {http://arxiv.org/abs/1703.09226} {arXiv:1703.09226
  [hep-ph]} \BibitemShut {NoStop}%
\bibitem [{\citenamefont {Altmannshofer}\ \emph {et~al.}(2017)\citenamefont
  {Altmannshofer}, \citenamefont {Bhupal~Dev},\ and\ \citenamefont
  {Soni}}]{Altmannshofer:2017poe}%
  \BibitemOpen
  \bibfield  {author} {\bibinfo {author} {\bibfnamefont {W.}~\bibnamefont
  {Altmannshofer}}, \bibinfo {author} {\bibfnamefont {P.}~\bibnamefont
  {Bhupal~Dev}}, \ and\ \bibinfo {author} {\bibfnamefont {A.}~\bibnamefont
  {Soni}},\ }\href {\doibase 10.1103/PhysRevD.96.095010} {\bibfield  {journal}
  {\bibinfo  {journal} {Phys. Rev.}\ }\textbf {\bibinfo {volume} {D96}},\
  \bibinfo {pages} {095010} (\bibinfo {year} {2017})},\ \Eprint
  {http://arxiv.org/abs/1704.06659} {arXiv:1704.06659 [hep-ph]} \BibitemShut
  {NoStop}%
\bibitem [{\citenamefont {Marzocca}(2018)}]{Marzocca:2018wcf}%
  \BibitemOpen
  \bibfield  {author} {\bibinfo {author} {\bibfnamefont {D.}~\bibnamefont
  {Marzocca}},\ }\href {\doibase 10.1007/JHEP07(2018)121} {\bibfield  {journal}
  {\bibinfo  {journal} {JHEP}\ }\textbf {\bibinfo {volume} {07}},\ \bibinfo
  {pages} {121} (\bibinfo {year} {2018})},\ \Eprint
  {http://arxiv.org/abs/1803.10972} {arXiv:1803.10972 [hep-ph]} \BibitemShut
  {NoStop}%
\bibitem [{\citenamefont {He}\ and\ \citenamefont
  {Valencia}(2013)}]{He:2012zp}%
  \BibitemOpen
  \bibfield  {author} {\bibinfo {author} {\bibfnamefont {X.-G.}\ \bibnamefont
  {He}}\ and\ \bibinfo {author} {\bibfnamefont {G.}~\bibnamefont {Valencia}},\
  }\href {\doibase 10.1103/PhysRevD.87.014014} {\bibfield  {journal} {\bibinfo
  {journal} {Phys. Rev.}\ }\textbf {\bibinfo {volume} {D87}},\ \bibinfo {pages}
  {014014} (\bibinfo {year} {2013})},\ \Eprint {http://arxiv.org/abs/1211.0348}
  {arXiv:1211.0348 [hep-ph]} \BibitemShut {NoStop}%
\bibitem [{\citenamefont {Greljo}\ \emph {et~al.}(2015)\citenamefont {Greljo},
  \citenamefont {Isidori},\ and\ \citenamefont {Marzocca}}]{Greljo:2015mma}%
  \BibitemOpen
  \bibfield  {author} {\bibinfo {author} {\bibfnamefont {A.}~\bibnamefont
  {Greljo}}, \bibinfo {author} {\bibfnamefont {G.}~\bibnamefont {Isidori}}, \
  and\ \bibinfo {author} {\bibfnamefont {D.}~\bibnamefont {Marzocca}},\ }\href
  {\doibase 10.1007/JHEP07(2015)142} {\bibfield  {journal} {\bibinfo  {journal}
  {JHEP}\ }\textbf {\bibinfo {volume} {07}},\ \bibinfo {pages} {142} (\bibinfo
  {year} {2015})},\ \Eprint {http://arxiv.org/abs/1506.01705} {arXiv:1506.01705
  [hep-ph]} \BibitemShut {NoStop}%
\bibitem [{\citenamefont {Boucenna}\ \emph {et~al.}(2016)\citenamefont
  {Boucenna}, \citenamefont {Celis}, \citenamefont {Fuentes-Martin},
  \citenamefont {Vicente},\ and\ \citenamefont {Virto}}]{Boucenna:2016wpr}%
  \BibitemOpen
  \bibfield  {author} {\bibinfo {author} {\bibfnamefont {S.~M.}\ \bibnamefont
  {Boucenna}}, \bibinfo {author} {\bibfnamefont {A.}~\bibnamefont {Celis}},
  \bibinfo {author} {\bibfnamefont {J.}~\bibnamefont {Fuentes-Martin}},
  \bibinfo {author} {\bibfnamefont {A.}~\bibnamefont {Vicente}}, \ and\
  \bibinfo {author} {\bibfnamefont {J.}~\bibnamefont {Virto}},\ }\href
  {\doibase 10.1016/j.physletb.2016.06.067} {\bibfield  {journal} {\bibinfo
  {journal} {Phys. Lett.}\ }\textbf {\bibinfo {volume} {B760}},\ \bibinfo
  {pages} {214} (\bibinfo {year} {2016})},\ \Eprint
  {http://arxiv.org/abs/1604.03088} {arXiv:1604.03088 [hep-ph]} \BibitemShut
  {NoStop}%
\bibitem [{\citenamefont {He}\ and\ \citenamefont
  {Valencia}(2018)}]{He:2017bft}%
  \BibitemOpen
  \bibfield  {author} {\bibinfo {author} {\bibfnamefont {X.-G.}\ \bibnamefont
  {He}}\ and\ \bibinfo {author} {\bibfnamefont {G.}~\bibnamefont {Valencia}},\
  }\href {\doibase 10.1016/j.physletb.2018.01.073} {\bibfield  {journal}
  {\bibinfo  {journal} {Phys. Lett.}\ }\textbf {\bibinfo {volume} {B779}},\
  \bibinfo {pages} {52} (\bibinfo {year} {2018})},\ \Eprint
  {http://arxiv.org/abs/1711.09525} {arXiv:1711.09525 [hep-ph]} \BibitemShut
  {NoStop}%
\bibitem [{\citenamefont {Kalinowski}(1990)}]{Kalinowski:1990ba}%
  \BibitemOpen
  \bibfield  {author} {\bibinfo {author} {\bibfnamefont {J.}~\bibnamefont
  {Kalinowski}},\ }\href {\doibase 10.1016/0370-2693(90)90134-R} {\bibfield
  {journal} {\bibinfo  {journal} {Phys. Lett.}\ }\textbf {\bibinfo {volume}
  {B245}},\ \bibinfo {pages} {201} (\bibinfo {year} {1990})}\BibitemShut
  {NoStop}%
\bibitem [{\citenamefont {Hou}(1993)}]{Hou:1992sy}%
  \BibitemOpen
  \bibfield  {author} {\bibinfo {author} {\bibfnamefont {W.-S.}\ \bibnamefont
  {Hou}},\ }\href {\doibase 10.1103/PhysRevD.48.2342} {\bibfield  {journal}
  {\bibinfo  {journal} {Phys. Rev.}\ }\textbf {\bibinfo {volume} {D48}},\
  \bibinfo {pages} {2342} (\bibinfo {year} {1993})}\BibitemShut {NoStop}%
\bibitem [{\citenamefont {Kosnik}(2012)}]{Kosnik:2012dj}%
  \BibitemOpen
  \bibfield  {author} {\bibinfo {author} {\bibfnamefont {N.}~\bibnamefont
  {Kosnik}},\ }\href {\doibase 10.1103/PhysRevD.86.055004} {\bibfield
  {journal} {\bibinfo  {journal} {Phys. Rev.}\ }\textbf {\bibinfo {volume}
  {D86}},\ \bibinfo {pages} {055004} (\bibinfo {year} {2012})},\ \Eprint
  {http://arxiv.org/abs/1206.2970} {arXiv:1206.2970 [hep-ph]} \BibitemShut
  {NoStop}%
\bibitem [{\citenamefont {Biswas}\ \emph
  {et~al.}(2018{\natexlab{b}})\citenamefont {Biswas}, \citenamefont {Shaw},\
  and\ \citenamefont {Swain}}]{Biswas:2018iak}%
  \BibitemOpen
  \bibfield  {author} {\bibinfo {author} {\bibfnamefont {A.}~\bibnamefont
  {Biswas}}, \bibinfo {author} {\bibfnamefont {A.}~\bibnamefont {Shaw}}, \ and\
  \bibinfo {author} {\bibfnamefont {A.~K.}\ \bibnamefont {Swain}},\ }\href@noop
  {} {\  (\bibinfo {year} {2018}{\natexlab{b}})},\ \Eprint
  {http://arxiv.org/abs/1811.08887} {arXiv:1811.08887 [hep-ph]} \BibitemShut
  {NoStop}%
\bibitem [{\citenamefont {Crivellin}\ \emph {et~al.}(2012)\citenamefont
  {Crivellin}, \citenamefont {Greub},\ and\ \citenamefont
  {Kokulu}}]{Crivellin:2012ye}%
  \BibitemOpen
  \bibfield  {author} {\bibinfo {author} {\bibfnamefont {A.}~\bibnamefont
  {Crivellin}}, \bibinfo {author} {\bibfnamefont {C.}~\bibnamefont {Greub}}, \
  and\ \bibinfo {author} {\bibfnamefont {A.}~\bibnamefont {Kokulu}},\ }\href
  {\doibase 10.1103/PhysRevD.86.054014} {\bibfield  {journal} {\bibinfo
  {journal} {Phys. Rev.}\ }\textbf {\bibinfo {volume} {D86}},\ \bibinfo {pages}
  {054014} (\bibinfo {year} {2012})},\ \Eprint {http://arxiv.org/abs/1206.2634}
  {arXiv:1206.2634 [hep-ph]} \BibitemShut {NoStop}%
\bibitem [{\citenamefont {Crivellin}\ \emph {et~al.}(2013)\citenamefont
  {Crivellin}, \citenamefont {Kokulu},\ and\ \citenamefont
  {Greub}}]{Crivellin:2013wna}%
  \BibitemOpen
  \bibfield  {author} {\bibinfo {author} {\bibfnamefont {A.}~\bibnamefont
  {Crivellin}}, \bibinfo {author} {\bibfnamefont {A.}~\bibnamefont {Kokulu}}, \
  and\ \bibinfo {author} {\bibfnamefont {C.}~\bibnamefont {Greub}},\ }\href
  {\doibase 10.1103/PhysRevD.87.094031} {\bibfield  {journal} {\bibinfo
  {journal} {Phys. Rev.}\ }\textbf {\bibinfo {volume} {D87}},\ \bibinfo {pages}
  {094031} (\bibinfo {year} {2013})},\ \Eprint {http://arxiv.org/abs/1303.5877}
  {arXiv:1303.5877 [hep-ph]} \BibitemShut {NoStop}%
\bibitem [{\citenamefont {Celis}\ \emph {et~al.}(2013)\citenamefont {Celis},
  \citenamefont {Jung}, \citenamefont {Li},\ and\ \citenamefont
  {Pich}}]{Celis:2012dk}%
  \BibitemOpen
  \bibfield  {author} {\bibinfo {author} {\bibfnamefont {A.}~\bibnamefont
  {Celis}}, \bibinfo {author} {\bibfnamefont {M.}~\bibnamefont {Jung}},
  \bibinfo {author} {\bibfnamefont {X.-Q.}\ \bibnamefont {Li}}, \ and\ \bibinfo
  {author} {\bibfnamefont {A.}~\bibnamefont {Pich}},\ }\href {\doibase
  10.1007/JHEP01(2013)054} {\bibfield  {journal} {\bibinfo  {journal} {JHEP}\
  }\textbf {\bibinfo {volume} {01}},\ \bibinfo {pages} {054} (\bibinfo {year}
  {2013})},\ \Eprint {http://arxiv.org/abs/1210.8443} {arXiv:1210.8443
  [hep-ph]} \BibitemShut {NoStop}%
\bibitem [{\citenamefont {Ko}\ \emph {et~al.}(2013)\citenamefont {Ko},
  \citenamefont {Omura},\ and\ \citenamefont {Yu}}]{Ko:2012sv}%
  \BibitemOpen
  \bibfield  {author} {\bibinfo {author} {\bibfnamefont {P.}~\bibnamefont
  {Ko}}, \bibinfo {author} {\bibfnamefont {Y.}~\bibnamefont {Omura}}, \ and\
  \bibinfo {author} {\bibfnamefont {C.}~\bibnamefont {Yu}},\ }\href {\doibase
  10.1007/JHEP03(2013)151} {\bibfield  {journal} {\bibinfo  {journal} {JHEP}\
  }\textbf {\bibinfo {volume} {03}},\ \bibinfo {pages} {151} (\bibinfo {year}
  {2013})},\ \Eprint {http://arxiv.org/abs/1212.4607} {arXiv:1212.4607
  [hep-ph]} \BibitemShut {NoStop}%
\bibitem [{\citenamefont {Crivellin}\ \emph {et~al.}(2016)\citenamefont
  {Crivellin}, \citenamefont {Heeck},\ and\ \citenamefont
  {Stoffer}}]{Crivellin:2015hha}%
  \BibitemOpen
  \bibfield  {author} {\bibinfo {author} {\bibfnamefont {A.}~\bibnamefont
  {Crivellin}}, \bibinfo {author} {\bibfnamefont {J.}~\bibnamefont {Heeck}}, \
  and\ \bibinfo {author} {\bibfnamefont {P.}~\bibnamefont {Stoffer}},\ }\href
  {\doibase 10.1103/PhysRevLett.116.081801} {\bibfield  {journal} {\bibinfo
  {journal} {Phys. Rev. Lett.}\ }\textbf {\bibinfo {volume} {116}},\ \bibinfo
  {pages} {081801} (\bibinfo {year} {2016})},\ \Eprint
  {http://arxiv.org/abs/1507.07567} {arXiv:1507.07567 [hep-ph]} \BibitemShut
  {NoStop}%
\bibitem [{\citenamefont {Dhargyal}(2016)}]{Dhargyal:2016eri}%
  \BibitemOpen
  \bibfield  {author} {\bibinfo {author} {\bibfnamefont {L.}~\bibnamefont
  {Dhargyal}},\ }\href {\doibase 10.1103/PhysRevD.93.115009} {\bibfield
  {journal} {\bibinfo  {journal} {Phys. Rev.}\ }\textbf {\bibinfo {volume}
  {D93}},\ \bibinfo {pages} {115009} (\bibinfo {year} {2016})},\ \Eprint
  {http://arxiv.org/abs/1605.02794} {arXiv:1605.02794 [hep-ph]} \BibitemShut
  {NoStop}%
\bibitem [{\citenamefont {Chen}\ and\ \citenamefont
  {Nomura}(2017)}]{Chen:2017eby}%
  \BibitemOpen
  \bibfield  {author} {\bibinfo {author} {\bibfnamefont {C.-H.}\ \bibnamefont
  {Chen}}\ and\ \bibinfo {author} {\bibfnamefont {T.}~\bibnamefont {Nomura}},\
  }\href {\doibase 10.1140/epjc/s10052-017-5198-6} {\bibfield  {journal}
  {\bibinfo  {journal} {Eur. Phys. J.}\ }\textbf {\bibinfo {volume} {C77}},\
  \bibinfo {pages} {631} (\bibinfo {year} {2017})},\ \Eprint
  {http://arxiv.org/abs/1703.03646} {arXiv:1703.03646 [hep-ph]} \BibitemShut
  {NoStop}%
\bibitem [{\citenamefont {Iguro}\ and\ \citenamefont
  {Tobe}(2017)}]{Iguro:2017ysu}%
  \BibitemOpen
  \bibfield  {author} {\bibinfo {author} {\bibfnamefont {S.}~\bibnamefont
  {Iguro}}\ and\ \bibinfo {author} {\bibfnamefont {K.}~\bibnamefont {Tobe}},\
  }\href {\doibase 10.1016/j.nuclphysb.2017.10.014} {\bibfield  {journal}
  {\bibinfo  {journal} {Nucl. Phys.}\ }\textbf {\bibinfo {volume} {B925}},\
  \bibinfo {pages} {560} (\bibinfo {year} {2017})},\ \Eprint
  {http://arxiv.org/abs/1708.06176} {arXiv:1708.06176 [hep-ph]} \BibitemShut
  {NoStop}%
\bibitem [{\citenamefont {Martinez}\ \emph {et~al.}(2018)\citenamefont
  {Martinez}, \citenamefont {Sierra},\ and\ \citenamefont
  {Valencia}}]{Martinez:2018ynq}%
  \BibitemOpen
  \bibfield  {author} {\bibinfo {author} {\bibfnamefont {R.}~\bibnamefont
  {Martinez}}, \bibinfo {author} {\bibfnamefont {C.~F.}\ \bibnamefont
  {Sierra}}, \ and\ \bibinfo {author} {\bibfnamefont {G.}~\bibnamefont
  {Valencia}},\ }\href@noop {} {\  (\bibinfo {year} {2018})},\ \Eprint
  {http://arxiv.org/abs/1805.04098} {arXiv:1805.04098 [hep-ph]} \BibitemShut
  {NoStop}%
\bibitem [{\citenamefont {Biswas}\ \emph
  {et~al.}(2018{\natexlab{c}})\citenamefont {Biswas}, \citenamefont {Ghosh},
  \citenamefont {Shaw},\ and\ \citenamefont {Patra}}]{Biswas:2018jun}%
  \BibitemOpen
  \bibfield  {author} {\bibinfo {author} {\bibfnamefont {A.}~\bibnamefont
  {Biswas}}, \bibinfo {author} {\bibfnamefont {D.~K.}\ \bibnamefont {Ghosh}},
  \bibinfo {author} {\bibfnamefont {A.}~\bibnamefont {Shaw}}, \ and\ \bibinfo
  {author} {\bibfnamefont {S.~K.}\ \bibnamefont {Patra}},\ }\href@noop {} {\
  (\bibinfo {year} {2018}{\natexlab{c}})},\ \Eprint
  {http://arxiv.org/abs/1801.03375} {arXiv:1801.03375 [hep-ph]} \BibitemShut
  {NoStop}%
\bibitem [{\citenamefont {Be\v{c}irevi\'{c}}\ \emph {et~al.}(2016)\citenamefont
  {Be\v{c}irevi\'{c}}, \citenamefont {Fajfer}, \citenamefont {Ko\v{s}nik},\
  and\ \citenamefont {Sumensari}}]{Becirevic:2016yqi}%
  \BibitemOpen
  \bibfield  {author} {\bibinfo {author} {\bibfnamefont {D.}~\bibnamefont
  {Be\v{c}irevi\'{c}}}, \bibinfo {author} {\bibfnamefont {S.}~\bibnamefont
  {Fajfer}}, \bibinfo {author} {\bibfnamefont {N.}~\bibnamefont {Ko\v{s}nik}},
  \ and\ \bibinfo {author} {\bibfnamefont {O.}~\bibnamefont {Sumensari}},\
  }\href {\doibase 10.1103/PhysRevD.94.115021} {\bibfield  {journal} {\bibinfo
  {journal} {Phys. Rev.}\ }\textbf {\bibinfo {volume} {D94}},\ \bibinfo {pages}
  {115021} (\bibinfo {year} {2016})},\ \Eprint
  {http://arxiv.org/abs/1608.08501} {arXiv:1608.08501 [hep-ph]} \BibitemShut
  {NoStop}%
\bibitem [{\citenamefont {Be\v{c}irevi\'{c}}\ \emph {et~al.}(2018)\citenamefont
  {Be\v{c}irevi\'{c}}, \citenamefont {Dor\v{s}ner}, \citenamefont {Fajfer},
  \citenamefont {Ko\v{s}nik}, \citenamefont {Faroughy},\ and\ \citenamefont
  {Sumensari}}]{Becirevic:2018afm}%
  \BibitemOpen
  \bibfield  {author} {\bibinfo {author} {\bibfnamefont {D.}~\bibnamefont
  {Be\v{c}irevi\'{c}}}, \bibinfo {author} {\bibfnamefont {I.}~\bibnamefont
  {Dor\v{s}ner}}, \bibinfo {author} {\bibfnamefont {S.}~\bibnamefont {Fajfer}},
  \bibinfo {author} {\bibfnamefont {N.}~\bibnamefont {Ko\v{s}nik}}, \bibinfo
  {author} {\bibfnamefont {D.~A.}\ \bibnamefont {Faroughy}}, \ and\ \bibinfo
  {author} {\bibfnamefont {O.}~\bibnamefont {Sumensari}},\ }\href {\doibase
  10.1103/PhysRevD.98.055003} {\bibfield  {journal} {\bibinfo  {journal} {Phys.
  Rev.}\ }\textbf {\bibinfo {volume} {D98}},\ \bibinfo {pages} {055003}
  (\bibinfo {year} {2018})},\ \Eprint {http://arxiv.org/abs/1806.05689}
  {arXiv:1806.05689 [hep-ph]} \BibitemShut {NoStop}%
\bibitem [{\citenamefont {Alonso}\ \emph {et~al.}(2014)\citenamefont {Alonso},
  \citenamefont {Jenkins}, \citenamefont {Manohar},\ and\ \citenamefont
  {Trott}}]{Alonso:2013hga}%
  \BibitemOpen
  \bibfield  {author} {\bibinfo {author} {\bibfnamefont {R.}~\bibnamefont
  {Alonso}}, \bibinfo {author} {\bibfnamefont {E.~E.}\ \bibnamefont {Jenkins}},
  \bibinfo {author} {\bibfnamefont {A.~V.}\ \bibnamefont {Manohar}}, \ and\
  \bibinfo {author} {\bibfnamefont {M.}~\bibnamefont {Trott}},\ }\href
  {\doibase 10.1007/JHEP04(2014)159} {\bibfield  {journal} {\bibinfo  {journal}
  {JHEP}\ }\textbf {\bibinfo {volume} {04}},\ \bibinfo {pages} {159} (\bibinfo
  {year} {2014})},\ \Eprint {http://arxiv.org/abs/1312.2014} {arXiv:1312.2014
  [hep-ph]} \BibitemShut {NoStop}%
\bibitem [{\citenamefont {Iguro}\ \emph {et~al.}(2019)\citenamefont {Iguro},
  \citenamefont {Kitahara}, \citenamefont {Omura}, \citenamefont {Watanabe},\
  and\ \citenamefont {Yamamoto}}]{Iguro:2018vqb}%
  \BibitemOpen
  \bibfield  {author} {\bibinfo {author} {\bibfnamefont {S.}~\bibnamefont
  {Iguro}}, \bibinfo {author} {\bibfnamefont {T.}~\bibnamefont {Kitahara}},
  \bibinfo {author} {\bibfnamefont {Y.}~\bibnamefont {Omura}}, \bibinfo
  {author} {\bibfnamefont {R.}~\bibnamefont {Watanabe}}, \ and\ \bibinfo
  {author} {\bibfnamefont {K.}~\bibnamefont {Yamamoto}},\ }\href {\doibase
  10.1007/JHEP02(2019)194} {\bibfield  {journal} {\bibinfo  {journal} {JHEP}\
  }\textbf {\bibinfo {volume} {02}},\ \bibinfo {pages} {194} (\bibinfo {year}
  {2019})},\ \Eprint {http://arxiv.org/abs/1811.08899} {arXiv:1811.08899
  [hep-ph]} \BibitemShut {NoStop}%
\bibitem [{\citenamefont {Greljo}\ \emph
  {et~al.}(2018{\natexlab{b}})\citenamefont {Greljo}, \citenamefont
  {Martin~Camalich},\ and\ \citenamefont {Ruiz-Álvarez}}]{Greljo:2018tzh}%
  \BibitemOpen
  \bibfield  {author} {\bibinfo {author} {\bibfnamefont {A.}~\bibnamefont
  {Greljo}}, \bibinfo {author} {\bibfnamefont {J.}~\bibnamefont
  {Martin~Camalich}}, \ and\ \bibinfo {author} {\bibfnamefont {J.~D.}\
  \bibnamefont {Ruiz-Álvarez}},\ }\href@noop {} {\  (\bibinfo {year}
  {2018}{\natexlab{b}})},\ \Eprint {http://arxiv.org/abs/1811.07920}
  {arXiv:1811.07920 [hep-ph]} \BibitemShut {NoStop}%
\bibitem [{\citenamefont {Greljo}(2018)}]{greljotalk}%
  \BibitemOpen
  \bibfield  {author} {\bibinfo {author} {\bibfnamefont {A.}~\bibnamefont
  {Greljo}},\ }\href {https://indico.cern.ch/event/701759} {\enquote {\bibinfo
  {title} {{$B$ anomalies vs. high-$p_T$ lepton tails}},}\ } (\bibinfo {year}
  {2018}),\ \bibinfo {note} {talk at \emph{Workshop on high-energy implications
  of flavor anomalies}, p.~23, CERN, 24 Oct 2018, {\tt
  https://indico.cern.ch/event/701759}}\BibitemShut {NoStop}%
\bibitem [{\citenamefont {Blanke}\ \emph {et~al.}(2019)\citenamefont {Blanke},
  \citenamefont {Crivellin}, \citenamefont {Kitahara}, \citenamefont {Moscati},
  \citenamefont {Nierste},\ and\ \citenamefont
  {Nišandžić}}]{Blanke:2019qrx}%
  \BibitemOpen
  \bibfield  {author} {\bibinfo {author} {\bibfnamefont {M.}~\bibnamefont
  {Blanke}}, \bibinfo {author} {\bibfnamefont {A.}~\bibnamefont {Crivellin}},
  \bibinfo {author} {\bibfnamefont {T.}~\bibnamefont {Kitahara}}, \bibinfo
  {author} {\bibfnamefont {M.}~\bibnamefont {Moscati}}, \bibinfo {author}
  {\bibfnamefont {U.}~\bibnamefont {Nierste}}, \ and\ \bibinfo {author}
  {\bibfnamefont {I.}~\bibnamefont {Nišandžić}},\ }\href@noop {} {\
  (\bibinfo {year} {2019})},\ \Eprint {http://arxiv.org/abs/1905.08253}
  {arXiv:1905.08253 [hep-ph]} \BibitemShut {NoStop}%
\end{thebibliography}%

\end{document}